\documentclass[aos,preprint]{imsart}
\usepackage[]{graphicx}
\usepackage[]{color}
\usepackage{xcolor}

\makeatletter
\def\maxwidth{ %
  \ifdim\Gin@nat@width>\linewidth
    \linewidth
  \else
    \Gin@nat@width
  \fi
}
\makeatother

\usepackage{framed}
\makeatletter
 {\par\unskip\endMakeFramed%
 \at@end@of@kframe}
\makeatother

\definecolor{shadecolor}{rgb}{.97, .97, .97}
\definecolor{messagecolor}{rgb}{0, 0, 0}
\definecolor{warningcolor}{rgb}{1, 0, 1}
\definecolor{errorcolor}{rgb}{1, 0, 0}

\usepackage{alltt} 
\usepackage[latin1]{inputenc}
\usepackage{enumitem}
\usepackage{todonotes}
\usepackage{latexsym}
\usepackage{amssymb}
\usepackage{epsfig}
\usepackage{amsmath}
\usepackage{xcolor}
\usepackage{float}
\usepackage{hyperref}
\usepackage{booktabs}
\usepackage{longtable}

\allowdisplaybreaks

\IfFileExists{upquote.sty}{\usepackage{upquote}}{}

\usepackage{algorithm} 
\usepackage{algpseudocode} 
\makeatletter
\newcommand{\algmargin}{\the\ALG@thistlm}
\makeatother
\newlength{\whilewidth}
\algnewcommand{\parState}[1]{\State%
	\parbox[t]{\dimexpr\linewidth-\algmargin}{\strut #1\strut}}
\usepackage[latin1]{inputenc}
\usepackage{enumitem}
\usepackage{todonotes}
\RequirePackage[OT1]{fontenc}
\RequirePackage{amsmath,amssymb,amsthm}
\RequirePackage[colorlinks,citecolor=blue,urlcolor=blue]{hyperref}
\usepackage{graphicx}
\usepackage[11pt]{moresize}

\startlocaldefs
\numberwithin{equation}{section}
\theoremstyle{plain}

\endlocaldefs

\theoremstyle{plain}
\long\def\comment#1{}

\theoremstyle{definition}

\numberwithin{definition}{section}

\numberwithin{remark}{section}

\renewcommand{\qed}{\hfill{\tiny \ensuremath{\blacksquare} }}%

\begin{document}

\begin{frontmatter}
\title{Credit Ratings: Heterogeneous Effect on Capital Structure}
\runtitle{Credit Ratings and Capital Structure}

\begin{aug}
\author{\fnms{Helmut} \snm{Wasserbacher*}\ead[label=e1]{}}
\and
\author{\fnms{Martin} \snm{Spindler}\ead[label=e2]{}}
\thankstext{T1}{The views and opinions expressed in this document are those of the first author and do not necessarily reflect the official policy or position of Novartis or any of its officers.}


\address{Helmut Wasserbacher\\
Novartis International AG\\
Novartis Campus\\
4002 Basel\\
Switzerland\\
E-mail: helmut.wasserbacher@novartis.com}

\address{Martin Spindler\\
University of Hamburg\\
Hamburg Business School\\
Moorweidenstr. 18\\
20148 Hamburg\\
Germany\\
E-mail: martin.spindler@uni-hamburg.de}
\end{aug}

\begin{abstract}
Why do companies choose particular capital structures? A compelling answer to this question remains elusive despite extensive research. In this article, we use double machine learning to examine the heterogeneous causal effect of credit ratings on leverage. Taking advantage of the flexibility of random forests within the double machine learning framework, we model the relationship between variables associated with leverage and credit ratings without imposing strong assumptions about their functional form. This approach also allows for data-driven variable selection from a large set of individual company characteristics, supporting valid causal inference. We report three findings: First, credit ratings causally affect the leverage ratio. Having a rating, as opposed to having none, increases leverage by approximately 7 to 9 percentage points, or 30\% to 40\% relative to the sample mean leverage. However, this result comes with an important caveat, captured in our second finding: the effect is highly heterogeneous and varies depending on the specific rating. For AAA and AA ratings, the effect is negative, reducing leverage by about 5 percentage points. For A and BBB ratings, the effect is approximately zero. From BB ratings onwards, the effect becomes positive, exceeding 10 percentage points. Third, contrary to what the second finding might imply at first glance, the change from no effect to a positive effect does not occur abruptly at the boundary between investment and speculative grade ratings. Rather, it is gradual, taking place across the granular rating notches (``+/-'') within the BBB and BB categories.

\end{abstract}



\end{frontmatter}


\section{Introduction}

The specific mix of debt and equity instruments a company uses to finance its operations represents its ``capital structure''. The ratio of debt to equity within this structure constitutes the company's financial ``leverage'' ratio.  Corporate finance theory suggests that the optimal capital structure is that which maximizes the company's market value \cite{Brealey2000, Ross2002}. Surprisingly, however, the question of why a company chooses a particular capital structure has remained the subject of extensive debate \cite{Graham2011, Frank2009} ever since Stewart Myers first highlighted the ``Capital Structure Puzzle'' in 1984 \cite{Myers1984a}. Additionally, there is no clear, unifying model for optimal leverage \cite{Swoboda1994, Barclay1999, Barclay2005, Kumar2017}. Empirical research has extensively examined firm and industry characteristics to determine which factors can explain observed leverage ratios \cite{Graham2011, Kumar2017}. In recent years, machine learning approaches have begun to complement traditional econometric methods \cite{Amini2021}, enabling researchers to apply tree-based models, which are more flexible than conventional linear frameworks. Moreover, by incorporating regularization (shrinkage) methods that perform automatic, data-driven variable selection \cite{Athey2018}, machine learning allows researchers to consider a larger set of potential explanatory factors. However, the primary focus of most machine learning methods is to maximize predictive performance, rather than uncovering causal relationships \cite{Athey2018, Taddy2022}. Causal machine learning is an emerging field that attempts to fill this gap. It is specifically concerned with uncovering causal mechanisms through the use of machine learning techniques. In this paper, we employ the double/debiased machine learning framework \cite{Chernozhukov2017, Chernozhukov2018} to investigate the causal effect of credit ratings on leverage. In applying this very recent methodology, we contribute to the literature by identifying the heterogeneity of this effect across the different credit rating levels in a complex, high-dimensional setting.\\

We structure this paper as follows. Section~\ref{CST} provides a brief overview of the predominant capital structure theories. Section \ref{CR} consists of an introduction to credit ratings. Section~\ref{LR} reviews recent publications on the use of machine learning models to predict leverage or credit ratings, as well as publications on the impact of credit ratings on leverage. Section~\ref{DML} introduces the double machine learning framework we will employ in Section~\ref{EM} to determine the causal effect of credit ratings on leverage ratios for a sample of companies from the Compustat database. Section~\ref{Conclusion} summarizes our findings, outlines the limitations of our work and suggests how these could form the basis for further research. Lastly, section \ref{APP} is the appendix, which contains further details related to several sections of the main text.

\section{Capital Structure Theories}
\label{CST}

The capital structure of a company refers to the specific mix of financial instruments it uses to finance its operations. Alongside decisions regarding investments, it constitutes a fundamental question for management \cite{Hawawini2002}: what is the source of funds for our investments? 
Capital structure analysis typically focuses on ``leverage'', which is the ratio of the total amount of debt (as a broad asset class) to total equity in the capital structure.\\

Different theories have been developed about the optimal capital structure, or the leverage that maximizes the overall market value of a company \cite{Graham2011, Kumar2017, Ross2002, Frank2009, Brealey2000, Bernstein1993}. Key theories include Modigliani and Miller's \textit{theory of irrelevance} \cite{Modigliani1958} (page 268, ``Proposition I''), the \textit{trade-off theory} \cite{Miller1977, Jensen1976, DeAngelo1980, Jensen1986, Lang1989}, the \textit{pecking order theory} \cite{Myers1984a, Ross1977, Ross2002} and the \textit{market timing theory} \cite{Baker2002, Welch2004}. Over time, many extensions and complementary perspectives have been proposed (e.g., \cite{Fischer1989} and \cite{Bertrand2003}). In the context of this article, the inclusion of credit ratings as a factor affecting a company's leverage is of particular interest. \cite{Kisgen2006} labelled this the \textit{credit rating - capital structure hypothesis}.\\


Among the theories concerning the optimal capital structure, there is no consensus, and the reasons why individual companies choose a particular capital structure remain largely unknown \cite{Swoboda1994, Barclay1999, Barclay2005, Kumar2017}. Thus, the ``Capital Structure Puzzle'' \cite{Myers1984a} remains unsolved. A second important aspect of most theories is that they do not explicitly specify the functional form through which the putative factors influence or determine capital structure. In particular, they neither postulate nor even imply a linear relationship between leverage and its determining factors.\\ 

At the same time, there is extensive empirical research on the potential determinants of observed leverage ratios, as witnessed by surveys such as \cite{Kumar2017, Graham2011} or \cite{Cantor2004}. The considered explanatory variables are mostly financial in nature and include elements from the balance sheet, income statement and cash flow statement. These variables are typically scaled by total assets or total sales and are accompanied by a rationale of what they measure \cite{Graham2011}.
However, the precise economic concepts that these measures are intended to proxy and, thus, the causal mechanisms involved are not always clear \cite{Frank2009}.\\

 Empirical studies sometimes also attempt to capture individual company attributes beyond financial characteristics. This typically involves the use of dummy variables to represent traits such as company maturity, ``uniqueness''  (often linked to sub-industry sectors) or operations in regulated industries such as utilities or railroads. More generally, dummy variables representing a company's industry at the two- to four-digit Standard Industrial Classification (SIC) code level or relying on the Fama and French industry classification \cite{Fama1997} are commonly included as covariates in empirical analyses. Examples of such approaches can be found in \cite{Frank2009, Amini2021} and \cite{Kisgen2009}.\\ 

Less frequently employed variables are those that attempt to capture concepts such as management skills \cite{Grunert2012, Matthies2013}, effective corporate governance mechanisms \cite{Chaganti1991, Zaid2020}, or the impact of the economic regime in which a company operates, such as the tax system \cite{Fonseca2020}. \cite{Kumar2017} refers to this group of explanatory factors as ``cognitive variables'' (page 115).\\

Most studies examining leverage include a subset of the aforementioned explanatory variables using variants of linear regression \cite{Kumar2017}. However, empirical research \cite{Graham2011, Amini2021} supports the view that ``the relation between leverage and many of these variables is nonlinear'' (\cite{Graham2011}, page 311) and that these nonlinearities persist even after excluding particular subgroups of companies, such as distressed firms. However, as highlighted by \cite{Graham2011} (page 337), few empirical analyses have explicitly taken account of these nonlinear dynamics. Machine learning methods, which are adaptable to complex, non-linear patterns, appear well-suited to address our research question in this environment \cite{Bishop2006}. \\

\section{Credit Ratings}
\label{CR}

\cite{SPFSL2022} (page 4) define credit ratings as ``opinions about credit risk'', that aim ``to provide investors and market participants with information about the relative credit risk of issuers and individual debt issues.'' Indeed, the main purpose of credit ratings is to reduce information asymmetries in financial markets, facilitated by credit agencies' access
to privileged information from company management.\\

The credit rating market is dominated by three big agencies: S\&P Global Ratings (formerly known as Standard \& Poor's Ratings Services), Moody's Investor Services and Fitch Ratings, with S\&P holding approximately 50\% of the market share \cite{Matthies2013}.\\

Ratings are typically assigned using a hierarchical, letter-based scale. For instance, the highest S\&P rating is denoted as ``AAA'', corresponding to ``[e]xtremely strong capacity to meet financial commitments'', while the lowest rating is denoted as ``D'', corresponding to ``[p]ayment default on a financial commitment or breach of an imputed promise; also used when a bankruptcy petition has been filed or similar action taken'' \cite{SPFSL2022} (page 9).
Further elements include ``+'' or ``-'' signs added to the rating to indicate the relative standing (``notch'') within a broad rating category, as well as the distinction between ``investment-grade'' (from AAA to BBB-) and ``speculative-grade'' (BB+ and below) ratings, ``outlooks'' for possible rating changes anticipated within six to 24 months, and ``watchlists'' for more immediate concerns (usually 90 days). Industry sources such as \cite{MIS2022, SPFSL2022, SPFSL2022a} and \cite{FitchRatings2022} provide further details on the codification of ratings.  \\


For this article, it is important to stress that the credit rating industry operates almost entirely under the ``issuer-pays model''. In this model, a company 
seeking a credit rating approaches a rating agency and pays for the service \cite{BAGHAI2014} (page 1961). This approach contrasts with the ``investor-pays model'', under which rating agencies are financed through fees charged to investors accessing the ratings; this model is now less common. Alternative models remain marginal, such as the ``public-utility model'' in China \cite{Hu2019}.
Thus, the issuance of a company rating is the result of an explicit decision by the company's management. For example, Fitch states that ``[t]he rating process usually begins when an issuer [...] contacts a member of Fitch's Business and Relationship Management (BRM) group with a request to engage Fitch to provide a credit rating'' \cite{FitchRatings2022a} (page 2). Similarly, ``Ratings request from issuer'' is the first box in S\&P's flowchart explaining ``Raising Capital Through Rated Securities'' \cite{SPFSL2022} (page 7). Put differently, companies self-select into having a rating.\footnote{At least theoretically, other situations are, of course, conceivable. For instance, a company might wish to obtain a rating, but the rating agency does not or cannot provide one (for whatever reason); or, after initial discussions about requesting a rating, a company withdraws its request. We believe, however, that such situations are limited in practice. \label{FN-RatingWithdrawal}}\\ 


The question of potential determinants of corporate credit ratings has been examined extensively in the empirical literature, with \cite{Matthies2013} providing a recent overview.
Similar to the determinants of leverage, financial ratios are widely employed in empirical research on credit ratings \cite{Huang2004, BAGHAI2014}; indeed, as \cite{Matthies2013} (page 10) asserts, ``Studies that omit these variables are almost always incomplete by definition''. Additionally, some authors indicate that corporate governance mechanisms also play a role in determining credit ratings \cite{AshbaughSkaife2006, Bhojraj2003}. Meanwhile, findings regarding the influence of macroeconomic variables on credit ratings have yielded mixed results \cite{Feng2008, BAGHAI2014}, corresponding with the assertion of credit rating agencies that they already include ``the anticipated ups and downs of the business cycle'' in their assessments \cite{SPFSL2022} (page 10).\\




A further similarity between leverage and rating studies is the predominant reliance on linear relationships between dependent and independent variables \cite{Matthies2013, Wallis2019}. \cite{Huang2004} (page 545) observe in their literature review that
the key advantage of these models is that they are ``succinct and [...] easy to explain.'' However, many authors are aware of likely non-linear effects and try to accommodate these. For instance, \cite{Amato2004} (page 2649) first transform interest coverage into a piecewise linear function and then create four distinct variables over four regions. \cite{BAGHAI2014} (page 1976) include squared and cubed versions of all explanatory variables.
Again, machine learning methods, which by design are flexible to adapt to complex, non-linear patterns, appear particularly appropriate in such scenarios \cite{Bishop2006}.\\

For further details on ratings, the references in \cite{Camanho2022, BAGHAI2014, Kemper2013, Matthies2013, White2013, DeHaan2011, Becker2011} and \cite{Partnoy2006} provide good starting points. \\

\section{Literature Review}
\label{LR}
\label{CR-CSH}

Our review of the literature focuses on the central topic of this paper, which is the \textit{causal effect} of credit ratings on capital structure. Here, we discuss the relatively few existing publications on this subject. In the appendix, we provide additional information by highlighting relevant findings from selected studies that employ machine learning methods to investigate leverage ratios or credit ratings.\\


A plausible case can be made that credit ratings influence capital structure decisions. For instance, companies sometimes mention credit rating objectives in the context of specific financing decisions. Surveys also consistently indicate that ratings are among the main factors when managers decide about leverage for their firms (e.g., \cite{Graham2001, Bancel2004}). Additionally, as early as 1936, federal regulations in the United States required banks to invest exclusively in investment-grade bonds (see \cite{White2013}, section 4, for a historical overview). It is therefore surprising that research on the determinants of capital structure took so long to consider the potential causal effect of credit ratings.\\

Given the possibility that ``in the social sciences often that is treated as important which happens to be accessible to measurement'' \cite{Hayek1975} (page 3), we hypothesize that the lack of early research on this topic was due to the initially limited availability of corporate credit rating data. First, prior to the late 1960s, credit rating agencies operated under the investor-pays model (see section \ref{CR}). Thus, credit ratings were private information purchased by investors. Only with the switch to the issuer-pays model did this information begin to be ``[distributed] to the general public at no charge'' \cite{White2013} (page 102). Second, ratings became available in popular databases substantially later than data such as balance sheet and income statement information. For instance, whereas Compustat\footnote{\url{https://www.marketplace.spglobal.com/en/datasets/compustat-financials-(8)\#dataset-overview} (accessed December 8, 2022)} was initiated and already had significant coverage as early as 1950, 1985 was, as \cite{Kisgen2006}  (page 1047) points out, the first year in which the S\&P long-term credit rating became available in Compustat. \\


Published in 2006, \cite{Kisgen2006} claims to be ``the first paper to examine the direct effect of credit ratings on capital structure decisions'' (page 1036). In particular, the author postulates the ``Credit Rating - Capital Structure Hypothesis'' (CR-CS) \cite{Kisgen2006} (page 1037), according to which credit ratings represent a material factor in capital structure decisions due to the discrete costs and benefits of different rating levels. Above all, changes in credit ratings trigger costs or benefits that influence the leverage decisions of companies according to the CR-CS.\\

The empirical test of the CR-CS theory \cite{Kisgen2006} relies on a sample of 12'336 firm-years from Compustat from 1986 to 2001. The sample is restricted to companies for which a Standard \& Poor's Long-Term Domestic Issuer Credit Rating is available. The fundamental idea of the test is to examine how managers' concerns about potential rating changes affect their decision to issue debt versus equity.
Using the presence of a plus or minus rating as a proxy for managerial concern about an impending rating change, the CR-CS theory predicts that such companies will issue relatively less debt. Indeed, \cite{Kisgen2006} finds that companies with a credit rating that includes a plus or minus (i.e., ``+'' or ``-'' notch qualification) issue approximately 0.5\% to 1\% less debt than companies with a straight rating (i.e., without a plus or minus qualification).
\\

In a subsequent paper on the relationship between ratings and capital structure, \cite{Kisgen2009} finds that companies reduce leverage
by approximately 1.5\% to 2.0\% of assets following a rating downgrade, whereas rating upgrades do not affect subsequent leverage levels. This asymmetry suggests that companies strive to achieve and maintain minimum rating levels. The hypothesized reason for this behavior is that certain ratings offer discrete benefits, such as the ability to issue commercial paper.\\


\cite{Faulkender2005} also explicitly consider the role of credit ratings in the context of capital structure decisions. However, their argument focuses on the supply side of capital, especially a company's access to the public bond market, as measured based on whether the company has a credit rating. The underlying reasoning is that a desired level of leverage might be unattainable for a company if lenders are rationing capital (see, for instance, \cite{Brealey2000}, pages 108-113).
Thus, the authors postulate a link between a company's source of capital and its leverage.
For their empirical analysis (described on pages 51-54), \cite{Faulkender2005} use Compustat data from 1986 to 2000, resulting in 77'659 firm-years and a dataset similar to that used in \cite{Kisgen2006} and \cite{Kisgen2009}.
\cite{Faulkender2005} find that the effect of having any credit rating (versus having none at all) increases a company's leverage by about 6 to 8 percentage points, corresponding to an approximate 35\% increase relative to the average leverage ratio of 22\%.
Because \cite{Faulkender2005} (page 54) use the existence of a debt rating as a proxy for access to the capital market, the authors conclude that companies with access to the public bond markets have significantly more leverage.\\

From an analytical perspective, we observe that while the vast majority of control variables in the five linear regression specifications of \cite{Faulkender2005} are statistically significant at the 1\% level, the $R^2$, even of model V, which includes 12 company control variables and a year dummy, does not exceed 37.3\%. This suggests that capital structures are difficult to predict with linear model specifications.\\


Adopting a different approach, \cite{Kisgen2019} took advantage of several changes made by the rating agency Moody's\footnote{We note that this is one of the very few papers identified in our literature search that rely on rating information from Moody's rather than S\&P Global Ratings (Standard and Poor's).} in 2006 to the calculation and reporting of leverage ratios in relation to pensions, operating leases, and hybrid securities. The author argues that because these changes were exogenous to company fundamentals, they provide a natural experiment (see for instance \cite{Rosenbaum2020}, page 75)
to determine their causal impact on capital structure and investment decisions. The findings across several analyses
support the view that changes to the rating adjustment methods
affected capital structure decisions.
\cite{Kisgen2019} therefore concludes that ``credit ratings have a significant impact on financial and real decisions of firms'' (page 581) and ``rating agencies have the power to affect corporate decisions'' (page 567).
\\


The results of the studies discussed so far suggest that credit ratings have a significant but very general effect on leverage. However, \cite{Kemper2013} provide a more nuanced view. The authors investigate the validity of the CR-CS model as proposed in \cite{Kisgen2006} and \cite{Kisgen2009} by testing four hypotheses about company-level attributes. The authors argue that for companies with these attributes, maintaining or achieving a certain rating is especially desirable. Thus, these attributes ``should proxy for management's inclination to adopt the CR-CS model'' (\cite{Kemper2013}, page 574). For instance, depending on their broad rating category, companies should behave differently because the relative costs of a change in ratings vary across categories; notably, companies on the verge of moving from investment-grade (BBB- on the S\&P rating scale) to speculative-grade (BB+ and below) ratings should be highly sensitive to the CR-CS logic due to the many negative regulatory implications of non-investment-grade status.\\

The sample period in \cite{Kemper2013} spans from 1986 to 2009, leading to a total of approximately 16'000 company-years.
Results across all four hypotheses do not support the view that credit ratings significantly affect capital structure decisions.
For instance, estimates of the effect of plus/minus ratings on leverage are generally not significant when companies are split by broad rating category or by investment- versus non-investment-grade ratings, with the sole exception of the minus-category of rating class B. \cite{Kemper2013} (page 574) infer that ``[\cite{Kisgen2006}'s] original findings appear to be driven by the subsample of firms with extremely low ratings.''
\cite{Kemper2013} conclude ``that the CR-CS model is not a good descriptor of how firms determine their marginal financing decision'' (page 594) and hypothesize that the ``marginal financing behavior [of B- rated firms] to avoid debt may be more an indication of lack of access to the debt market than an indication of a conscious attempt to decrease debt financing.''\\

In summary, while existing studies have estimated the average effect of credit ratings on leverage, the average effect may mask significant heterogeneity. Indeed, \cite{Kemper2013} has already provided preliminary evidence that such heterogeneity exists. In the present paper, we aim to go one step further and determine the presence, pattern and extent of this effect heterogeneity. To do so, we employ double machine learning, a modern machine learning approach. The next section will provide an introduction to double machine learning.\\

\section{Double Machine Learning}
\label{DML}

We have seen from the previous sections that there is no general consensus regarding the determinants of leverage and how they interact at the company level. Nevertheless, it is likely that many factors play a role and the mechanisms by which they influence capital structure are complex. Given the lack of a strong theoretical framework, isolating the causal effect of credit ratings poses a formidable challenge. Additionally, we need to consider that this effect may be heterogeneous. Double machine learning \cite{Chernozhukov2017, Chernozhukov2018, Belloni2013, Belloni2014} is a recently developed methodology that can help solve questions of causal inference in such settings by harnessing what \cite{Halevy2009} calls ``the unreasonable effectiveness of data.'' Among the key advantages of double machine learning are the following characteristics. First, there is the ability to handle high feature dimensionality, i.e., the presence of many potential influencing factors in addition to the treatment variable of interest, and to provide valid inference on treatment effects in such high-dimensional, complex data environments. Second, it employs a data-driven approach to select among these influencing factors. Third, it facilitates the use of various machine learning algorithms with flexible function-fitting capabilities. Fourth, there is double-robustness with respect to nuisance functions.\\

``Partialling-out'', ``Neyman orthogonality'' and ``cross-fitting'' are three important concepts enabling the ``doubly robust'' nature of the double machine learning approach. We will briefly discuss each of these concepts in this section and refer readers to the appendix of this paper and the literature referenced in this section for further details.

\subsection{Partialling-out}
\label{DML-PO}

Double machine learning builds on the concept of Frisch-Waugh-Lovell (FWL) ``partialling out'' \cite{Lovell2008,Chernozhukov2017a}. According to the FWL theorem, a parameter of interest $\theta$ in a linear model such as:

\begin{align}
\label{eqn:FWL-one}
Y=\theta D+\beta X +\epsilon 
\end{align}

with $\mathbb{E}(\mathcal{\epsilon} | D,X) = 0$
\\

can be estimated with linear regression using, for example, ordinary least squares taking either of two approaches. Under the first approach, $\theta$ can be directly estimated by regressing $Y$ on $D$ and $X$. Under the second approach, $\theta$ is determined in the last step of a three-step procedure: first, $Y$ is regressed on $X$, and the corresponding residuals $\epsilon_{Y}$ are determined. Second, $D$ is regressed on $X$ and again, the corresponding residuals $\epsilon_{D}$ are determined. Third, the residuals $\epsilon_{Y}$ from the first step are regressed on the residuals $\epsilon_{D}$ from the second step. The regression coefficient obtained from this third step corresponds to $\theta$, the parameter of interest.
This latter approach is employed for double machine learning with machine learning algorithms and even ensemble methods combining different machine learning methods being used for the first and second step. We underline that machine learning methods cannot be used to ``directly'' estimate equation \ref{eqn:FWL-one} as per the first approach described above. Such a ``naive approach'' \cite{Belloni2014} (page 36) entails a high risk of yielding a severely biased estimator for the treatment parameter \cite{Belloni2013, Belloni2014, Wasserbacher2022}, hence leading to invalid inference.\\

\subsection{Neyman orthogonality}
\label{DML-NO}

 Following the general outline of \cite{Bach2022}, we illustrate the approach using a ``partially linear model'' \cite{Robinson1988, Haerdle2000}, which we will also employ in our empirical analysis in section \ref{EM}. The usual form of a partially linear regression model is:
 
 \begin{align}
\label{eqn:PLR general model main equation}
Y=\theta_{0}D+g_{0}(X) +\zeta 
\end{align}

with $\mathbb{E}(\mathcal{\zeta} | D,X) = 0$
\\

and

\begin{align}
\label{eqn:PLR geneal model confounding}
D=m_{0}(X) +\mathcal{V}
\end{align}

with $\mathbb{E}(\mathcal{V} | X) = 0$,
\\

where $Y$ is the outcome variable, $D$ is the treatment (policy) variable of interest, and $X$ is a (potentially high-dimensional) vector of confounding covariates. $\mathcal{\zeta}$ and $\mathcal{V}$ are error terms. The regression coefficient $\theta_{0}$ is the parameter of interest. We can interpret $\theta_{0}$ as a causal parameter, i.e. the causal effect of treatment $D$ on outcome $Y$, provided that $D$ is ``as good as randomly assigned'' \cite{Chernozhukov2022} (page 73), conditional on the covariates X, thus rendering D exogenous conditionally on X.\\

Applying the partialling-out procedure to equations \ref{eqn:PLR general model main equation} and \ref{eqn:PLR geneal model confounding} removes the confounding effect of X. Afterwards, by regressing the residuals on each other, the regularization bias
introduced by machine learning methods with a penalty or regularization mechanism has no first-order effect on the target parameter \cite{Chernozhukov2018}.\\

Technically, a method-of-moment estimator for the parameter of interest $\theta_{0}$ is employed:

 \begin{align}
\label{eqn:MoM estimator}
\mathbb{E}[\mathcal{\psi}(W;\theta_{0}, \eta_{0})] = 0
\end{align}

where $\psi$ represents the score function, $W = (Y,D,X)$ is the set (data triplet) of outcome, treatment, and confounding variables, $\theta_{0}$ is the parameter of interest as already indicated above, and $\eta_{0}$ are nuisance functions (for instance, $g_{0}$ and $m_{0}$, which we will employ later in our empirical application).\\

For the double machine learning inference procedure, the score function $\psi(W;\theta_{0}, \eta_{0})$ from equation \ref{eqn:MoM estimator} (with $\theta_{0}$ as the unique solution) needs to satisfy the Neyman orthogonality \cite{Neyman1979, Bera2001} condition:

 \begin{align}
\label{eqn:Neymann orthogonality}
\partial_{\eta}\mathbb{E}[\mathcal{\psi}(W;\theta_{0}, \eta)]|_{\eta=\eta_0} = 0,
\end{align}

where the derivative $\partial_{\eta}$ denotes the pathwise Gateaux derivative operator. Intuitively, Neyman orthogonality in equation \ref{eqn:Neymann orthogonality} ensures that the moment condition $\psi(W;\theta_{0}, \eta_{0})$ from equation \ref{eqn:MoM estimator} is insensitive to small errors
in the estimation of the nuisance function $\eta$ (around its ``true'' full population value $\eta_{0}$). Thus, it removes the bias arising from using a machine learning-based estimator for $\eta_{0}$.\\

\subsection{Cross-fitting}
\label{DML-CF}

 A second point to consider is that machine learning methods usually rely on sample splitting to avoid bias introduced by overfitting \cite{James2013, Hastie2009}. 
 Under double machine learning, a similar data splitting methodology applies in the case of a partially linear model with two nuisance functions as described in the next section with equations \ref{eqn:PLR main equation} and \ref{eqn:PLR confounding equation}. Only one subset of the data is used to estimate the nuisance functions, which are partialled-out, while the other subset is used to estimate the parameter of interest (i.e., the treatment effect). Of course, such a limited use of the data  implies a loss of efficiency.\\
 
 To overcome this efficiency loss from data splitting, double machine learning employs a technique called ``cross-fitting'' \cite{Chernozhukov2018} (page C6). In this procedure, the roles of the two data subsets are swapped, and two estimates for the parameter of interest are obtained. Because these two estimators are approximately independent, they can simply be averaged to make use of the full data set \cite{Chernozhukov2018} (Figure 2, page C7). The cross-fitting procedure can be expanded beyond two data sets into a K-fold version to further increase robustness; \cite{Bach2022} (page 13) reports that four to five folds appear to work well in practice.\\

\subsection{Double robustness}
\label{DML-DR}

 ``Double machine learning'' is so named because it applies machine learning methods to estimate both equation \ref{eqn:PLR general model main equation} and equation \ref{eqn:PLR geneal model confounding}. However, the estimated treatment effect is also ``doubly robust'' thanks to the partialling-out procedure described previously. This robustness means that potential ``mistakes in either of the two prediction problems'' \cite{Taddy2022} (page 221) (i.e., equations \ref{eqn:PLR general model main equation} or \ref{eqn:PLR geneal model confounding}) do not invalidate the effect estimate as long as at least one equation is sufficiently well estimated. In other words, while it is necessary ``to do a good job on at least one of these two prediction problems'' \cite{Taddy2022} (page 221), it does not matter which one is more accurately modeled. Although this feature should not encourage lax model specifications, it underscores another attractive property of double machine learning, particularly when uncertainties remain about the precise model characteristics. As noted by \cite{Belloni2014} (page 34), ``Because model selection mistakes seem inevitable in realistic settings, it is important to develop inference procedures that are robust to such mistakes.''\\
 
 Finally, double machine learning exhibits a general robustness irrespective of the particular machine learning (ML) algorithm employed. \cite{Chernozhukov2018}  comment regarding their empirical results that ``the choice of the ML method used in estimating nuisance functions does not substantively change the conclusions`` (page C45). Of course, the machine learning methods need to be of sufficient quality for given task.
 Considering the broad spectrum of available machine learning models, however, this typically does not present a major hurdle, and even ensemble models are suitable \cite{Chernozhukov2018} (pages C22-C23). \\

\section{Empirical Methodology and Findings}
\label{EM}

In this section, we employ the double machine learning framework to estimate the causal effect of ratings on the leverage ratio. We first describe our study design and the data we employ. Subsequently, we report and discuss the results, including those of several robustness checks. Our main finding is the presence of heterogeneous effects of ratings on leverage.\\

\subsection{Empirical Design}
\label{EM-Design}

To estimate the effect of ratings on the leverage ratio, we employ a partially linear regression model (see for instance \cite{Haerdle2000}) of the following form:\\

\begin{align}
\label{eqn:PLR main equation}
LDA_{i,t}=\theta' D_{i,t}+g_{0}(X_{i,t}) +\zeta_{i,t} 
\end{align}

with $\mathbb{E}(\mathcal{\zeta}_{i,t} | D_{i,t},X_{i,t}) = 0$
\\

and

\begin{align}
\label{eqn:PLR confounding equation}
D_{i,t}=m_{0}(X_{i,t}) +\mathcal{V}_{i,t},
\end{align}

with $\mathbb{E}(\mathcal{V}_{i,t} | X_{i,t}) = 0$
\\

where $LDA_{i,t}$ is the outcome variable representing the leverage ratio for company $i$ in year $t$, defined as the book value of total debt (short-term and long-term) divided by the book value of total assets, $X_{i,t}$ corresponds to a vector of covariates for company $i$ at time $t$, which are assumed to be statistically associated with the outcome and treatment variables, and $\mathcal{\zeta}$ and $\mathcal{V}$ are stochastic error terms.\\

$D$ represents the ``treatment'', i.e., the policy variable of interest in our study of rating effects. Specifically, $D_{i,t}$ represents the vector of $p$ binary treatment variables for company $i$ in year $t$. To allow for heterogeneity, $D$ can contain the rating variable and, for example, interactions with other potentially relevant variables or a refined rating category coded as dummy variables (we will use such a strategy later on). In the initial setting, we consider only one treatment variable: the presence ($D=1$) or absence ($D=0$) of a rating for a given company. Subsequently, we will expand the scope of analysis to conduct simultaneous inference on different treatment variables, for example, by considering each distinct rating category (such as AAA, AA+, A, AA-) as a separate treatment. Equation \ref{eqn:PLR main equation} contains the parameter vector of interest, $\theta$, which corresponds to the $p$ causal effect measure(s) of the $p$ treatment(s) on the outcome variable - i.e., in our case, the effect of rating on leverage. This causal interpretation is valid if the treatment D is ``as good as randomly assigned'' (\cite{Chernozhukov2022} page 73) conditional on the covariates X, making D exogenous conditionally on X. In other words, the initially non-random treatment assignment can be ignored if controlling for the correct set of X \cite{Rosenbaum2020}, because the selection bias towards different treatment types ``disappears'' (\cite{Angrist2008}, page 54) in this case. Thus, the selection of the covariates X and the modeling of their relationship with the outcome and treatment variables are critical for the validity of the analysis. This is precisely where the machine learning approach proves invaluable, because it is able to perform data-driven selection from a large number of candidate covariates and can flexibly model the form of their influence on the outcome variable \cite{Taddy2022}.\\

$g_{0}$ and $m_{0}$ are two vector-valued functions that capture the relationship of the covariates $X$ with the outcome $LDA$ and the treatment $D$, respectively. These two ``learner'' functions do not need to be linear; in fact, we will use random forests \cite{Breiman2001} for our analyses because of their general strength in capturing non-linear, complex interactions and relationships, even in high dimensions and with large datasets \cite{Marsland2015}. The appropriateness of random forests specifically for empirical analyses of capital structures and ratings has been further confirmed by recent publications, which found random forests to perform better than other machine learning methods (see for instance \cite{Amini2021} and \cite{Karlsen2021} for leverage and \cite{Wallis2019} for company credit ratings).\\

While the learner functions $g_{0}$ and $m_{0}$ do not need to be linear and can be flexibly adapted by the machine learning algorithm, it is important to remind ourselves that our specification in equation \ref{eqn:PLR main equation} corresponds to a linear effect of D on the outcome. We refer interested readers to the literature on non-linear response models (e.g., \cite{Fedorov2013}).
\\

\subsection{Data}
\label{EM-Data}

We use Compustat data for North American companies for the years 2005 to 2015.\footnote{Financial data were extracted from ``Compustat Daily Updates - Fundamentals Annual'' and rating data from ``Compustat Daily Updates - Ratings'' on September 13, 2022, via Wharton Research Data Services (WRDS). \label{FNCompustat}} Similar to the extant literature,\footnote{See for instance: \cite{Faulkender2005} (page 51, page 54), \cite{Kisgen2009} (page 1329), \cite{Kisgen2006} (page 1047), \cite{Kemper2013} (page 583), \cite{Amini2021} (Table 1, page 8). We do not exclude utilities (unlike \cite{Kisgen2009}, page 1329) or winsorize data (unlike \cite{Amini2021}, Table 1, page 8).} we exclude companies from the financial and public sectors (SIC codes 6xxx and 9xxx), observations with negative shareholder equity or negative total debt, and observations involving sales or assets smaller than one million US dollars. For unreported balance sheet, income, and cash flow statement items, missing values are replaced by zero, while non-financial metrics, such as CEO/CFO SOX certification codes or the company's auditor, are explicitly coded as ``missing''. Following these criteria, we arrive at a sample of 57'832 company-year observations.\\ 

A common characteristic of the publications examining the impact of credit ratings on leverage described in our literature review in section \ref{CR-CSH} is their reliance on an a priori selection of variables deemed related to the leverage ratio or credit rating. This selection, such as in \cite{Faulkender2005} (page 57) or \cite{Kisgen2006} (page 1056), is based on capital structure theories or the results of previous research. We by no means wish to criticize this approach of relying on previous research findings, which we follow ourselves (as evidenced by our own use of random forests as learner functions in equations \ref{eqn:PLR main equation} and \ref{eqn:PLR confounding equation}). What we intend to highlight, however, is that our machine-learning approach does not require rigid a priori decisions about the inclusion or exclusion of specific variables for predicting leverage and the presence of a rating.\\

Moreover, the ability of random forests to automatically reflect complex interactions and nonlinearities implies that we only need a very basic level of researcher-driven ``feature engineering''. In fact, in our model, we only include three transformations. First, we scale balance sheet, income, and cash flow statement items by sales and total assets (e.g., PP\&E as a percentage of sales and of total assets). Second, as a measure of company size, we add the logarithm of sales and the logarithm of total assets to the variable set. Third, we transform certain data items into dummy variables, such as creating dummies for three-digit SIC codes and for the adoption of certain accounting changes.\footnote{This corresponds to the field ``ACCTCHG'' in Compustat. For instance, the adoption of the FASB accounting standard SFAS 157 effective during 2007 is coded as ``FS157''. SFAS 157 concerns measurement and disclosure principles of ``fair value'' in generally accepted accounting principles, mainly in illiquid markets. See \url{https://www.fasb.org/page/PageContent?pageId=/reference-library/superseded-standards/summary-of-statement-no-157.html&bcpath=tff} (accessed January 20, 2023).}\\

Given our deliberate use of random forests, a highly flexible method capable of learning complex interactions, we must prevent the algorithm from ``back-calculating'' total debt or equity, which are key determinants of the leverage ratio. We achieve this by removing all data items from the liabilities side of the balance sheet and excluding debt-related items from the income and cash flow statements, such as data items related to interest expenses. This clearly distinguishes the capital structure decision (liability side) from the investment decision (asset side) in alignment with the two fundamental managerial decisions discussed in section \ref{CST}.\\  

Employing this strategy, we compile a total of 1'840 features that comprise our set X of covariates. We provide a full list of the covariates in the appendix.\\

As already mentioned in subsection \ref{EM-Design}, we define the outcome variable, the leverage ratio $LDA$, as the ratio of total debt to total assets for company $i$ in year $t$:
\\

\begin{align}
\label{eqn:LDA}
LDA_{i,t}=(\textit{Long-term Debt}_{i,t} + \textit{Short-term Debt}_{i,t}) / \textit{Total Assets}_{i,t}
\end{align}

We measure $LDA$ (shorthand for ``Leverage Debt to Assets'') in terms of book values, because these reflect the actions of company managers more directly than do market values \cite{Kisgen2006}. However, we also verify the main results of our paper using a market value\footnote{Because the market value of debt is approximated by its book value in this definition, this corresponds to a ``quasi-market value measure'' \cite{Graham2001} (page 316), a concept employed by most extant research (see e.g., \cite{Amini2021, Kisgen2009, Kisgen2006, Faulkender2005}).} definition of leverage ($LDMA$), defined in line with previous research (e.g., \cite{Faulkender2005, Kisgen2009}) as: 

\begin{align}
\begin{split}
\label{eqn:LDMA}
LDMA_{i,t}=(\textit{Long-term Debt}_{i,t} + \textit{Short-term Debt}_{i,t}) / \\
(\textit{Total Assets}_{i,t} - \textit{Book Value of Equity}_{i,t} + \textit{Market Value of Equity}_{i,t})
\end{split}
\end{align}

Finally, for the treatment variable D, we use ``Standard \& Poor's Long-Term Domestic Issuer Credit Rating'' from Compustat\footnote{See footnote \ref{FNCompustat} for the date of data extraction.} to determine whether a rating is present for a given company-year and, if so, which rating it is. We acknowledge that a company may not have an S\&P rating but could instead hold ratings from other agencies such as Moody's Investor Services or Fitch Ratings. However, considering S\&P's dominant market share of approximately 50\% (see section \ref{CR}) and the fact that the majority of companies in the US have ratings from at least two leading rating agencies \cite{LIVINGSTON2010}, we see this as a minor concern for our analyses. Another concern when measuring the impact of individual rating categories is that companies may have what is called a ``split-rating'' \cite{Bowe2014}.\footnote{For instance, the pharmaceutical company Novartis reports a split rating in their presentation dated January 19, 2023, with AA- from S\&P one notch higher than A1 from Moody's \cite{Novartis2023} (page 19).} A split-rating describes a situation in which a company is rated differently by two different agencies. While this affects up to 50\% of rated companies, differences are typically at the ``notch level'', i.e. concerning the most granular plus and minus sub-levels within a given broader rating category \cite{LIVINGSTON2010}. We address this issue in our analyses by examining the causal impact of specific ratings at different levels of granularity. The corresponding results, which we report in the appendix, support the findings presented in the main paper.\\

Table \ref{table_LDA_summary} provides an overview of the outcome variable, leverage (LDA), with a split by broad rating category.\\

We observe several general facts from table \ref{table_LDA_summary}, which are broadly in line with the extant literature \cite{BAGHAI2014, Kisgen2006, Faulkender2005}. First, median and mean leverage are significantly higher for observations that have any kind of rating compared to observations that have no rating. Leverage increases as the rating category decreases, except for the particular rating classes ``SD'' (which signifies that selective default on a particular debt instrument has occurred, but it is believed that the company will honor its other obligations) and ``D'' (default).\footnote{Most authors do not include the rating categories ``SD'' and ``D''. Also, the absence of rating class ``C'' in our sample is consistent with its rarity in other empirical analyses; for instance, \cite{BAGHAI2014} (Table 1, page 1966) reports only three instances of ``C''-ratings out of a total of almost 30'000 ratings for their 1985-2009 sample.} Overall, roughly a quarter (26\%) of observations have a rating, out of which slightly more than half (14\% of 26\%) are investment-grade ratings (better than BB). \\

\newpage

\begin{longtable}{lrcccrc}
    \toprule
     \multicolumn{7}{l}{\textbf{Summary statistics for LDA (in \%) by rating category}} \\
    \midrule
Rating & 1st & Median & Mean & 3rd & Obser- & \% of  \\ 
category & quart. &  & & quart. & vations & total\\
    \midrule
AAA	&	5.0	&	12.1	&	17.8	&	20.0	&	86 & 0.1  \\  
AA	&	11.9	&	20.4	&	20.5	&	26.7	&	391 & 0.7 \\
A	&	18.4	&	26.4	&	26.9	&	34.4	&	2'458 & 4.3 \\
BBB	&	20.4	&	28.9	&	28.9	&	36.7	&	5'174  & 8.9 \\
BB	&	22.8	&	33.3	&	34.0	&	44.7	&	3'732 & 6.5 \\
B	&	33.1	&	45.0	&	44.9	&	56.9	&	2'981 & 5.2 \\
CCC	&	35.5	&	46.0	&	45.5	&	60.4	&	148 & 0.3  \\
CC	&	30.4	&	51.9	&	48.6	&	67.2	&	15 & 0.0  \\
SD	&	26.0	&	35.5	&	32.1	&	41.6	&	4 & 0.0  \\
D	&	16.7	&	31.1	&	28.8	&	41.4	&	41 & 0.1  \\
\midrule
Total ratings &	21.7	&	31.6	&	32.9	&	42.5	&	15'030 & 26.0  \\
No rating &	0.1	&	11.1	&	17.1	&	28.2	&	42'802	&	74.0  \\
\midrule
\textbf{Grand total} &	\textbf{1.8}	&	\textbf{18.5}	&	\textbf{21.2}	&	\textbf{34.1}	&	\textbf{57'832}	&	\textbf{100.0}  \\

    \midrule
    \bottomrule

\caption{Summary statistics for the outcome variable, leverage (LDA), by rating category. LDA values (total debt divided by total assets) are displayed as \%. ``1st quart.'' and ``3rd quart.'' correspond to the 25th- and 75th-percentile, respectively. ``Observations'' refers to the number of company-years over the 2005-2015 timespan. ``\% of total'' represents the share of observations of a particular rating class relative to all company-year observations. Values in this column are displayed as \%. The (single) C-rating category is absent because no firm-year had such a rating over the sample period.
}
\label{table_LDA_summary}
\end{longtable}

\subsection{Results}
\label{EM-Results}

In this section, we describe the key results from our analyses of the causal effect of credit ratings on leverage. We begin with the most fundamental question: Does having a credit rating affect the leverage ratio? Subsequently, we examine the individual effects of the 22 most granular rating categories. For the sake of conciseness, we delegate the more gradual exploration of our research question and its results to the appendix, where we first assess the difference in effect between investment-grade and speculative-grade ratings, followed by a delineation among the 10 broad rating categories. Additionally, we support our findings with robustness tests, including a second metric for leverage, different model specifications within the machine learning framework, and a different sample period.  \\

\subsubsection{Effect of having any rating versus having no rating}
\label{EM-Results-AnyRating}

For our first analysis, we estimate the causal effect on the leverage ratio of having any rating, regardless of the specific rating category, versus not having a rating.
\\

As mentioned previously, we use regression trees out of the machine learning toolbox as learners for the two functions $g_{0}$ and $m_{0}$, which capture the relationship of the covariates $X$ with the outcome, $LDA$, and the treatment, $D$, respectively. The literature sometimes refers to $g_{0}$ and $m_{0}$ as ``nuisance functions'' and their parameters as ``nuisance parameters'' \cite{Chernozhukov2018, Belloni2014} because their estimation is not the primary aim of the causal analysis (which is the estimation of the causal parameter $\theta$). We therefore limit the scope of their discussion to a brief description here and refer to the appendix for technical details. For $g_{0}$, we specify the random forest so that it has 500 trees, each with a maximum depth of seven levels, to predict $LDA$. This achieves an out-of-sample prediction accuracy of approximately 53\% for the $R^2$.
This level of accuracy is in line with existing literature (see for instance \cite{Amini2021}, Table 2, page 11 or \cite{Karlsen2021}, Table 2, page 23). For $m_{0}$, we also specify the random forest so that it has 500 trees, but with a slightly lower maximum depth of five levels each.
Out-of-sample, we achieve a correct classification rate of approximately 87\%, again in line with the literature (see for instance \cite{BAGHAI2014}, Table III, Panel A, page 1972).\\

With the learners $g_{0}$ and $m_{0}$ defined, we can apply the double machine learning framework described in section \ref{DML}
to determine the parameter of interest $\theta$. We follow the practical recommendation from \cite{Bach2022} (page 13) and use a five-fold split as well as two repetitions to arrive at aggregated parameter estimates and standard errors.
Table \ref{table_Res_General_Rating} summarizes the results. For an immediate robustness check, we also include here the results based on the alternative (quasi-market value) leverage definition (LDMA) from \ref{eqn:LDMA}; however, as motivated previously, the focus of our paper remains leverage as measured by book values (LDA). \\

\newpage

\begin{longtable}{lcc}
    \toprule
\textbf{Rating effect on leverage} & LDA & LDMA \\
    \midrule
$\theta$ (rating yes/no) & 0.0878 &	0.0655	  \\  
Std. error	&	0.0021 & 0.0020	 \\
t-value & 41.8 & 32.9 \\
p-value	&	0.000 &	0.000 \\
\midrule
\textit{Memo: mean LDA/LDMA} & \textit{0.212} & \textit{0.202} \\
\textit{Rating effect ($\theta$) vs. mean} & \textit{41\%}	& \textit{32\%} \\
    \bottomrule

\caption{Results for the estimated causal effect $\theta$ on LDA (book value leverage) and LDMA (quasi-market value leverage) of having or not having a rating. Parameter estimates and standard errors are aggregated over a five-fold split with two repetitions. Subsections \ref{EM-Design} and \ref{EM-Data} describe the analytical approach and the data sample.}
\label{table_Res_General_Rating}
\end{longtable}

Our estimate of the general rating effect summarized in table \ref{table_Res_General_Rating} is both statistically\footnote{A discussion of the relevance and validity of significance levels, including the controversy of the ``5\% p-value'' and the general topic of the ``replication crisis'', go beyond the scope of this paper. We refer interested readers to sources such as \cite{Amrhein2019, McShane2019, Ioannidis2005}. However, we underline the general relevance of this topic by highlighting that \cite{Kemper2013} (a paper we discussed in section \ref{CR-CSH}) mention that they are unable to replicate the results from \cite{Kisgen2006}. See \cite{Kemper2013} (footnote 13, page 584). \label{FN5Percent}} and economically significant. On average, having a rating increases LDA by roughly 9 percentage points (pps).
Compared to the sample average leverage ratio of approximately 21\%, this represents an increase of 41\%. Using LDMA as the outcome variable to check the robustness of the results, the effect estimate is roughly 6.5pps (32\% increase versus mean LDMA) and also highly significant. Thus, our results at this very general yes/no rating level corroborate  the finding in \cite{Faulkender2005} that firms with a credit rating have more debt. Moreover, the order of magnitude is very comparable: \cite{Faulkender2005} conduct their analysis for market leverage (LDMA) and, in fact, our own LDMA effect estimate of 32\% versus the mean is very close to the 35\% they report (page 1).\\

It is important to put our rating effect estimate of 9pps into perspective with purely descriptive statistics. Table \ref{table_LDA_summary} shows that, before adjusting for any confounding company characteristics via the double machine learning approach, the average leverage ratio for companies with a rating is nearly 16pps  (32.9\%-17.1\%) higher than that for companies without a rating. Thus, this figure would overstate the rating effect by 7pps, i.e., by close to 80\% above the value we estimated.\\

For a second robustness check, we employed different learner specifications for $g_{0}$ and $m_{0}$. The effect estimates across the different model alternatives are very consistent, as can be seen from table \ref{table_Res_General_Rating_Robust}. A detailed description of the alternative learner specifications and a discussion of the results are provided in the appendix. Here, we simply remind readers of the ``double robustness'' discussed in subsection \ref{DML-DR}: as long as one of the two nuisance functions is accurately specified within the double machine learning framework, the overall outcome for the causal parameter of interest is correct \cite{Taddy2022}.

\begin{longtable}{lccccc}
    \toprule
    \multicolumn{6}{l}{\textbf{Robustness check: alternative model specifications}} \\
  \midrule
\textbf{Rating effect} & MM & AM1 & AM2 & AM3 & AM4 \\
\textbf{on LDA} & (RF/RF) & (DML2) & (LASSO/RF) & (Ridge/RF) & (Restr.) \\
    \midrule
$\theta$ (rating yes/no) & 0.0878 &	0.0878 & 0.0925	& 0.0935 & 0.0942    \\  
Std. error	&	0.0021 & 0.0021 & 0.0023 & 0.0369 & 0.0021  \\
t-value & 41.8 & 41.8 & 40.0 & 2.5 & 45.2 \\
p-value	&	0.000 &	0.000 & 0.000 & 0.011 & 0.000  \\
\midrule
\textit{Effect ($\theta$) vs. mean} & \textit{41\%}	& \textit{41\%} & \textit{44\%} & \textit{44\%} & \textit{44\%} \\
    \bottomrule

\caption{Results for the estimated causal effect $\theta$ 
 on LDA of having or not having a rating, according to alternative model (AM) specifications. ``MM'' refers to the main model specification used throughout the paper. Please refer to the appendix for details of the specifications for the different AMs.
}
\label{table_Res_General_Rating_Robust}
\end{longtable}

\subsubsection{Effect of rating by individual, granular rating category}
\label{EM-Results-GranCat}

In the above analysis of having a rating versus having no rating, we implicitly assume that it does not matter which rating a company has: all rating types represent the same ``treatment'' for leverage; because ratings are opinions about credit risk, this implies that the type of opinion does not matter. However, it is easy to argue that different ratings, i.e. different opinions, may in reality represent different treatments, and thus, different versions of the treatment exist. Put differently, our initial analysis may suffer from incorrectly assuming that there are ``no hidden variations of treatments'' \cite{Imbens2015} (pages 10-13). This is one of the assumptions included in the ``stable unit treatment value assumption'' (SUTVA) \cite{Rubin1980}, which provides a fundamental framework for causal analysis (\cite{Holland1986, Rubin2005, Pearl2009, Morgan2015, Peters2017} or \cite{Chernozhukov2022}).\\

We therefore sequentially investigate different levels of rating granularity. The first two of these analyses are reported in the appendix: in the first, we examine whether the rating effect differs between the two very broad categories of ``investment-grade'' and ``speculative-grade'' (non-investment grade, ``junk bond''). In the second analysis, we look at whether the rating effect differs by broad rating categories as defined by one to three letters (such as AAA, AA, A, BBB). In the main body of this paper, we report the effect estimates for the most granular rating categories, i.e., those that include plus/minus notch qualifications (such as A+, A, and A-) within the broad rating categories. To avoid confusion, we label the granular sub-category ratings without a plus or minus sign as ``straight" (e.g. ``AA\textsuperscript{straight}'') and the broad categories as ``broad'' (e.g. ``AA\textsuperscript{broad}''). Taking AA as an example, ``AA+'', ``AA\textsuperscript{straight}'' and ``AA-'' ratings exist within the broad category of AA\textsuperscript{broad}.  At this level of detail, we simultaneously test 22 granular rating categories.\\

We note that moving from a single binary treatment variable to two or more treatment variables requires some technical adaptations to ensure valid statistical inference. The ``multiplicity problem'' is especially relevant in our case: the possibility of falsely identifying an effect as ``significant'' increases with the number of treatments tested. We therefore report multiplier bootstrap (MB) standard errors and p-values, as well as, for comparison, the corresponding Romano-Wolf (RoWo) and Bonferroni (Bonf) p-values to account for simultaneous inference on multiple parameters. We discuss these different methods briefly in the appendix. \\


Table \ref{table_Res_Granular_Rating_Categories} summarizes the effect estimates for the 22 granular rating categories.
For ease of direct comparison and to assess the robustness of our results, we also include the effect estimates from a less granular analysis based on the 10 ``broad'' rating categories as \textit{``memo''} in this table. As a reminder, these broad rating categories are defined by one to three letters (such as AAA, AA, A, BBB) without considering the plus and minus qualifications. Details from this broad analysis are included in the appendix (see table \ref{table_Res_Broad_Rating_Categories}).\\

We first compare results for the four rating categories without notch qualifications. For AAA, the effect estimates from the granular versus the broad analysis differ only at the fourth decimal place. The p-values are very similar, too, albeit slightly higher for AAA\textsuperscript{straight} in the granular analysis versus AAA\textsuperscript{broad} in the broad analysis. Intuitively, this should be expected because we are simultaneously testing 22 treatment variables in the granular versus only 10 in the broad analysis; thus, the risk of falsely identifying a treatment effect as ``non-zero'' increases. The same characteristic generally holds for the p-values of the other three rating categories without notch qualifications (only the Romano-Wolf p-value for D\textsuperscript{broad} is slightly higher than for D\textsuperscript{straight}). For CC, the effect estimates differ by approximately 1pp. 
We consider this small difference to be reassuring. We again find very consistent p-values, supporting in both cases the conclusion of a non-zero treatment effect. For SD, the effect estimates are also very consistent. They differ by 1pp and are both close to zero, with all p-values consistently indicating that the null hypothesis should not be rejected. Finally, the effect estimates for D differ by 1.5pps; however, both estimates are again very close to zero and the p-values consistently indicate that the null hypothesis of no treatment effect should not be rejected. In summary, we interpret the very consistent results for these four rating categories without notch qualifications as evidence supporting the robustness of our approach.\\

Next, we turn to the other six categories with notch qualifications. The pattern of the estimated rating effect within AA follows the rating scale, with AA+ displaying the largest (negative) effect. The coefficient estimate is also consistently negative for all three granular AA-ratings. For AA+ and AA\textsuperscript{straight}, the p-values are highly significant. However, AA- has the smallest (negative) coefficient estimate and high p-values, indicating that the null hypothesis of no rating effect should not be directly rejected (MB is still slightly below 0.05, while RoWo at 0.24 and Bonferroni at 0.91 are clearly above 0.05). This situation is consistent when considering the next rating category, A+. A+ has a smaller (albeit still negative) effect estimate coupled with higher p-values compared to AA-. The ``disappearing'' of a clear rating effect as one moves from AA\textsuperscript{straight} to the subsequent rating categories AA- and A+ thus appears to be gradual. Considering the p-values for A+, the null hypothesis of no effect should definitely not be rejected for A+.\\

Similar to the transition from AA- to A+, the results for A- in conjunction with those for BBB+ are consistent with the view that there is a gradual, smooth change in effect across these granular rating categories. BBB+ also has an effect estimate of approximately -1pp with p-values that are very similar to those for A-. Within the broad BBB rating category, we observe the same ``concave'' treatment heterogeneity as in the broad A category: the BBB+ and BBB- coefficient estimates are negative, while the one for BBB\textsuperscript{broad} in the middle of the category is positive, or more precisely, very close to zero with p-values that suggest non-rejection of the null hypothesis.
A possible ad-hoc interpretation for this phenomenon is that BBB+ companies may try to achieve at least A- status to benefit from the ``better letter''.\footnote{For instance, the European Banking Authority (EBA) maintains mapping tables that match ratings to certain rules and requirements, such as regulatory capital rules. In this context, AAA and AA are within the same ``credit quality steps'' (1), whereas A (2), BBB (3), BB (4), and B (5) are each in different step categories. From CCC downward, no distinction applies, and all categories are summarized in credit quality step 6. Thus, A and BBB ratings have different implications in this context, even though both are investment-grade ratings \cite{EBA2015}.\label{FNEBA}} On the other side of the spectrum, companies rated BBB- and thus at risk of a downgrade from BBB to BB may preemptively take actions that also lead to lower leverage.
The distinction between BBB and BB is particularly important because this represents the dividing line between investment- and speculative-grade ratings with the corresponding economic implications (see section \ref{CR}). Nevertheless, we strongly caution against overinterpreting these findings and qualify our ad-hoc interpretation as speculative. While it is true that both the A and BBB categories display concavity, their neighboring categories AA, BB and B do not. This is in line with the findings from \cite{Kemper2013} discussed in the literature review in section \ref{CR-CSH}, which show that the general effects attributed to plus/minus ratings stem from specific sub-samples of low-rated firms. In particular, \cite{Kemper2013} find these effects only in the B category  - a category in which we do not find these effects. Thus, the concept of a general plus/minus rating effect remains doubtful.\\

Within the BB and B rating categories, the pattern of the estimated effects follows the rating scale, with BB+ displaying the smallest (positive) effect and B- the largest. The coefficient estimates are consistently positive for all granular ratings within these two categories, and the p-values are generally highly significant; even for BB+, the multiplier bootstrap p-value is below 0.01. Here, we highlight two points. First, the granular analysis refines our understanding of the boundary regarding rating impact. While the analysis based on 10 broad rating categories (reported in the appendix) identifies the first inflection point of the treatment effect between BBB\textsuperscript{broad} and BB\textsuperscript{broad}, marking the transition from investment- to speculative-grade ratings, the granular analysis within the BB\textsuperscript{broad}-category reveals a more gradual rise; starting at approximately 1\% for BB+, moving to 2\% for BB\textsuperscript{straight}, and peaking at 6\% for BB-. Second, and in contrast to what we have just described for the BB rating category, the effect estimates within the B rating category are all consistently above 10\%. In particular, the increase from BB- to B+ is immediate and not gradual. Taking these two observations together, we hypothesize that companies that are still very close to an investment-grade rating have lower leverage, potentially in anticipation of regaining investment-grade status. Those that are much farther away and thus probably not anticipating an upgrade have markedly higher leverage. The fact that the differences in rating effects between the categories B and CCC are small provides additional support for this hypothesis (approximately 13\% for both B\textsuperscript{broad} and CCC\textsuperscript{broad}).\\ 

Finally, in the CCC category, the rating effects are similar for all notch categories, 
leading to a highly significant overall CCC\textsuperscript{broad} effect of close to 13\%. Of note, the sample size for company-years in this category is limited with only 148 observations, of which the majority (107) are concentrated in CCC+.\\

\newpage

\begin{longtable}{lcccccrr}

 \toprule
    \multicolumn{8}{l}{\textbf{Granular rating categories}}  \\
  \midrule
\textbf{Rating effect} & Coef. & MB & MB & RoWo & Bonf & Obser- & \% of \\
\textbf{(on LDA)} & estim. & Std. error & p-val. & p-val. & p-val. & vations & total\\
    \midrule
\endfirsthead

 \toprule
    \multicolumn{6}{l}{\textbf{Granular rating categories}}  \\
  \midrule
\textbf{Rating effect} & Coef. & MB & MB & RoWo & Bonf & Obser- & \% of\\
\textbf{(on LDA)} & estim. & Std. error & p-val. & p-val. & p-val. & vations & total \\
    \midrule
\endhead

\multicolumn{6}{r}{\textit{Continued on next page}}\\ \hline
\endfoot

\hline \hline
\endlastfoot

$\theta^{AAA \ straight}$ & -0.0588 &	0.0191 & 0.002	& 0.020 & 0.046 & 86 & 0.1    \\ 
\textit{memo:} $\theta^{AAA \ broad}$ & -0.0582 &	0.0189 & 0.002	& 0.015 & 0.021 & 86 & 0.1\\    
    \midrule
$\theta^{AA+}$ & -0.0683 &	0.0183 & 0.000	& 0.003 & 0.004 & 26 & 0.0 \\
$\theta^{AA \ straight}$ & -0.0547 &	0.0098 & 0.000	& 0.000 & 0.000 & 151 & 0.3  \\
$\theta^{AA-}$ & -0.0169 &	0.0083 & 0.041	& 0.243 & 0.911 & 214 & 0.4  \\
    \midrule
\textit{memo:} $\theta^{AA \ broad}$ & -0.0385 &	0.0068 & 0.000	& 0.000 & 0.000 & 391 & 0.7   \\  
    \midrule
$\theta^{A+}$ & -0.0060 & 0.0051 & 0.242	& 0.663 & 1.000 & 433 & 0.7   \\  
$\theta^{A \ straight}$ & 0.0086 &	0.0041 & 0.035	& 0.244 & 0.761 & 906 & 1.6    \\  
$\theta^{A-}$ & -0.0115 &	0.0033 & 0.000	& 0.007 & 0.009 & 1'119 & 1.9   \\  
    \midrule
\textit{memo:} $\theta^{A \ broad}$ & 0.0001 &	0.0027 & 0.956	& 0.950 & 1.000 & 2'458 & 4.3   \\  
    \midrule
$\theta^{BBB+}$ & -0.0106 &	0.0028 & 0.000	& 0.002 & 0.003 & 1'589 & 2.7    \\ 
$\theta^{BBB \ straight}$ & 0.0019 & 0.0026 & 0.463	& 0.759 & 1.000 & 2'105 & 3.6     \\ 
$\theta^{BBB-}$ & -0.0105 &	0.0031 & 0.001	& 0.010 & 0.016 & 1'480 & 2.6   \\ 
    \midrule
\textit{memo:} $\theta^{BBB \ broad}$ & -0.0009 &	0.0021 & 0.677	& 0.942 & 1.000 & 5'174 & 8.9    \\  
    \midrule
$\theta^{BB+}$ & 0.0110 &	0.0042 & 0.009	& 0.063 & 0.191 & 946 & 1.6   \\ 
$\theta^{BB \ straight}$ & 0.0235 &	0.0036 & 0.000	& 0.000 & 0.000 & 1'248 & 2.2    \\ 
$\theta^{BB-}$ & 0.0568 &	0.0036 & 0.000	& 0.000 & 0.000 & 1'538 & 2.7   \\ 
    \midrule
\textit{memo:} $\theta^{BB \ broad}$ & 0.0512 &	0.0024 & 0.000	& 0.000 & 0.000 & 3'732 & 6.5     \\ 
    \midrule
$\theta^{B+}$ & 0.1010 &	0.0041 & 0.000	& 0.000 & 0.000 & 1'413 & 2.5   \\ 
$\theta^{B \ straight}$ & 0.1069 &	0.0048 & 0.000	& 0.000 & 0.000 & 1'124 & 1.9    \\ 
$\theta^{B-}$ & 0.1128 &	0.0079 & 0.000	& 0.000 & 0.000 & 444 & 0.8   \\ 
    \midrule
\textit{memo:} $\theta^{B \ broad}$ & 0.1301 &	0.0031 & 0.000	& 0.000 & 0.000 & 2'981 & 5.2    \\
    \midrule
$\theta^{CCC+}$ & 0.1034 &	0.0160 & 0.000	& 0.000 & 0.000 & 107 & 0.2   \\ 
$\theta^{CCC \ straight}$ & 0.1497 &	0.0345 & 0.000	& 0.000 & 0.000 & 32 & 0.1   \\ 
$\theta^{CCC-}$ & 0.1108 &	0.0595 & 0.062	& 0.269 & 1.000 & 9 & 0.0   \\ 
   \midrule
\textit{memo:} $\theta^{CCC \ broad}$ & 0.1284 &	0.0144 & 0.000	& 0.000 & 0.000 & 148 & 0.3    \\ 
    \midrule
$\theta^{CC \ straight}$ & 0.1367 &	0.0437 & 0.002	& 0.020 & 0.040 & 15 & 0.0    \\ 
\textit{memo:} $\theta^{CC \ broad}$ & 0.1471 &	0.0044 & 0.001	& 0.004 & 0.008 & 15 & 0.0  \\ 
    \midrule
$\theta^{SD \ straight}$ & 0.0482 &	0.0553 & 0.383	& 0.759 & 1.000 & 4 & 0.0    \\ 
\textit{memo:} $\theta^{SD \ broad}$ & 0.0597 &	0.0531 & 0.261	& 0.689 & 1.000 & 4 & 0.0    \\
    \midrule
$\theta^{D \ straight}$ & 0.0291 &	0.0289 & 0.920	& 0.917 & 1.000 & 41 & 0.1    \\ 
\textit{memo:} $\theta^{D \ broad}$ & 0.0141 &	0.0294 & 0.632	& 0.942 & 1.000 & 41 & 0.1    \\ 
\midrule

Total ratings & - &	- & -	& - & - & 15'030 & 26.0    \\ 
No rating & - &	- & -	& - & - & 42'802 & 74.0    \\
\midrule
\textbf{Grand total} & - &	- & -	& - & - & 57'832 & 100.0    \\

    \bottomrule

\caption{Estimated causal effects on leverage by granular rating category (i.e., split by the plus and minus notch qualifications within a broad category) versus the baseline of having no rating. ``Straight'' indicates the categories in between the plus and minus notches; \textit{pro memoria (``memo:'')} and for ease of comparison, effect estimates from the broad rating analysis summarized in table \ref{table_Res_Broad_Rating_Categories} in the appendix have been added in this table as ``broad''. The rating categories ``AAA'', ``CC'', ``SD'' and ```D'' do not have notch qualifications. The rating category C is absent because no firm-year had such a rating over the sample period.
The empirical design (\ref{EM-Design}), data (\ref{EM-Data}), and random forest characteristics (\ref{EM-Results-AnyRating}) are described in the main text. Standard errors and corresponding p-values are corrected for simultaneous multiple inference: ``MB'' refers to the multiplier bootstrapping method, ``RoWo'' to the Romano-Wolf procedure, and ``Bonf'' to the Bonferroni-correction. ``Observations'' refers to the number of company-years from 2005 to 2015. ``\% of total'' represents the share of observations of each rating category relative to all company-year observations. Values in this column are displayed as \%.}
\label{table_Res_Granular_Rating_Categories}
\end{longtable}

\begin{figure}[htbp]
	\centering
		\includegraphics[width=1.00\textwidth]{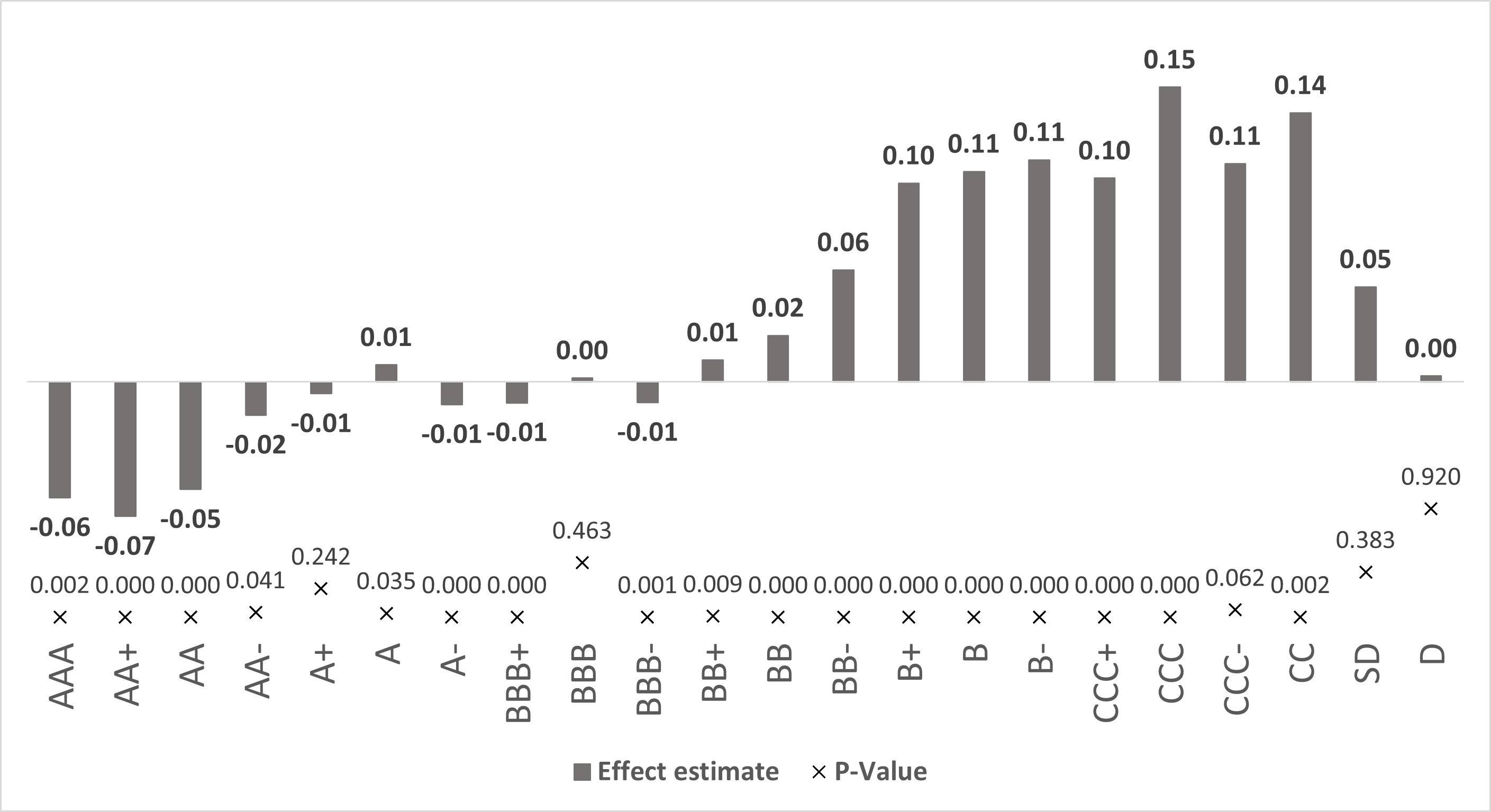}
	\caption{Graphical representation of table  \ref{table_Res_Granular_Rating_Categories} illustrating the heterogeneity of the treatment effect estimates for the 22 granular rating categories (gray bars). The numbers have been rounded to two decimal places. For instance, -0.07 for AA+ corresponds to -0.0683 in table \ref{table_Res_Granular_Rating_Categories} and indicates that the leverage for the granular rating category AA+ is roughly 7pps lower. The values next to the black crosses indicate the respective multiplier bootstrap (MB) p-values (rounded to three decimal places). The position of the black crosses has been selected to provide an intuition about the magnitude of the p-values. Note that for ease of reading, we have not added "straight" to the rating category labels that do not have a plus/minus notch qualification.}
	\label{fig:GranularRatingChart}
\end{figure}

Having presented the individual effect estimates for the granular ratings, we now conclude our analysis by taking a holistic view of the pattern of the effects across the full spectrum of 22 granular ratings. Figure \ref{fig:GranularRatingChart} provides this visual summary.
Initially, for the highest rating categories, the effect estimates are negative, ranging from -5\% to -7\%. Values for the middle rating categories hover around what can be seen as neutral to small effects, from -2\% to +2\%. Finally, from BB- onward, the effect estimates become significantly positive and reach double-digit values throughout the B+ to CC categories. CCC- stands out somewhat because its effect is smaller than those of CCC\textsuperscript{straight} and CC\textsuperscript{straight} and its MB p-value reaches 0.062. However, there are only nine company-year observations in this particular class.  \\

In summary, our analysis of granular rating categories yields three important insights. First, treatment effects are heterogeneous across the rating spectrum. Second, they follow a distinct pattern along the rating scale, with initially negative effects on leverage for the highest rating categories, no or very limited effects for the middle rating categories and large positive effects towards the lower end of the rating scale. For the two default categories at the very end of the rating scale, the effect vanishes. Third, the transition from no/very limited effects to clearly positive effects does not precisely coincide with the boundary between investment- and speculative-grade ratings, as the results from the broad analysis would suggest. Rather, it occurs gradually over the granular ratings within the two categories BBB and BB, which represent the boundary between investment- and speculative-grade ratings. \\

Before closing the empirical section of our article, we report highly summarized results from two further robustness checks. The appendix contains full details for each of these analyses. First, we apply our analytical approach to a data sample from a different time frame.
Second, we partially loosen the restriction of interest-related expense categories from the income statement. We had initially excluded these items to prevent the random forest learners from ``back-calculating'' the leverage ratio. Because interest coverage is believed to be a decisive factor for credit ratings (\cite{Koller2020}, pages 645-650), we will include this metric as a covariate.\\

\subsubsection{Rating effects in a different sample period}
\label{EM-Results-Different-Sample}

For this robustness check, we consider a second data sample from a different time period. Employing double machine learning with the same analytical methodology and data sources described in subsections \ref{EM-Design}, \ref{EM-Data}, and \ref{EM-Results-AnyRating}, we use data from the years 2000 to 2004 to arrive at a sample of 32'162 company-year observations.\\

Table \ref{table_Res_General_Rating_11vs5years} compares the results of our main analysis (as per table \ref{table_Res_General_Rating}) in the left column with the results from the second sample period in the right column. The rating effect estimate amounts to 9.6pps, which is 0.8pps higher than the parameter estimate of 8.8pps from the main sample. Compared to the mean leverage of the sample, this corresponds to an impact of 43\% versus 41\% from the main sample. Again, the rating effect is highly significant, both statistically and economically. We interpret this result as further evidence in support of the presence of a rating effect.\\

\begin{figure}[htbp]
	\centering
		\includegraphics[width=1.00\textwidth]{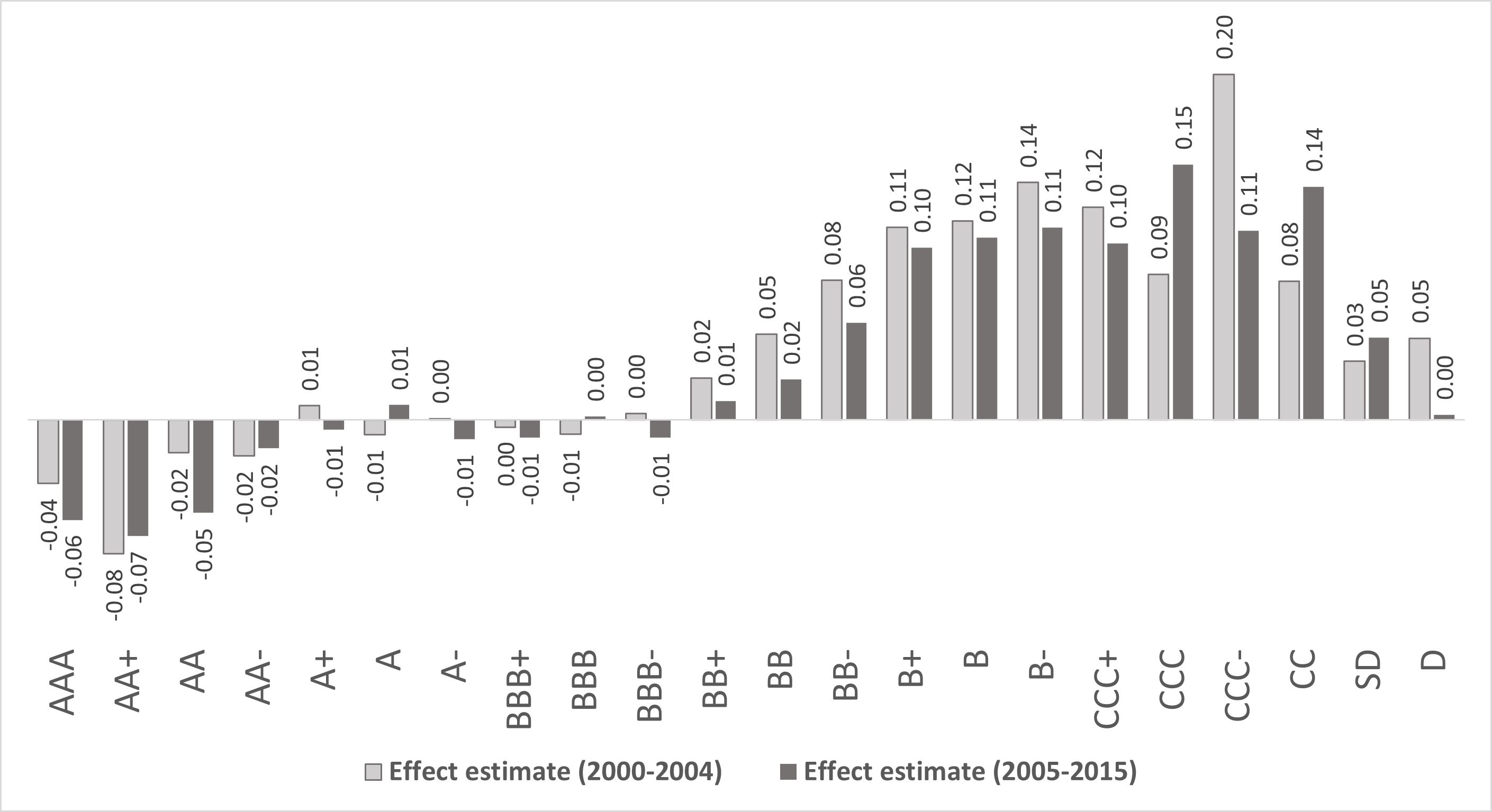}
	\caption{Graphical comparison of the results for the 2005-2015 period from the main analysis (dark gray bars) in this paper with the results from the 2000-2004 period (light gray bars) used as a robustness check for the ``effect shape'' of the 22 granular rating categories. Effect estimates have been rounded to two decimal places. Values below the x-axis indicate negative values (e.g., -0.0043 displayed as 0.00 for BBB+ in the 2000-2004 sample). For ease of reading, the chart does not repeat the  respective multiplier bootstrap (MB) p-values (already displayed in previous charts). Moreover, we have not added ``straight'' to the rating category labels that do not have a plus/minus notch qualification.}
	\label{fig:CompGranularRatingChartElevenFiveYears}
\end{figure}

Similarly, the results from the second sample period support the results from the main analysis for the broadest down to the most granular rating categories. Full details are provided in the appendix. We restrict ourselves here in the main text to plotting the effect estimates from the main analysis next to those from the second data sample for the granular rating categories (figure  \ref{fig:CompGranularRatingChartElevenFiveYears}). The similarity of the shapes from the main and the second data sample is compelling.\\

\begin{longtable}{lcc}
    \toprule
\textbf{Rating effect} & 2005-2015 & 2000-2004 \\
\textbf{on leverage (LDA)} & n=57'832 & n=32'162 \\
    \midrule
$\theta$ (rating yes/no) & 0.0878 &	0.0962	  \\  
Std. error	&	0.0021 & 0.0029	 \\
t-value & 41.8 & 32.9 \\
p-value	&	0.000 &	0.000 \\
\midrule
\textit{Memo: mean leverage} & \textit{0.212} & \textit{0.224} \\
\textit{Rating effect ($\theta$) vs. mean} & \textit{41\%}	& \textit{43\%} \\
    \bottomrule

\caption{Comparison of results for the estimated causal effect $\theta$ on leverage (LDA) of having or not having a rating for the main data sample from 2005 to 2015 with 57'832 company-year observations compared to a second, different data sample for the years 2000 to 2004 with 32'162 company-year observations. The methodology for the second data sample is the same as for the main one (as described in previous sections), including aggregation of parameter estimates and standard errors over a five-fold split with two repetitions.}
\label{table_Res_General_Rating_11vs5years}
\end{longtable}

\subsubsection{Rating effects when including interest coverage as a covariate}
\label{EM-Results-IntCov}

As described in subsection \ref{EM-Data}, we excluded data items that would allow the random forest to back-calculate total debt or equity. However, we still want to verify that the rating effect estimates hold when including selected items that determine credit ratings (or are at least strongly believed to do so). \cite{Koller2020} (pages 645-650) explain that ``credit ratings are primarily related to two financial indicators'' (page 647). One of these is size, which we have already included via items such as the logarithm of sales, the logarithm of assets or the number of employees in our set of covariates. The second is coverage, which measures ``a company's ability to comply with its debt service obligations'' (page 648). We therefore include interest coverage ($IntCov$) as defined in \cite{Koller2020} (Exhibit 33.8, left panel, page 649):

\begin{align}
\label{eqn:IntCov}
IntCov_{i,t}=\textit{EBITDA}_{i,t}  / \textit{Interest expenses}_{i,t}
\end{align}

where $EBITDA$ represents earnings before interest, taxes, depreciation and amortization and $interest$ $expenses$ represents the expenses for servicing a company's total financial debt.\footnote{In Compustat, this is the item with code ``xint'' (``Interest and Related Expense - Total'').}\\

We use our empirical sample as described in subsection \ref{EM-Data} and remove company-years with interest expenses of less than USD ten thousand in a given year and arrive at 48'585 company-year observations. We make no change to the double machine learning model described in subsections \ref{EM-Design} and \ref{EM-Results-AnyRating}. Table \ref{table_Res_General_Rating_IntCov} compares the results for the estimate of the general rating effect from our main analysis (as per table \ref{table_Res_General_Rating}) with those from the approach in this subsection, which includes interest coverage (``IntCov'') as a feature in the set of covariates. The effect estimate amounts to approximately 7pps including IntCov, or 29\% versus the sample mean leverage of roughly 25\%. This effect estimate is 1.5pps lower than the one from the main analysis, which translates into a drop of 10pps in the relative effect magnitude versus the mean leverage (29\% versus 41\% in the main analysis). Nevertheless, the rating effect remains clearly present.\\

\begin{longtable}{lcc}
    \toprule
\textbf{Rating effect} & Excl. IntCov & Incl. IntCov \\
\textbf{on leverage (LDA)} & n=57'832 & n=48'585 \\
    \midrule
$\theta$ (rating yes/no) & 0.0878 &	0.0731	  \\  
Std. error	&	0.0021 & 0.0021	 \\
t-value & 41.8 & 35.3 \\
p-value	&	0.000 &	0.000 \\
\midrule
\textit{Memo: mean leverage} & \textit{0.212} & \textit{0.249} \\
\textit{Rating effect ($\theta$) vs. mean} & \textit{41\%}	& \textit{29\%} \\
    \bottomrule

\caption{Comparison of results for the estimated causal effect $\theta$ on leverage (LDA) of having or not having a rating, depending on whether interest coverage (``IntCov'') as defined in equation \ref{eqn:IntCov} is excluded or included in the set X of covariates as per equations \ref{eqn:PLR main equation} and \ref{eqn:PLR confounding equation}. The general methodology for ``Incl. IntCov'' is the same as for the main model used throughout this paper (``Excl. IntCov'', as described in previous sections), including aggregation of parameter estimates and standard errors over a five-fold split with two repetitions.}
\label{table_Res_General_Rating_IntCov}
\end{longtable}

\begin{figure}[htbp]
	\centering
		\includegraphics[width=1.00\textwidth]{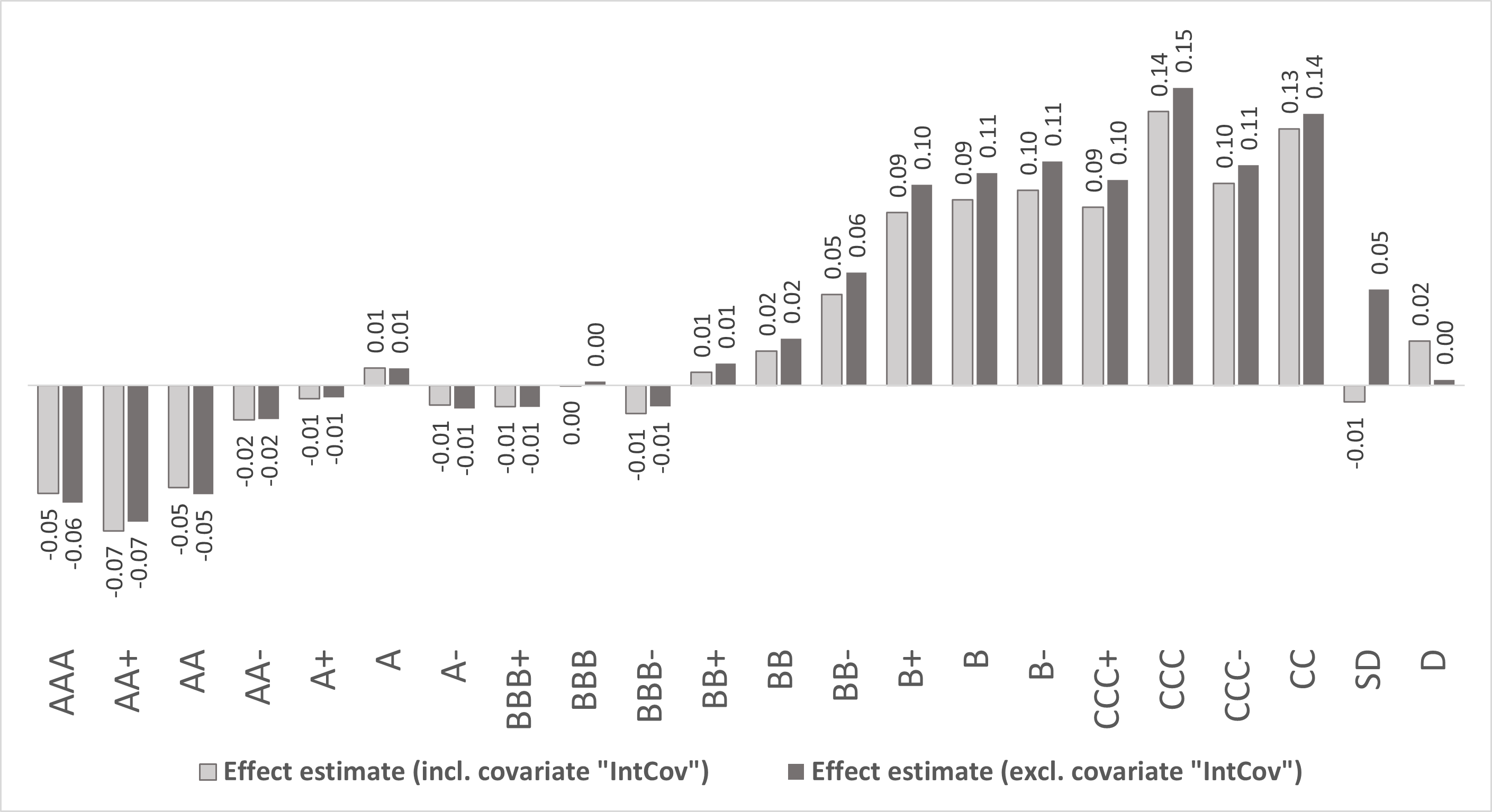}
	\caption{Graphical comparison for the effect estimate of the 22 granular rating categories on leverage (LDA) including ``IntCov'' as a covariate feature (light gray bars) versus the results from the main analyses of this paper, in which ``IntCov'' was not included (dark gray bars). Effect estimates have been rounded to two decimal places. Values below (above) the x-axis indicate a negative (positive) effect (e.g., for BBB incl. IntCov, the effect estimate is -0.0002, while it is +0.0019 for BBB excl. IntCov; both are displayed as 0.00 in the figure). Note that for ease of reading, we have not added ``broad'' to the rating category labels.}
	\label{fig:IntCovGranularComparison}
\end{figure}

Similarly, the results from the analyses including interest coverage support the results from the main analysis for the broadest down to the most granular rating categories. Full details are provided in the appendix. We provide here in the main section the graphical comparison of results for the granular rating categories in figure \ref{fig:IntCovGranularComparison}; the figure displaying the effect estimates together with the corresponding multiplier bootstrap p-values is located in the appendix. As can be seen, the results including interest coverage are very similar to those from the main analysis without interest coverage, both in terms of magnitude and overall shape. In particular, we also see the gradual rise in effect size over the notch-ratings within the BBB and BB rating classes, which supports our previous finding that there is no abrupt divide in effect between investment-grade and speculative-grade ratings. \\

In summary, our robustness checks reinforce our three main conclusions regarding the effects of credit ratings on leverage: first, ratings affect the leverage ratio. Second, this effect is heterogeneous and depends on the rating category. Third, the change in effect size is gradual across the individual, granular categories within BBB and BB and thus does not occur abruptly at the boundary between investment- and speculative-grade ratings. \\

\section{Conclusion}
\label{Conclusion}

To date, the literature has not provided a definitive explanation for why individual companies choose particular capital structures. In the absence of a consensus model and considering the large number of potential influencing factors, we employed double machine learning to investigate the causal effect of credit ratings on leverage. This approach allowed us to use random forests, which are highly flexible models capable of discerning the relationship between company characteristics, leverage, and ratings from the data, without the need for assuming linear relationships, pre-selecting a limited set of variables, or undertaking extensive feature engineering. We were able to perform valid inference and to estimate the heterogeneity of the treatment effect using machine learning methods that, without double machine learning, would have led to bias in the estimated coefficients. As a result, we were able to document for our empirical sample three important facts about the effect of ratings on leverage.\\

First, ratings have a causal effect on the leverage ratio. Holding all else equal, having a rating increases the book leverage ratio by approximately 7 to 9 percentage points, or roughly 30\% to 40\% compared to the mean leverage ratio of our sample. However, this effect exhibits a significant degree of heterogeneity, captured in our second finding. To use colorful language, consider a cocktail bar where drinks have an average alcohol content of 9\%; one could remain completely sober or get completely drunk from just one drink depending on what one is served. Applying this analogy to our context, the impact of ratings on leverage varies significantly across different rating categories. For the two highest categories, AAA and AA, the rating effect is negative, leading to lower leverage. For the next two categories, A and BBB, the effect is approximately zero. However, beginning with BB, the effect turns distinctly positive, leading to higher leverage, and then stays high or increases even further for the last three non-default categories B, CCC, and CC. Third, and in contrast to what the second point would seem to suggest at first glance, the transition in the direction of the effect is gradual over the individual, granular categories within the broad categories of BBB and BB, and especially over straight BBB, BBB-, BB+, and straight BB. Thus, the shift from no effect to a positive effect does not occur abruptly at the boundary between investment- and speculative-grade ratings. \\

Several different robustness tests corroborate these findings. Nevertheless, as with most empirical research, our work has a number of limitations, each of which we see as a potential area for complementary and future study. First, our empirical data could be enriched in various ways. Obvious dimensions would be longer or different time periods and additional covariates, including metrics related to environmental, social, and governance (ESG) criteria, or company officer characteristics. Second, an interesting next step would be to understand the mechanisms underlying the relationship between different ratings and different capital structures: Why do different ratings lead to different leverage ratios? Third, expectations about the future often play an important role in economics. As noted by \cite{Mintzberg2009} (page 123): ``[I]n business, there is usually no before, during, or after.'' Thus, it is highly likely that ratings are influenced by expectations about company characteristics, including the leverage ratio itself. Disentangling the effect of expectations on the relationship between ratings and leverage represents another formidable research challenge.\\

\cite{Halevy2009} (page 8) advise that we sometimes need to refrain from devising ``extremely elegant theories, and instead embrace complexity and make use of the best ally we have: the unreasonable effectiveness of data.'' We hope that our paper, with double machine learning and data as our allies, has illuminated the heterogeneous effects of credit ratings on leverage. \\

\newpage

\section{Appendix}
\label{APP}

In this appendix, we provide additional information pertaining to the main part of our paper. We begin with an addition to the literature review from the main section and discuss leverage and rating studies that used machine learning methods. We then provide an overview of double machine learning which is more extended than the one in the main paper. For the empirical part, we provide the full list of covariates used in our analyses, the specifications of the learner functions for the main model as well as the specifications for the alternative models used for robustness checks. The appendix also contains the analyses of the rating effect on leverage at two levels of granularity not reported for sake of brevity in the main part of the paper. First, the rating effect split by investment- and speculative-grade rating and second, the effect by individual broad rating category. Finally, we present here also the details of two further robustness checks: results from a different sample period and results including interest coverage as an additional covariate.\\

\subsection{Literature review: machine learning-based leverage and rating studies}
\label{APP-LR-MLforLR}

As indicated in sections \ref{CST} and \ref{CR} in the main text, machine learning methods are very flexible to adapt to complex, non-linear patterns. Additionally, they can handle data that do not follow well-behaved distributions (such as the normal or at least symmetrical distributions) and can cope with multicollinearity between covariates \cite{Bishop2006}.\\

In the context of forecasting credit ratings, machine learning first appeared in the computer science literature (e.g., \cite{Garavaglia1991} and \cite{Kwon1997} with neural networks or \cite{Huang2004} with support vector machines) rather than in economic research publications \cite{Matthies2013}. We hypothesize that one contributing factor is the fact that the primary objective of machine learning methods has been to maximize predictive performance, and not to identify causal patterns \cite{Athey2018, Taddy2022}, while ``[t]he goal of most empirical studies in economics and other social sciences is to determine whether a change in one variable [...] causes a change in another variable'' \cite{Wooldridge2010} (page 3). For instance, \cite{Kim2020} suggest in their recent review of machine learning-based corporate default predictions that models should be able to suggest the cause(s) of default in order to increase their usefulness.\\

Still, machine learning models have the advantage that they can easily incorporate large numbers of financial covariates (predictive variables, features). For instance, \cite{Wallis2019} use 27 covariates in their credit rating study and \cite{Golbayani2020} find that their models perform better when the full set is provided as input and the various neural networks themselves perform feature selection in the training process. Indeed, the selection of the relevant covariates, and thus the resulting model, is data-driven with machine learning algorithms. ``This approach contrasts with economics, where (in principle, though rarely in reality) the researcher picks a model based on principles and estimates it once'' \cite{Athey2018} (page 508). Given the absence of a predominant capital structure theory and the fact that different theories ``lead to such different, an in some ways diametrically opposed, decisions and outcomes'' \cite{Barclay2005} (page 8), this feature of machine learning appears especially attractive for analyzing leverage ratios.\\

Also, the predictive power of machine learning models is generally strong. For instance, \cite{Wallis2019} (Table 2, page 194) find in their comparison of model accuracy for a large set of S\&P 500 company ratings that random forests 
and support vector models 
improve prediction accuracy by two to three percentage points versus linear discriminant analysis, 
the best-performing non-machine learning method. The improvements in prediction accuracy by machine learning versus benchmark statistical methods (predominantly logistic regression and multiple discriminant analysis) in the rating studies listed by \cite{Huang2004} (Table 1, page 547) are even higher, surpassing ten percentage points in several instances.   \\

\cite{Amini2021} is a recent paper comparing non-linear machine learning models with linear models to predict leverage one year in advance. Out of six different models, the random forest performs best to predict the leverage and improves out-of-sample $R^2$ by 16 percentage points compared to standard linear models, with an $R^2$ for the random forest of 56\% versus 40\% for linear approaches (\cite{Amini2021}, Table 2, page 11). The models rely on 34 covariates as input, of which eight are dummy variables. 
\cite{Amini2021} (page 2) thus ``challenge the conventional wisdom that the standard set of firm and macroeconomic determinants has limited ability to explain firms' leverage choices.''\\


We highlight that none of the eight dummy variables figures among the key determinants of leverage in the best performing models (random forest and gradient boosting). Specifically, the binary ``debt rating'' dummy separating very low and unrated companies from the others has one of the lowest measures of variable importance \cite{Amini2021} (Figure 3, page 12). We hypothesize that since both the z-score as a measure of bankruptcy probability \cite{Altman1968, Altman2013} and the rating dummy attempt to measure the general construct of ``the ability to meet debt obligations'' (see section \ref{CR}), the information contained in both covariates is highly overlapping and thus, the random forest and the gradient boosting machine selectively use only the one with the higher predictive power. However, from this observation, we obviously can ``merely make associational claims'' \cite{Neal2020} (page 2). We  cannot make a causal statement which of these two factors causes leverage (if at all), as opposed to merely predicting it (for an example of using machine learning for forecasting (prediction) versus planning (causation) in the corporate finance function, see for instance \cite{Wasserbacher2022}).\\

Finally, even when \cite{Amini2021} augment linear models with common non-linear transformations of the input variables (e.g., squared or cubed values) as well as a full set of interaction effects between the covariates, the random forest (excl. transformed and interaction covariates) continues to predict leverage significantly better 
(\cite{Amini2021}, internet appendix, Table A4, page 8). Additionally, the authors find in untabulated results that the predictive performance of the random forest does not improve when interaction terms are included. In summary, out of the machine learning toolbox, random forests appear to be a very powerful tool, requiring only limited feature pre-selection or feature engineering. \\

\subsection{Double Machine Learning}
\label{APP-DML}

We have seen from the sections in the main text that there is no general consensus regarding the determinants of leverage and how they interact at the company level. Nevertheless, it is likely that many factors play a role and the mechanisms by which they influence capital structure are complex. Given the lack of a strong theoretical framework, isolating the causal effect of credit ratings poses a formidable challenge. Additionally, we need to consider that this effect may be heterogeneous. Double machine learning \cite{Chernozhukov2017, Chernozhukov2018, Belloni2013, Belloni2014} is a recently developed methodology that can help solve questions of causal inference in such settings by harnessing what \cite{Halevy2009} calls ``the unreasonable effectiveness of data.'' Among the key advantages of double machine learning are the following characteristics. First, there is the ability to handle high feature dimensionality, i.e., the presence of many potential influencing factors in addition to the treatment variable of interest, and to provide valid inference on treatment effects in such high-dimensional, complex data environments. Second, it employs a data-driven approach to select among these influencing factors. Third, it facilitates the use of various machine learning algorithms with flexible function-fitting capabilities. Fourth, there is double-robustness with respect to nuisance functions. ``Partialling-out'', ``Neyman orthogonality'' and ``cross-fitting'' are three important concepts enabling the ``doubly robust'' double machine learning approach. We will discuss each of these terms in this section.\\


\subsubsection{Partialling-out}
\label{APP-DML-PO}

Double machine learning builds on the concept of Frisch-Waugh-Lovell (FWL) ``partialling out'' \cite{Lovell2008,Chernozhukov2017a}. According to the FWL theorem, a parameter of interest $\theta$ in a linear model such as:

\begin{align}
\label{eqn:APP-FWL-one}
Y=\theta D+\beta X +\epsilon 
\end{align}

with $\mathbb{E}(\mathcal{\epsilon} | D,X) = 0$
\\

can be estimated with linear regression, using e.g., ordinary least squares, in either of two ways. Under the first approach, $\theta$ can be directly estimated by regressing $Y$ on $D$ and $X$. Under the second approach, $\theta$ is determined in the last step of a three-step procedure: first, $Y$ is regressed on $X$, and the corresponding residuals $\epsilon_{Y}$ are determined. Second, $D$ is regressed on $X$ and again, the corresponding residuals $\epsilon_{D}$ are determined. Third, the residuals $\epsilon_{Y}$ from the first step are regressed on the residuals $\epsilon_{D}$ from the second step. The regression coefficient from this third step corresponds to $\theta$, the parameter of interest. Both approaches will yield the same estimate for $\theta$. Throughout this paper, we will continue to use the term ``partialling-out'' for the second approach, which is usually employed in the economics literature. ``Residualization" represents another term for the same technique \cite{Taddy2022} (pages 219-220).  \\

It would be convenient if machine learning methods could be used instead of OLS-based linear regression to determine $\theta$. Machine learning has traditionally emphasized predictive performance, and principally predictive performance on the validation (hold-out) data sample, which is intentionally not used for model estimation. Machine learning thus represents the ``algorithmic modeling culture'' described by \cite{Breiman2001} and comes with several advantages. These include a high flexibility with respect to the model choice and design. Many machine learning methods do not impose strong assumptions on the functional forms, but learn those from the data. This constitutes a valuable safeguard against incorrect model specifications, which also lead to biased parameter estimates, even if there are no unmeasured confounding variables. This quality is particularity useful for the empirical analysis in this paper, because, as we described in section \ref{CST}, no consensus about a unifying model for capital structure exists. Additionally, most machine learning algorithms allow us to rely on a largely ``automatic'', data-driven variable selection process. Again, this is a welcome feature in the absence of a consensus model and the presence of many potentially influencing factors. Finally, with the data-driven variable selection approach, machine learning methods are able to handle high-dimensional settings, in which the number of potential predictive variables is large compared to the number of observations. We refer interested readers to the vast literature in this context, for instance \cite{Hastie2009, James2013, Taddy2022} or \cite{Murphy2022}.\\

 The downside of this focus on predictive performance is that inference on the model parameters, the core of causal inference, is generally not possible with machine learning methods. Yet, empirical research in many domains, including economics, is predominantly concerned with causal questions \cite{Wooldridge2010, Athey2018}, and e.g., \cite{Pearl2019} sees the general inability of machine learning methods to uncover causal relationships as a fundamental obstacle to further expand their applications. In particular, machine learning methods generally produce biased parameter estimates. The bias stems from the fact that machine learning methods use regularization penalties in their data-driven variable selection procedure. The intuition behind this regularization bias is that the parameter estimates for covariates that are highly correlated with the treatment variable will get severely ``shrunk'' versus their true value (e.g., in the case of Ridge regression)  or even set to zero (e.g., for LASSO), because the treatment variable by itself has sufficient predictive power. Correspondingly, the parameter value of the treatment variable will get inflated, because it will incorporate the effect of correlated covariates. Of course, the reverse situation with inflated covariate parameters and a significantly shrunk or even zero treatment parameter is also possible. This is the reason that machine learning methods cannot be used to ``directly'' estimate equation \ref{eqn:APP-FWL-one} as per the first approach described above. Such a ``naive approach'' \cite{Belloni2014} (page 36) incurs a high risk of yielding a severely biased estimator for the treatment parameter \cite{Belloni2013, Belloni2014, Wasserbacher2022}.\\
 
 However, machine learning methods can be employed following the second approach, i.e. by partialling-out. This leads to the double machine learning approach for causal analysis. For this approach, Neyman orthogonality plays an central role which we will detail in the next subsection.
 
\subsubsection{Neyman orthogonality}
\label{APP-DML-NO}

 Following the general outline of \cite{Bach2022}, we illustrate the approach using a ``partially linear regression'' model \cite{Robinson1988, Haerdle2000}, which we will also employ in our empirical analysis in section \ref{EM}. The usual form of a partially linear regression model is:
 
 \begin{align}
\label{eqn:APP-PLR general model main equation}
Y=\theta_{0}D+g_{0}(X) +\zeta 
\end{align}

with $\mathbb{E}(\mathcal{\zeta} | D,X) = 0$
\\

and

\begin{align}
\label{eqn:APP-PLR geneal model confounding}
D=m_{0}(X) +\mathcal{V}
\end{align}

with $\mathbb{E}(\mathcal{V} | X) = 0$,
\\

where $Y$ is the outcome variable, $D$ is the treatment (policy) variable of interest, and $X$ is a (potentially high-dimensional) vector of confounding covariates. $\mathcal{\zeta}$ and $\mathcal{V}$ are error terms. The regression coefficient $\theta_{0}$ is the parameter of interest. We can interpret $\theta_{0}$ as a causal parameter, i.e. the causal effect of treatment $D$ on outcome $Y$, if $D$ is ``as good as randomly assigned'' \cite{Chernozhukov2022} (page 73) conditional on the covariates X and thus, D is exogenous conditionally on X. Of course, the other standard assumptions of causal inference need to hold as well, for instance consistency, conditional exchangeability, and positivity \cite{Hernan2020}.\\

Applying the partialling-out procedure on equations \ref{eqn:APP-PLR general model main equation} and \ref{eqn:APP-PLR geneal model confounding} removes both the confounding effect of X and the regularization bias\footnote{More precisely, the first-order effect of the regularization bias is removed. Removing the first-order effect is usually enough to produce a high-quality, low-bias estimator for the parameter of interest \cite{Bach2022}. \cite{Mackey2018} expand this to k-th order orthogonality but show that for partially linear regressions (as employed in our empirical analysis in section \ref{EM}), first-order orthogonality is the limit of robustness when treatment residuals are normally distributed.} introduced by a machine learning method with a penalty or regularization mechanism \cite{Chernozhukov2018}. Within the partialling-out procedure, cross-validation remains required for the determination of the residuals to avoid bias from overfitting. \\

Technically, a method-of-moment estimator for the parameter of interest $\theta_{0}$ is employed:

 \begin{align}
\label{eqn:APP-MoM estimator}
\mathbb{E}[\mathcal{\psi}(W;\theta_{0}, \eta_{0})] = 0
\end{align}

where $\psi$ represents the score function, $W = (Y,D,X)$ is the set (data triplet) of outcome, treatment, and confounding variables, $\theta_{0}$ is the parameter of interest as already indicated above and $\eta_{0}$ are nuisance functions (for instance, $g_{0}$ and $m_{0}$, which we will employ later in our empirical application).\\

For the double machine learning inference procedure, the score function $\psi(W;\theta_{0}, \eta_{0})$ from equation \ref{eqn:APP-MoM estimator} (with $\theta_{0}$ as the unique solution) needs to satisfy the Neyman orthogonality \cite{Neyman1979, Bera2001} condition:

 \begin{align}
\label{eqn:APP-Neymann orthogonality}
\partial_{\eta}\mathbb{E}[\mathcal{\psi}(W;\theta_{0}, \eta)]|_{\eta=\eta_0} = 0,
\end{align}

where the derivative $\partial_{\eta}$ denotes the pathwise Gateaux derivative operator. Intuitively, Neyman orthogonality in equation \ref{eqn:APP-Neymann orthogonality} ensures that the moment condition $\psi(W;\theta_{0}, \eta_{0})$ from equation \ref{eqn:APP-MoM estimator} is insensitive to small errors\footnote{Technically, this concerns the ``speed'' of the convergence rates. We refer interested  readers to \cite{Chernozhukov2018}.} in the estimation of the nuisance function $\eta$ (around its ``true'' full population value $\eta_{0}$). Thus, it removes the bias arising from using a machine learning based estimator for $\eta_{0}$. As a further consequence, Neyman orthogonality ensures ``adaptivity'' of the estimator for $\theta_{0}$: its approximate distribution does not depend on the fact that the machine learning based estimate for $\eta_{0}$ contains errors, if the latter are ``mild'' (as described in \cite{Chernozhukov2018}).  \\

\subsubsection{Cross-fitting}
\label{APP-DML-CF}

 A second point to consider is that machine learning methods usually rely on sample splitting in order to avoid bias introduced by overfitting. Overfitting occurs when models follow the data that they are trained on ``too closely'', thus picking up not only the true underlying pattern, but also the noise contained in the (sample) data. The more complex and flexible a model, the higher the risk for this behavior \cite{James2013, Hastie2009}. Thus, as mentioned previously, machine learning typically divides the data into two distinct sub-sets: one training data set, used to determine the model, and one validation (hold-out) data set to evaluate the model (but not used to train the model). A similar data splitting methodology applies in the case of a partially linear model with two nuisance functions as described in equations \ref{eqn:APP-PLR general model main equation} and \ref{eqn:APP-PLR geneal model confounding}. Only one part of the data is used to estimate the nuisance functions which are partialled-out, while the other part of the data is used to estimate the parameter of interest (i.e., the treatment effect). Of course, such a limited use of the data  implies a loss of efficiency. To overcome this efficiency loss due to the necessary data splitting, double machine learning employs a technique called ``cross-fitting'' \cite{Chernozhukov2018} (page C6).\\
 
 Under cross-fitting, the roles of the two data sets are swapped and two estimates for the parameter of interest are obtained. Since these two estimators are approximately independent, they can simply be averaged to make use of the full data set \cite{Chernozhukov2018} (Figure 2, page C7). The cross-fitting procedure can be expanded beyond two data sets into a K-fold version to further increase robustness; \cite{Bach2022} (page 13) reports that four to five folds appear to work well in practice. Furthermore, the cross-fitting procedure can be repeated to enhance robustness of the estimator with respect to potential effects of a particular random split of the data in the K-folds. While the specific sample partition has no impact on results asymptotically \cite{Chernozhukov2018} (page C30), it is recommended in practice to repeat the estimation procedure \cite{Bach2022} (page 13). \\

\subsubsection{Double robustness}
\label{APP-DML-DR}

 ``Double Machine Learning'' derives its name from the fact that machine learning methods are used to estimate both equation \ref{eqn:APP-PLR general model main equation} and equation \ref{eqn:APP-PLR geneal model confounding}. However, the estimated treatment effect is also ``doubly robust'' thanks to the partialling-out procedure described previously. This means that potential ``mistakes in either of the two prediction problems'' \cite{Taddy2022} (page 221) (i.e., equations \ref{eqn:APP-PLR general model main equation} or \ref{eqn:APP-PLR geneal model confounding}) do not invalidate the effect estimate as long as at least one of these two is sufficiently well estimated. In other words, while it is necessary to ``to do a good job on at least one of these two prediction problems'' \cite{Taddy2022} (page 221), it does not matter on which one. While we caution practitioners against interpreting this as an invitation to careless model specifications, we believe that this represents another attractive property of double machine learning whenever doubts about the precise model characteristics persist. \cite{Belloni2014} (page 34) remarks: ``Because model selection mistakes seem inevitable in realistic settings, it is important to develop inference procedures that are robust to such mistakes.''\\
 
 Finally, a general robustness of double machine learning with respect to the particular machine learning (ML) algorithm that is employed has been observed. For instance,  \cite{Chernozhukov2018} (page C45) comment on their empirical results that ``the choice of the ML method used in estimating nuisance functions does not substantively change the conclusions.`` Of course, the machine learning methods employed need to be of sufficient quality for the problem at hand.
 Considering the large choice of machine learning models, this is typically not an important hurdle, and even ensemble models are suitable \cite{Chernozhukov2018} (pages C22-C23). \\

\label{APP-LRD}

\subsection{Covariates}
\label{APP-Cov}

Tables \ref{table_Covariates_Financials_Overview} and \ref{table_Covariates_Dummies_Overview} provide an overview of the 1'840 covariates (features) used throughout the empirical analysis. The vector X in equations \ref{eqn:PLR main equation} and \ref{eqn:PLR confounding equation} is composed of these variables. In addition to the variables displayed in the two tables, only two variables have been ``engineered'' to provide a potentially better measure for size, which can be useful for purely linear models such as LASSO and Ridge regression. These two variables are the logarithm of sales (code ``sale'' in table \ref{table_Covariates_Financials_Overview}) and the logarithm of total assets (code ``at'').\\

\begin{longtable}{ll}

 \toprule
    \multicolumn{2}{l}{\textbf{Covariates: financial data (absolute and as \% of assets and \% of sales)}}  \\
  \midrule
\textbf{Code} & \textbf{Long text}  \\
    \midrule
\endfirsthead

 \toprule
    \multicolumn{2}{l}{\textbf{Covariates: financial data (absolute and as \% of assets and \% of sales)}}  \\
  \midrule
\textbf{Code} & \textbf{Long text}  \\
    \midrule
\endhead

\multicolumn{2}{r}{\textit{Continued on next page}}\\ \hline
\endfoot

\hline \hline
\endlastfoot

acchg	&	Accounting Changes - Cumulative Effect	\\
accrt	&	ARO Accretion Expense	\\
acdo	&	Current Assets of Discontinued Operations	\\
aco	&	Current Assets - Other - Total	\\
acodo	&	Other Current Assets Excl Discontinued Operations	\\
acominc	&	Accumulated Other Comprehensive Income (Loss)	\\
acox	&	Current Assets - Other - Sundry	\\
acqao	&	Acquired Assets $>$ Other Long-Term Assets	\\
acqcshi	&	Shares Issued for Acquisition	\\
acqgdwl	&	Acquired Assets - Goodwill	\\
acqic	&	Acquisitions - Current Income Contribution	\\
acqintan	&	Acquired Assets - Intangibles	\\
acqinvt	&	Acquired Assets - Inventory	\\
acqppe	&	Acquired Assets $>$ Property, Plant \& Equipment	\\
acqsc	&	Acquisitions - Current Sales Contribution	\\
act	&	Current Assets - Total	\\
adjex\_c	&	Cumulative Adjustment Factor by Ex-Date - Calendar	\\
adjex\_f	&	Cumulative Adjustment Factor by Ex-Date - Fiscal	\\
afudcc	&	Allowance for Funds Used During Construction (Cash Flow)	\\
afudci	&	Allowance for Funds Used During Construction	\\
ajex	&	Adjustment Factor (Company) - Cumulative by Ex-Date	\\
ajp	&	Adjustment Factor (Company) - Cumulative byPay-Date	\\
aldo	&	Long-term Assets of Discontinued Operations	\\
am	&	Amortization of Intangibles	\\
amc	&	Amortization (Cash Flow) - Utility	\\
ano	&	Assets Netting \& Other Adjustments	\\
ao	&	Assets - Other	\\
aocidergl	&	Accum Other Comp Inc - Derivatives Unrealized Gain/Loss	\\
aociother	&	Accum Other Comp Inc - Other Adjustments	\\
aocipen	&	Accum Other Comp Inc - Min Pension Liab Adj	\\
aocisecgl	&	Accum Other Comp Inc - Unreal G/L Ret Int in Sec Assets	\\
aodo	&	Other Assets excluding Discontinued Operations	\\
aol2	&	Assets Level2 (Observable)	\\
aoloch	&	Assets and Liabilities - Other - Net Change	\\
aox	&	Assets - Other - Sundry	\\
aqa	&	Acquisition/Merger After-tax	\\
aqc	&	Acquisitions	\\
aqd	&	Acquisition/Merger Diluted EPS Effect	\\
aqeps	&	Acquisition/Merger Basic EPS Effect	\\
aqi	&	Acquisitions - Income Contribution	\\
aqp	&	Acquisition/Merger Pretax	\\
aqpl1	&	Assets Level1 (Quoted Prices)	\\
aqs	&	Acquisitions - Sales Contribution	\\
arce	&	As Reported Core - After-tax	\\
arced	&	As Reported Core - Diluted EPS Effect	\\
arceeps	&	As Reported Core - Basic EPS Effect	\\
at	&	Assets - Total	\\
aul3	&	Assets Level3 (Unobservable)	\\
bastr	&	Average Short-Term Borrowings Rate	\\
billexce	&	Billings in Excess of Cost \& Earnings	\\
capsft	&	Capitalized Software	\\
capx	&	Capital Expenditures	\\
capxv	&	Capital Expend Property, Plant and Equipment Schd V	\\
cb	&	Compensating Balance	\\
cdvc	&	Cash Dividends on Common Stock (Cash Flow)	\\
ceiexbill	&	Cost \& Earnings in Excess of Billings	\\
ch	&	Cash	\\
che	&	Cash and Short-Term Investments	\\
chech	&	Cash and Cash Equivalents - Increase/(Decrease)	\\
ci	&	Comprehensive Income - Total	\\
cibegni	&	Comp Inc - Beginning Net Income	\\
cicurr	&	Comp Inc - Currency Trans Adj	\\
cidergl	&	Comp Inc - Derivative Gains/Losses	\\
cimii	&	Comprehensive Income - Noncontrolling Interest	\\
ciother	&	Comp Inc - Other Adj	\\
cipen	&	Comp Inc - Minimum Pension Adj	\\
cisecgl	&	Comp Inc - Securities Gains/Losses	\\
citotal	&	Comprehensive Income - Parent	\\
cogs	&	Cost of Goods Sold	\\
cshfd	&	Common Shares Used to Calc Earnings Per Share - Fully Diluted	\\
cshi	&	Common Shares Issued	\\
csho	&	Common Shares Outstanding	\\
cshpri	&	Common Shares Used to Calculate Earnings Per Share - Basic	\\
cshr	&	Common/Ordinary Shareholders	\\
cshtr\_c	&	Common Shares Traded - Annual - Calendar	\\
cshtr\_f	&	Common Shares Traded - Annual - Fiscal	\\
cstke	&	Common Stock Equivalents - Dollar Savings	\\
currtr	&	Currency Translation Rate	\\
curuscn	&	US Canadian Translation Rate	\\
datadate	&	Data Date	\\
dc	&	Deferred Charges	\\
depc	&	Depreciation and Depletion (Cash Flow)	\\
derac	&	Derivative Assets - Current	\\
deralt	&	Derivative Assets Long-Term	\\
derhedgl	&	Gains/Losses on Derivatives and Hedging	\\
diladj	&	Dilution Adjustment	\\
dilavx	&	Dilution Available - Excluding Extraordinary Items	\\
dlcch	&	Current Debt - Changes	\\
dltis	&	Long-Term Debt - Issuance	\\
do	&	Discontinued Operations	\\
donr	&	Nonrecurring Disc Operations	\\
dp	&	Depreciation and Amortization	\\
dpacre	&	Accumulated Depreciation of RE Property	\\
dpact	&	Depreciation, Depletion and Amortization (Accumulated)	\\
dpc	&	Depreciation and Amortization (Cash Flow)	\\
dpret	&	Depr/Amort of Property	\\
dpvieb	&	Depreciation (Accumulated) - Ending Balance (Schedule VI)	\\
drlt	&	Deferred Revenue - Long-term	\\
dv	&	Cash Dividends (Cash Flow)	\\
dvc	&	Dividends Common/Ordinary	\\
dvintf	&	Dividends \& Interest Receivable (Cash Flow)	\\
dvp	&	Dividends - Preferred/Preference	\\
dvpsp\_c	&	Dividends per Share - Pay Date - Calendar	\\
dvpsp\_f	&	Dividends per Share - Pay Date - Fiscal	\\
dvpsx\_c	&	Dividends per Share - Ex-Date - Calendar	\\
dvpsx\_f	&	Dividends per Share - Ex-Date - Fiscal	\\
dvt	&	Dividends - Total	\\
ebit	&	Earnings Before Interest and Taxes	\\
ebitda	&	Earnings Before Interest	\\
emp	&	Employees	\\
epsfi	&	Earnings Per Share (Diluted) - Including Extraordinary Items	\\
epsfx	&	Earnings Per Share (Diluted) - Excluding Extraordinary Items	\\
epspi	&	Earnings Per Share (Basic) - Including Extraordinary Items	\\
epspx	&	Earnings Per Share (Basic) - Excluding Extraordinary Items	\\
esub	&	Equity in Earnings - Unconsolidated Subsidiaries	\\
esubc	&	Equity in Net Loss - Earnings	\\
exre	&	Exchange Rate Effect	\\
fatb	&	Property, Plant, and Equipment - Buildings at Cost	\\
fatc	&	Property, Plant, and Equipment - Construction in Progress at Cost	\\
fate	&	Property, Plant, and Equipment - Machinery and Equipment at Cost	\\
fatl	&	Property, Plant, and Equipment - Leases at Cost	\\
fatn	&	Property, Plant, and Equipment - Natural Resources at Cost	\\
fato	&	Property, Plant, and Equipment - Other at Cost	\\
fatp	&	Property, Plant, and Equipment - Land and Improvements at Cost	\\
fca	&	Foreign Exchange Income (Loss)	\\
ffo	&	Funds From Operations (REIT)	\\
fiao	&	Financing Activities - Other	\\
finaco	&	Finance Division Other Current Assets, Total	\\
finao	&	Finance Division Other Long-Term Assets, Total	\\
fincf	&	Financing Activities - Net Cash Flow	\\
finch	&	Finance Division - Cash	\\
finivst	&	Finance Division  Short-Term Investments	\\
finrecc	&	Finance Division  Current Receivables	\\
finreclt	&	Finance Division  Long-Term Receivables	\\
finrev	&	Finance Division Revenue	\\
finxopr	&	Finance Division Operating Expense	\\
fopo	&	Funds from Operations - Other	\\
fopox	&	Funds from Operations - Other excluding Option Tax Benefit	\\
fopt	&	Funds From Operations - Total	\\
fsrco	&	Sources of Funds - Other	\\
fsrct	&	Sources of Funds - Total	\\
fuseo	&	Uses of Funds - Other	\\
fuset	&	Uses of Funds - Total	\\
fyear	&	Data Year - Fiscal	\\
gdwl	&	Goodwill	\\
gdwlam	&	Goodwill Amortization	\\
gdwlia	&	Impairments of Goodwill After-tax	\\
gdwlid	&	Impairments of Goodwill Diluted EPS Effect	\\
gdwlieps	&	Impairments of Goodwill Basic EPS Effect	\\
gdwlip	&	Impairments of Goodwill Pretax	\\
gla	&	Gain/Loss After-tax	\\
glcea	&	Gain/Loss on Sale (Core Earnings Adjusted) After-tax	\\
glced	&	Gain/Loss on Sale (Core Earnings Adjusted) Diluted EPS	\\
glceeps	&	Gain/Loss on Sale (Core Earnings Adjusted) Basic EPS Effect	\\
glcep	&	Gain/Loss on Sale (Core Earnings Adjusted) Pretax	\\
gld	&	Gain/Loss Diluted EPS Effect	\\
gleps	&	Gain/Loss Basic EPS Effect	\\
gliv	&	Gains/Losses on investments	\\
glp	&	Gain/Loss Pretax	\\
gp	&	Gross Profit (Loss)	\\
hedgegl	&	Gain/Loss on Ineffective Hedges	\\
ib	&	Income Before Extraordinary Items	\\
ibadj	&	Income Before Extraordinary Items - Adjusted for Common Stock	\\
ibc	&	Income Before Extraordinary Items (Cash Flow)	\\
ibcom	&	Income Before Extraordinary Items - Available for Common	\\
ibmii	&	Income before Extraordinary Items and Noncontrolling Interests	\\
intan	&	Intangible Assets - Total	\\
intano	&	Other Intangibles	\\
intc	&	Interest Capitalized	\\
invch	&	Inventory - Decrease (Increase)	\\
invfg	&	Inventories - Finished Goods	\\
invo	&	Inventories - Other	\\
invrm	&	Inventories - Raw Materials	\\
invt	&	Inventories - Total	\\
invwip	&	Inventories - Work In Process	\\
irent	&	Rental Income	\\
itcb	&	Investment Tax Credit (Balance Sheet)	\\
itcc	&	Investment Tax Credit - Net (Cash Flow) - Utility	\\
itci	&	Investment Tax Credit (Income Account)	\\
ivaco	&	Investing Activities - Other	\\
ivch	&	Increase in Investments	\\
ivncf	&	Investing Activities - Net Cash Flow	\\
ivst	&	Short-Term Investments - Total	\\
ivstch	&	Short-Term Investments - Change	\\
lifr	&	LIFO Reserve	\\
lifrp	&	LIFO Reserve - Prior	\\
lno	&	Liabilities Netting \& Other Adjustments	\\
mib	&	Noncontrolling Interest (Balance Sheet)	\\
mibn	&	Noncontrolling Interests - Nonredeemable - Balance Sheet	\\
mibt	&	Noncontrolling Interests - Total - Balance Sheet	\\
mii	&	Noncontrolling Interest (Income Account)	\\
mkvalt	&	Market Value - Total - Fiscal	\\
msa	&	Marketable Securities Adjustment	\\
ni	&	Net Income (Loss)	\\
niadj	&	Net Income Adjusted for Common/Ordinary Stock	\\
nipfc	&	Pro Forma Net Income - Current	\\
nipfp	&	Pro Forma Net Income - Prior	\\
nopi	&	Nonoperating Income (Expense)	\\
nopio	&	Nonoperating Income (Expense) - Other	\\
nrtxt	&	Nonrecurring Income Taxes After-tax	\\
nrtxtd	&	Nonrecurring Income Tax Diluted EPS Effect	\\
nrtxteps	&	Nonrecurring Income Tax Basic EPS Effect	\\
oancf	&	Operating Activities - Net Cash Flow	\\
ob	&	Order Backlog	\\
oiadp	&	Operating Income After Depreciation	\\
oibdp	&	Operating Income Before Depreciation	\\
opeps	&	Earnings Per Share from Operations	\\
oprepsx	&	Earnings Per Share - Diluted - from Operations	\\
optca	&	Options - Cancelled (-)	\\
optdr	&	Dividend Rate - Assumption (\%)	\\
optex	&	Options Exercisable (000)	\\
optexd	&	Options - Exercised (-)	\\
optgr	&	Options - Granted	\\
optlife	&	Life of Options - Assumption (\# yrs)	\\
optosby	&	Options Outstanding - Beg of Year	\\
optosey	&	Options Outstanding - End of Year	\\
optprcby	&	Options Outstanding Beg of Year - Price	\\
optprcca	&	Options Cancelled - Price	\\
optprcex	&	Options Exercised - Price	\\
optprcey	&	Options Outstanding End of Year - Price	\\
optprcgr	&	Options Granted - Price	\\
optprcwa	&	Options Exercisable - Weighted Avg Price	\\
optrfr	&	Risk Free Rate - Assumption (\%)	\\
optvol	&	Volatility - Assumption (\%)	\\
pddur	&	Period Duration	\\
pdvc	&	Cash Dividends on Preferred/Preference Stock (Cash Flow)	\\
pi	&	Pretax Income	\\
pidom	&	Pretax Income - Domestic	\\
pifo	&	Pretax Income - Foreign	\\
pnca	&	Core Pension Adjustment	\\
pncad	&	Core Pension Adjustment Diluted EPS Effect	\\
pncaeps	&	Core Pension Adjustment Basic EPS Effect	\\
pncia	&	Core Pension Interest Adjustment After-tax	\\
pncid	&	Core Pension Interest Adjustment Diluted EPS Effect	\\
pncieps	&	Core Pension Interest Adjustment Basic EPS Effect	\\
pncip	&	Core Pension Interest Adjustment Pretax	\\
pncwia	&	Core Pension w/o Interest Adjustment After-tax	\\
pncwid	&	Core Pension w/o Interest Adjustment Diluted EPS Effect	\\
pncwieps	&	Core Pension w/o Interest Adjustment Basic EPS Effect	\\
pncwip	&	Core Pension w/o Interest Adjustment Pretax	\\
pnrsho	&	Nonred Pfd Shares Outs (000)	\\
ppegt	&	Property, Plant and Equipment - Total (Gross)	\\
ppenc	&	Property, Plant, and Equipment - Construction in Progress (Net)	\\
ppent	&	Property, Plant and Equipment - Total (Net)	\\
ppevbb	&	Property, Plant and Equipment - Beginning Balance (Schedule V)	\\
ppeveb	&	Property, Plant, and Equipment - Ending Balance (Schedule V)	\\
prca	&	Core Post Retirement Adjustment	\\
prcad	&	Core Post Retirement Adjustment Diluted EPS Effect	\\
prcaeps	&	Core Post Retirement Adjustment Basic EPS Effect	\\
prcc\_c	&	Price Close - Annual - Calendar	\\
prcc\_f	&	Price Close - Annual - Fiscal	\\
prch\_c	&	Price High - Annual - Calendar	\\
prch\_f	&	Price High - Annual - Fiscal	\\
prcl\_c	&	Price Low - Annual - Calendar	\\
prcl\_f	&	Price Low - Annual - Fiscal	\\
prsho	&	Redeem Pfd Shares Outs (000)	\\
prstkc	&	Purchase of Common and Preferred Stock	\\
prstkcc	&	Purchase of Common Stock (Cash Flow)	\\
prstkpc	&	Purchase of Preferred/Preference Stock (Cash Flow)	\\
rca	&	Restructuring Costs After-tax	\\
rcd	&	Restructuring Costs Diluted EPS Effect	\\
rceps	&	Restructuring Costs Basic EPS Effect	\\
rcp	&	Restructuring Costs Pretax	\\
rdip	&	In Process R\&D Expense	\\
rdipa	&	In Process R\&D Expense After-tax	\\
rdipd	&	In Process R\&D Expense Diluted EPS Effect	\\
rdipeps	&	In Process R\&D Expense Basic EPS Effect	\\
recch	&	Accounts Receivable - Decrease (Increase)	\\
recco	&	Receivables - Current - Other	\\
recd	&	Receivables - Estimated Doubtful	\\
rect	&	Receivables - Total	\\
ret	&	Total RE Property	\\
revt	&	Revenue - Total	\\
rra	&	Reversal - Restructruring/Acquisition Aftertax	\\
rrd	&	Reversal - Restructuring/Acq Diluted EPS Effect	\\
rreps	&	Reversal - Restructuring/Acq Basic EPS Effect	\\
rrp	&	Reversal - Restructruring/Acquisition Pretax	\\
rstche	&	Restricted Cash \& Investments - Current	\\
rstchelt	&	Long-Term Restricted Cash \& Investments	\\
sale	&	Sales/Turnover (Net)	\\
salepfc	&	Pro Forma Net Sales - Current Year	\\
salepfp	&	Pro Forma Net Sales - Prior Year	\\
scstkc	&	Sale of Common Stock (Cash Flow)	\\
seta	&	Settlement (Litigation/Insurance) After-tax	\\
setd	&	Settlement (Litigation/Insurance) Diluted EPS Effect	\\
seteps	&	Settlement (Litigation/Insurance) Basic EPS Effect	\\
setp	&	Settlement (Litigation/Insurance) Pretax	\\
siv	&	Sale of Investments	\\
spce	&	S\&P Core Earnings	\\
spced	&	S\&P Core Earnings EPS Diluted	\\
spceeps	&	S\&P Core Earnings EPS Basic	\\
spi	&	Special Items	\\
spid	&	Other Special Items Diluted EPS Effect	\\
spieps	&	Other Special Items Basic EPS Effect	\\
spioa	&	Other Special Items After-tax	\\
spiop	&	Other Special Items Pretax	\\
sppe	&	Sale of Property	\\
sppiv	&	Sale of Property, Plant and Equipment and Investments - Gain (Loss)	\\
spstkc	&	Sale of Preferred/Preference Stock (Cash Flow)	\\
sret	&	Gain/Loss on Sale of Property	\\
sstk	&	Sale of Common and Preferred Stock	\\
stkco	&	Stock Compensation Expense	\\
stkcpa	&	After-tax stock compensation	\\
tdc	&	Deferred Income Taxes - Net (Cash Flow)	\\
tfva	&	Total Fair Value Assets	\\
tfvce	&	Total Fair Value Changes including Earnings	\\
tfvl	&	Total Fair Value Liabilities	\\
tlcf	&	Tax Loss Carry Forward	\\
txach	&	Income Taxes - Accrued - Increase/(Decrease)	\\
txbco	&	Excess Tax Benefit Stock Options - Cash Flow Operating	\\
txbcof	&	Excess Tax Benefit of Stock Options - Cash Flow Financing	\\
txc	&	Income Taxes - Current	\\
txdb	&	Deferred Taxes (Balance Sheet)	\\
txdba	&	Deferred Tax Asset - Long Term	\\
txdbca	&	Deferred Tax Asset - Current	\\
txdbcl	&	Deferred Tax Liability - Current	\\
txdc	&	Deferred Taxes (Cash Flow)	\\
txdfed	&	Deferred Taxes-Federal	\\
txdfo	&	Deferred Taxes-Foreign	\\
txdi	&	Income Taxes - Deferred	\\
txditc	&	Deferred Taxes and Investment Tax Credit	\\
txds	&	Deferred Taxes-State	\\
txfed	&	Income Taxes - Federal	\\
txfo	&	Income Taxes - Foreign	\\
txndb	&	Net Deferred Tax Asset (Liab) - Total	\\
txndba	&	Net Deferred Tax Asset	\\
txndbl	&	Net Deferred Tax Liability	\\
txndbr	&	Deferred Tax Residual	\\
txo	&	Income Taxes - Other	\\
txp	&	Income Taxes Payable	\\
txpd	&	Income Taxes Paid	\\
txr	&	Income Tax Refund	\\
txs	&	Income Taxes - State	\\
txt	&	Income Taxes - Total	\\
txtubadjust	&	Other Unrecog Tax Benefit Adj.	\\
txtubbegin	&	Unrecog. Tax Benefits - Beg of Year	\\
txtubend	&	Unrecog. Tax Benefits - End of Year	\\
txtubmax	&	Chg. In Unrecog. Tax Benefits - Max	\\
txtubmin	&	Chg. In Unrecog. Tax Benefits - Min	\\
txtubposdec	&	Decrease- Current Tax Positions	\\
txtubposinc	&	Increase- Current Tax Positions	\\
txtubpospdec	&	Decrease- Prior Tax Positions	\\
txtubpospinc	&	Increase- Prior Tax Positions	\\
txtubsettle	&	Settlements with Tax Authorities	\\
txtubsoflimit	&	Lapse of Statute of Limitations	\\
txtubtxtr	&	Impact on Effective Tax Rate	\\
txtubxintbs	&	Interest \& Penalties Accrued - B/S	\\
txtubxintis	&	Interest \& Penalties Reconized - I/S	\\
txw	&	Excise Taxes	\\
uaoloch	&	Other Assets and Liabilities - Net Change (Statement of Cash Flows)	\\
uaox	&	Other Assets - Utility	\\
uapt	&	Accounts Payable - Utility	\\
uccons	&	Contributions in Aid of Construction	\\
ucustad	&	Customer Advances for Construction	\\
udcopres	&	Deferred Credits and Operating Reserves - Other	\\
udfcc	&	Deferred Fuel - Increase (Decrease) (Statement of Cash Flows)	\\
udpfa	&	Depreciation of Fixed Assets	\\
udvp	&	Preferred Dividend Requirements	\\
ugi	&	Gross Income (Income Before Interest Charges)	\\
uinvt	&	Inventories - Utility	\\
ulcm	&	Current Liabilities - Miscellaneous	\\
ulco	&	Current Liabilities - Other - Utility	\\
uniami	&	Net Income before Extraordinary Items	\\
unopinc	&	Nonoperating Income (Net) - Other	\\
uois	&	Other Internal Sources - Net (Cash Flow)	\\
uopi	&	Operating Income - Total - Utility	\\
uopres	&	Operating Reserves	\\
updvp	&	Preference Dividend Requirements*	\\
upstksf	&	Preferred/Preference Stock Sinking Fund Requirement	\\
urect	&	Receivables (Net)	\\
urectr	&	Accounts Receivable - Trade - Utility	\\
urevub	&	Accrued Unbilled Revenues (Balance Sheet)	\\
uspi	&	Special Items	\\
usubdvp	&	Subsidiary Preferred Dividends	\\
utme	&	Maintenance Expense - Total	\\
utxfed	&	Current Taxes - Federal (Operating)	\\
wcap	&	Working Capital (Balance Sheet)	\\
wcapc	&	Working Capital Change - Other - Increase/(Decrease)	\\
wcapch	&	Working Capital Change - Total	\\
wda	&	Writedowns After-tax	\\
wdd	&	Writedowns Diluted EPS Effect	\\
wdeps	&	Writedowns Basic EPS Effect	\\
wdp	&	Writedowns Pretax	\\
xacc	&	Accrued Expenses	\\
xad	&	Advertising Expense	\\
xi	&	Extraordinary Items	\\
xido	&	Extraordinary Items and Discontinued Operations	\\
xidoc	&	Extraordinary Items and Discontinued Operations (Cash Flow)	\\
xintopt	&	Implied Option Expense	\\
xlr	&	Staff Expense - Total	\\
xopr	&	Operating Expenses - Total	\\
xoptd	&	Implied Option EPS Diluted	\\
xopteps	&	Implied Option EPS Basic	\\
xpp	&	Prepaid Expenses	\\
xpr	&	Pension and Retirement Expense	\\
xrd	&	Research and Development Expense	\\
xrdp	&	Research \& Development - Prior	\\
xrent	&	Rental Expense	\\
xsga	&	Selling, General and Administrative Expense	\\

    \bottomrule

\caption{Overview of data items sourced from Compustat (``Compustat Daily Updates - Fundamentals Annual'') and employed as continuous-valued covariates throughout the empirical analysis (see section \ref{EM}). Data items are sorted in alphabetical order of their code. The code and long text are as per the Compustat Data Guide, accessible via the Wharton Research Data Service (WRDS). Please see subsection \ref{EM-Data} for further details. Data items are used as absolute values (as sourced from Compustat) and as scaled values, once by total assets (code ``at'') and once by total sales (code ``sales''). Scaling was not performed in selected cases where this appeared to be meaningless, for instance for the fiscal year (code ``fyear''). The long text for the codes ``afudci'' (``Allowance for Funds Used During Construction (Investing) (Cash Flow)''), ``ibad'' (``Income Before Extraordinary Items - Adjusted for Common Stock Equivalents''), ``niadj'' (``Net Income Adjusted for Common/Ordinary Stock (Capital) Equivalents'') and ``uniami'' (``Net Income before Extraordinary Items and after Noncontrolling Interest'') has been shortened in the table to limit the breadth of the second column.}
\label{table_Covariates_Financials_Overview}
\end{longtable}

\newpage

\begin{longtable}{llr}

 \toprule
    \multicolumn{3}{l}{\textbf{Covariates: dummy variables}}  \\
  \midrule
\textbf{Code} & \textbf{Long text} & \textbf{Count} \\
    \midrule
\endfirsthead

 \toprule
    \multicolumn{3}{l}{\textbf{Covariates: dummy variables}}  \\
  \midrule
\textbf{Code} & \textbf{Long text} & \textbf{Count}  \\
    \midrule
\endhead

\multicolumn{3}{r}{\textit{Continued on next page}}\\ \hline
\endfoot

\hline \hline
\endlastfoot

acctchg	&	Adoption of Accounting Changes	&	5	\\
acctstd	&	Accounting Standard	&	3	\\
acqmeth	&	Acquisition Method	&	6	\\
adrr	&	ADR Ratio	&	54	\\
au	&	Auditor	&	21	\\
auop	&	Auditor Opinion	&	5	\\
auopic	&	Auditor Opinion - Internal Control	&	3	\\
bspr	&	Balance Sheet Presentation	&	2	\\
ceoso	&	Chief Executive Officer SOX Certification	&	3	\\
cfoso	&	Chief Financial Officer SOX Certification	&	3	\\
cik	&	CIK Number	&	1	\\
compst	&	Comparability Status	&	10	\\
costat	&	Active/Inactive Status Marker	&	1	\\
curcd	&	ISO Currency Code	&	1	\\
curncd	&	Native Currency Code	&	33	\\
cusip	&	CUSIP	&	1	\\
dldte	&	Research Company Deletion Date	&	1	\\
dlrsn	&	Research Co Reason for Deletion	&	1	\\
exchg	&	Stock Exchange Code	&	12	\\
fax	&	Fax Number	&	1	\\
fic	&	Current ISO Country Code - Incorporation	&	58	\\
final	&	Final Indicator Flag	&	1	\\
fyr	&	Fiscal Year-end Month	&	11	\\
fyrc	&	Current Fiscal Year End Month	&	11	\\
idbflag	&	International, Domestic, Both Indicator	&	1	\\
incorp	&	Current State/Province of Incorporation Code	&	52	\\
ipodate	&	Company Initial Public Offering Date	&	1	\\
ismod	&	Income Statement Model Number	&	2	\\
loc	&	Current ISO Country Code - Headquarters	&	65	\\
ltcm	&	Long Term Contract Method	&	3	\\
ogm	&	OIL \& GAS METHOD	&	2	\\
phone	&	Phone Number	&	1	\\
prican	&	Current Primary Issue Tag - Canada	&	1	\\
prirow	&	Primary Issue Tag - Rest of World	&	1	\\
priusa	&	Current Primary Issue Tag - US	&	1	\\
rank	&	Rank - Auditor	&	1	\\
scf	&	Cash Flow Format	&	3	\\
sic	&	Standard Industry Classification Code	&	306	\\
src	&	Source Document	&	7	\\
stalt	&	Status Alert	&	2	\\
state	&	State/Province	&	61	\\
stko	&	Stock Ownership Code	&	3	\\
tic	&	Ticker Symbol	&	1	\\
udpl	&	Utility - Liberalized Depreciation Code	&	3	\\
upd	&	Update Code	&	1	\\
weburl	&	Web URL	&	1	\\

    \bottomrule

\caption{Overview of data items sourced from Compustat (``Compustat Daily Updates - Fundamentals Annual'') and transformed into dummy variables throughout the empirical analysis (see section \ref{EM}). Data items are sorted in alphabetical order of their code. The code and long text are as per the Compustat Data Guide, accessible via the Wharton Research Data Service (WRDS). Please see subsection \ref{EM-Data} for further details. ``Count'' refers to the number of dummirized variables into which one particular data item was transformed. For instance, there are six types for ``acctchg'' in our empirical data set (i.e., whether or not a company has adopted a particular new accounting standard), which translates into five dummy variables. For data items with ``count'' = 1 (i.e, one single dummy variable), dummy coding corresponds to presence or absence of the data item. For instance, a company may have or may not have in a given year a central index key (CIK number, code ``cik'') from the FDA, or a fax number (code ``fax''), displayed in Compustat. For the Standard Industry Classification (code ``sic''), dummies have been created at the first (7), second (58) and third (241) level for a total count of 306.}
\label{table_Covariates_Dummies_Overview}
\end{longtable}

\newpage

\subsection{Learner specifications: technical details}
\label{APP-Learners-Specifications}

We provide here technical details for the learners (``nuisance functions'', \cite{Chernozhukov2018, Belloni2014}) $g_{0}$ and $m_{0}$ from subsection \ref{EM-Results-AnyRating} which capture the relationship of the covariates $X$ with the outcome $LDA$ and the treatment $D$, respectively.\\

For $g_{0}$, we specify the random forest to consist of 500 trees, each with a maximum depth of seven levels, to predict $LDA$. This achieves an out-of-sample prediction accuracy of approximately 53\% for the $R^2$. Specifically, we use the ``regr.ranger'' function in R with the following parameters: $num.trees = 500$, $mtry = 50$, $min.node.size = 10$, $max.depth = 7$; we refer interested readers to the corresponding R package documentation \cite{Wright2017}. We tuned these parameters based on a 30\% training - 70\% testing sample split. For reference, the out-of-bag (OOB) $R^2$ in the training data is 49\%.\\

For $m_{0}$, we specify the random forest to consist also of 500 trees, but with a slightly lower maximum depth of five levels each. Specifically, we used the ``classif.ranger'' function in R with the following parameters: $num.trees = 500$, $mtry = 50$, $min.node.size = 10$, $max.depth = 5$; we refer interested readers to the corresponding R package documentation \cite{Wright2017}. We tuned these parameters based on a 30\% training - 70\% testing sample split. For reference, the out-of-bag (OOB) correct classification rate is 87\% in the training data. Out-of-sample, we also achieve a correct classification rate of approximately 87\%.\\

With these learner specifications and a five-fold split as well as two repetitions (following the recommendation in \cite{Bach2022}, page 13) total run time with this set-up was approx. 35 minutes on a standard personal computer (Intel Core i7, 8 cores) for the analysis with one binary treatment variable.

\subsection{Robustness check: alternative model specifications}
\label{APP-AM-Specifications}

As mentioned in the main section of this paper, we have employed different learner specifications for $g_{0}$ and $m_{0}$ as a robustness check. The effect estimates across the different model alternatives are very consistent as can be seen from table \ref{table_Res_General_Rating_Robust} reported in the main part of this paper, which we repeat for ease of reading with an extended legend in table \ref{APP-table_Res_General_Rating_Robust}. We also provide here a detailed description of the alternative learner specifications and a discussion of results.\\

For the first alternative model (AM1), we used the specifications of our main model (MM) described in the main section of our paper but changed the cross-fitting algorithm to``DML2'' instead of ``DML1'' (differences between the two algorithms are specified in \cite{Bach2022}, page 12). In a second alternative model (AM2), we changed the machine learning method for learner $g_{0}$ to LASSO \cite{Hastie2009, Buehlmann2011, Taddy2022} while keeping the random forest from the main model for $m_{0}$. For AM3, we switched to Ridge regression \cite{Hastie2009, Buehlmann2011} for learner $g_{0}$, again keeping the random forest from the MM for $m_{0}$. For AM4, we maintained the random forest method for both learners, but restrained the trees by limiting their maximum depth to five ($g_{0}$) and three ($m_{0}$) levels (versus seven and five in MM).\\

The effect estimates across the different model alternatives are very consistent: AM1 results are virtually indistinguishable from MM. The effect estimate only differs in the sixth digit after the decimal point (not shown in the table). Of course, this should be expected, since only the cross-fitting algorithm was changed, while the learner models and parametrization were identical. However, also AM2 and AM3, where the machine learning approach for $g_{0}$ were changed from random forest to the LASSO and Ridge regression, respectively, yield causal effect estimates that differ only by 0.5pps to 0.6pps from MM. Similarly, AM4, where both learner functions were ``held back from learning'' by restricting their tree depth, yields an effect estimate that differs only by 0.6pps from the main model employed in this paper. In terms of p-values, all models are highly significant with p-values of 0.000; only for AM3 (Ridge regression), the p-value is different with 0.011, but of course still clearly below the usual cut-off value of 0.05.\\

\newpage

\begin{longtable}{lccccc}
    \toprule
    \multicolumn{6}{l}{\textbf{Robustness check: alternative model specifications}} \\
  \midrule
\textbf{Rating effect} & MM & AM1 & AM2 & AM3 & AM4 \\
\textbf{on LDA} & (RF/RF) & (DML2) & (LASSO/RF) & (Ridge/RF) & (Restr.) \\
    \midrule
$\theta$ (rating yes/no) & 0.0878 &	0.0878 & 0.0925	& 0.0935 & 0.0942    \\  
Std. error	&	0.0021 & 0.0021 & 0.0023 & 0.0369 & 0.0021  \\
t-value & 41.8 & 41.8 & 40.0 & 2.5 & 45.2 \\
p-value	&	0.000 &	0.000 & 0.000 & 0.011 & 0.000  \\
\midrule
\textit{Effect ($\theta$) vs. mean} & \textit{41\%}	& \textit{41\%} & \textit{44\%} & \textit{44\%} & \textit{44\%} \\
    \bottomrule

\caption{Results for the estimated causal effect $\theta$ of having a rating (or not) on LDA, according to alternative model (AM) specifications. ``MM'' refers to the main model specification used throughout the paper. The main characteristics are random forests for both learners with the specifications and tuning parameters detailed in the main text. ``AM1'' differs from MM only by using a different aggregation procedure (``DML2'' versus ``DML1'' \cite{Bach2022}) for the score function; results are virtually indistinguishable from MM. ``AM2'' (``AM3'') uses the LASSO (Ridge) as learner for $g_{0}$, while retaining the random forest from MM for $m_{0}$.``AM4'' is set up like MM, except that the two random forests learners are ``restrained" by limiting the maximum depth to five ($g_{0}$) and three ($m_{0}$) levels versus respectively seven and five in the MM specification. The ``Effect ($\theta$) vs. mean'' is calculated versus the mean LDA value of 0.212. Parameter estimates and standard errors (bootstrap procedure) are aggregated over a five-fold split with two repetitions for all models. Subsection \ref{EM-Data} describes the data sample.}
\label{APP-table_Res_General_Rating_Robust}
\end{longtable}

\subsection{Effect of investment-grade rating and speculative grade rating (versus having no rating)}
\label{EM-Results-InvGSpeG}

As mentioned in the main text, the initial analysis of having a rating versus having no rating implicitly assumes that it does not matter which rating a company has: all rating types are the same ``treatment'' for leverage; since ratings are opinions about credit risk, this implies that the type of opinion would not matter. However, it is easy to argue that different ratings, i.e. different opinions, may in reality represent different treatments, and thus, different versions of the treatment exist. Put differently, our initial analysis may suffer from the fact that it incorrectly assumes that there are ``no hidden variations of treatments'' \cite{Imbens2015} (pages 10-13). This is one of the assumptions included in the ``stable unit treatment value assumption'' (SUTVA) \cite{Rubin1980}, which provides a fundamental framework for causal analysis. Interested readers can access a vast literature on this topic, for instance \cite{Holland1986, Rubin2005, Pearl2009, Morgan2015, Peters2017} or \cite{Chernozhukov2022}).\\

We therefore investigate in a second analysis whether the rating effect differs between the two very broad categories of ``investment-grade'' and ``speculative-grade'' (non-investment grade, ``junk bonds''). Rating agencies themselves explicitly categorize their different ratings into these two broad groups \cite{SPFSL2022} and the distinction has significant implications for regulatory purposes as mentioned in section \ref{LR}.\\

The partially linear model described in equations \ref{eqn:PLR main equation} and \ref{eqn:PLR confounding equation} can thus be written to contain two different binary treatment variables, $D^{InvGR}_{i,t}$ and $D^{SpeGR}_{i,t}$ and their corresponding causal parameters $\theta^{InvGR}$ and $\theta^{SpeGR}$. These two treatment variables specify whether a given company $i$ had in year $t$ an \underline{inv}estment-\underline{g}rade \underline{r}ating ($D^{InvGR}_{i,t} = 1$, $D^{SpeGR}_{i,t} = 0$) or a \underline{spe}culative-\underline{g}rade \underline{r}ating ($D^{InvGR}_{i,t} = 0$, $D^{SpeGR}_{i,t} = 1$), or no rating at all ($D^{InvGR}_{i,t} = 0$, $D^{SpeGR}_{i,t} = 0$):\footnote{We remind ourselves that investment-grade ratings include rating categories from AAA to BBB-, speculative-grade ratings include BB+ and below and that the three categories (investment-grade, speculative-grade, no rating) are mutually exclusive and collectively exhaustive at any given point in time for each company. Of course, the (granular) rating for a company can change within a given year; however, the likelihood of change across these three very broad categories is small.}

\begin{align}
\label{eqn:PLR main equation InvSpe}
LDA_{i,t}=\theta^{InvGR} D^{InvGR}_{i,t}+\theta^{SpeGR} D^{SpeGR}_{i,t}+g_{0}(X_{i,t}) +\zeta_{i,t} 
\end{align}

with $\mathbb{E}(\mathcal{\zeta}_{i,t} | D^{InvGR}_{i,t},D^{SpeGR}_{i,t},X_{i,t}) = 0$.
\\

Equation \ref{eqn:PLR confounding equation} is defined accordingly to reflect two different binary treatment variables. By considering two treatment variables, we are now conducting (causal) inference on multiple parameters at the same time. Therefore, we need to take into account the ``multiplicity problem'': the possibility of falsely identifying an effect as ``significant'' increases with the number of treatments tested. Several methods have been proposed to account for this (see \cite{Bach2018} for a condensed review and applications in high-dimensional settings). The classical method to control the ``family-wise error rate'' (i.e., the probability of at least one false rejection of the null hypothesis of no causal effect) is the Bonferroni correction; it is considered as very conservative \cite{VanderWeele2018}. As an alternative, the Benjamini-Hochberg false discovery rate control \cite{Benjamini1995}, which targets the expected share of falsely rejected null hypotheses, relies on independence between tests, which is often an unrealistic assumption \cite{Taddy2022}. Other approaches attempt to maintain the concept of the family-wise error rate while reducing its conservatism. These include step-down methods, such as the step-down method of Holm \cite{Holm1979} or the more recent Romano-Wolf step-down procedure \cite{Romano2005, Romano2005a}, which also takes the dependence structure of test statistics into consideration. Another approach for valid simultaneous inference relies on the multiplier bootstrap procedure proposed by \cite{Chernozhukov2013, Chernozhukov2014}. This procedure iterates over the set of treatment variables and selects each of them to individually estimate its effect on the outcome variable; the other, currently not selected treatment variables are included in the nuisance functions.\\

Table \ref{table_Res_General_Rating_InvSpec} reports results for the effect estimate of investment-grade ratings and the effect of speculative-grade ratings on leverage (versus the baseline of no rating). Taking into account the multiplicity problem of simultaneous inference on multiple parameters described above, we report multiplier bootstrap (MB) standard errors and p-values, as well as, for comparison, the corresponding Romano-Wolf (RoWo) and Bonferroni (Bonf) p-values.\\

\begin{longtable}{lrcccc}
    \toprule
    \multicolumn{6}{l}{\textbf{Investment- versus speculative-grade rating category}}  \\
  \midrule
\textbf{Rating effect (on LDA)} & Coef. & MB & MB & RoWo & Bonf \\
\textbf{ } & estim. & Std. error & p-val. & p-val. & p-val. \\
    \midrule
$\theta^{InvGR}$ (investment-grade) & -0.0030 &	0.0024 & 0.209	& 0.204 & 0.417     \\  
$\theta^{SpeGR}$ (speculative-grade) & 0.1045 &	0.0022 & 0.000	& 0.000 & 0.000     \\  
\midrule

    \bottomrule

\caption{Results for the estimated causal effect on leverage of having an investment-grade rating ($\theta^{InvGR}$) or a speculative-grade rating ($\theta^{SpeGR}$) versus the baseline of having no rating. The empirical design (\ref{EM-Design}), the data (\ref{EM-Data}) and the random forest characteristics (\ref{EM-Results-AnyRating}) are described in the main text. Standard errors and corresponding p-values are corrected for simultaneous multiple inference: ``MB'' refers to the multiplier bootstrapping method, ``RoWo'' to the Romano-Wolf procedure and ``Bonf'' to the Bonferroni-correction.}
\label{table_Res_General_Rating_InvSpec}
\end{longtable}

Our estimates show that speculative-grade ratings have a large effect on leverage: on average, having a speculative-grade rating increases leverage by nearly 10.5pps. Also, p-values for $\theta^{SpeGR}$ are highly significant across the three reported methods. However, the coefficient estimate for investment-grade ratings $\theta^{InvGR}$ is close to zero with -0.3pps. It is hardly relevant from an economic perspective and p-values are not significant, surpassing 0.20 for the multiplier bootstrap and Romano-Wolf procedure and even 0.40 for the (more conservative) Bondferroni-corrected one. Thus, it is in reality the speculative-grade rating category that drives the (apparent) general rating effect (having any rating versus having no rating) identified in the initial analysis. In contrast, having an investment-grade rating does not affect leverage.   \\

At this stage, the result of our analysis refines the understanding of the rating effect proposed by \cite{Faulkender2005}, who had concluded that firms with a rating, i.e., any rating, have more debt. Rather, our data suggest that firms with low ratings, i.e., speculative-grade ratings, have more debt, while the effect from investment-grade ratings on leverage is approximately zero. Considering these very different results between investment- and speculative-grade ratings, we explore in the following subsection the rating effect by individual broad rating category.\\

\subsection{Effect of rating by individual broad rating category}
\label{EM-Results-IndCat}

The analysis in the previous section yielded a highly heterogeneous treatment effect for the two very general groups of investment- and speculative-grade rating. In this section, we explore whether treatment effects are also heterogeneous at finer levels.\\ 

We remind ourselves from section \ref{CR} that the ``broad'' rating categories are defined by one to three letters (such as AAA, AA, A, BBB). Within the broad categories from AA to CCC, three more granular sub-categories (``notches'') exist, separated by ``+'' and ``-'' signs, for instance AA+, AA and AA-. To add clarity, we will label the granular sub-category ratings without a ``+'' or ``-'' sign as ``straight" (e.g. ``AA\textsuperscript{straight}'') and the broad categories as ``broad'' (e.g. ``AA\textsuperscript{broad}''). Thus, AA\textsuperscript{broad} is comprised of AA+, AA\textsuperscript{straight} and AA-.\\

Applying the same approach as in the investment versus speculative grade rating analysis, we can determine the causal effect estimate for the ten different broad rating categories, again accounting in the methodology for the standard errors and p-value for the fact that we test multiple hypotheses.\\

The results in table \ref{table_Res_Broad_Rating_Categories} provide an interesting picture: effects are highly heterogeneous across the broad rating categories, but follow a distinct pattern. The effect estimates for the two highest-quality ratings (AAA\textsuperscript{broad} and AA\textsuperscript{broad}) are negative and highly significant. The AAA\textsuperscript{broad} rating reduces leverage by approximately -6pps, and the AA\textsuperscript{broad} rating by approximately -4pps.\\

The effect of the next two categories (A\textsuperscript{broad} and BBB\textsuperscript{broad}) can be considered zero, both in terms of the parameter estimate itself (0.01pps and -0.09pps, respectively) and in terms of their p-values, which suggest by their values of 0.956 and 0.677 (for the multiplier bootstrapping method) that the null hypothesis of no effect can hardly be rejected based on the observed data.\\

The coefficient estimates for the next four categories, BB\textsuperscript{broad} to CC\textsuperscript{broad}, are all positive and the corresponding p-values highly significant. Thus, these ratings increase the leverage ratio between approx. 5pps (BB\textsuperscript{broad}) and up to 15pps (CC\textsuperscript{broad}).\\

Coefficient estimates for the categories corresponding to (partial) default, SD\textsuperscript{broad} and D\textsuperscript{broad}, are still positive, albeit of much smaller magnitude; however, their p-values suggest that the null hypothesis of no effect can hardly be rejected.\\

With ten different treatments tested simultaneously, the difference in p-values between the three methods employed to account for simultaneous inference becomes also more pronounced in table \ref{table_Res_Broad_Rating_Categories} as compared to the situation with only two treatment variables in table \ref{table_Res_General_Rating_InvSpec}. However, the results of the three methods are very consistent in their general direction, especially if customary cutoffs (e.g., 0.01 or 0.05) are used for p-values.\footnote{Please see footnote \ref{FN5Percent} in the main part of this paper regarding the p-value controversy.} The results also support the previously indicated view that the Bonferroni correction is more conservative than the two other methods.

\begin{longtable}{lrcccc}
    \toprule
    \multicolumn{6}{l}{\textbf{Broad rating categories}}  \\
  \midrule
\textbf{Rating effect} & Coef. & MB & MB & RoWo & Bonf \\
\textbf{(on LDA)} & Estim. & Std. Error & p-val. & p-val. & p-val. \\
    \midrule
\endfirsthead

 \toprule
   \multicolumn{6}{l}{\textbf{Broad rating categories}}  \\
  \midrule
\textbf{Rating effect} & Coef. & MB & MB & RoWo & Bonf \\
\textbf{(on LDA)} & estim. & Std. error & p-val. & p-val. & p-val. \\
    \midrule
\endhead

\multicolumn{6}{r}{\textit{Continued on next page}}\\ \hline
\endfoot

\hline \hline
\endlastfoot

$\theta^{AAA \ broad}$ & -0.0582 &	0.0189 & 0.002	& 0.015 & 0.021     \\  
$\theta^{AA \ broad}$ & -0.0385 &	0.0068 & 0.000	& 0.000 & 0.000     \\  
$\theta^{A \ broad}$ & 0.0001 &	0.0027 & 0.956	& 0.950 & 1.000     \\  
$\theta^{BBB \ broad}$ & -0.0009 &	0.0021 & 0.677	& 0.942 & 1.000     \\  
$\theta^{BB \ broad}$ & 0.0512 &	0.0024 & 0.000	& 0.000 & 0.000     \\ 
$\theta^{B \ broad}$ & 0.1301 &	0.0031 & 0.000	& 0.000 & 0.000     \\ 
$\theta^{CCC \ broad}$ & 0.1284 &	0.0144 & 0.000	& 0.000 & 0.000     \\ 
$\theta^{CC \ broad}$ & 0.1471 &	0.0044 & 0.001	& 0.004 & 0.008     \\ 
$\theta^{SD \ broad}$ & 0.0597 &	0.0531 & 0.261	& 0.689 & 1.000     \\ 
$\theta^{D \ broad}$ & 0.0141 &	0.0294 & 0.632	& 0.942 & 1.000     \\ 
\midrule

    \bottomrule

\caption{Results for the estimated causal effect on leverage by broad rating category versus the baseline of having no rating. Broad rating categories comprise the ``+'' and ``-'' notch qualifications for those categories within which they exist (e.g., ``AA\textsuperscript{broad}'' includes the S\&P rating categories AA+, AA and AA-). The ``C''-rating category is absent as no firm-year had such a rating over the sample period. The ``SD'' rating indicates that while a ``selective'' default on a particular debt instrument occurred, the company is believed to honor the other obligations. 
Standard errors and corresponding p-values are corrected for simultaneous multiple inference: ``MB'' refers to the multiplier bootstrapping method, ``RoWo'' to the Romano-Wolf procedure and ``Bonf'' to the Bonferroni-correction.}
\label{table_Res_Broad_Rating_Categories}
\end{longtable}

\begin{figure}[htbp]
	\centering
		\includegraphics[width=1.00\textwidth]{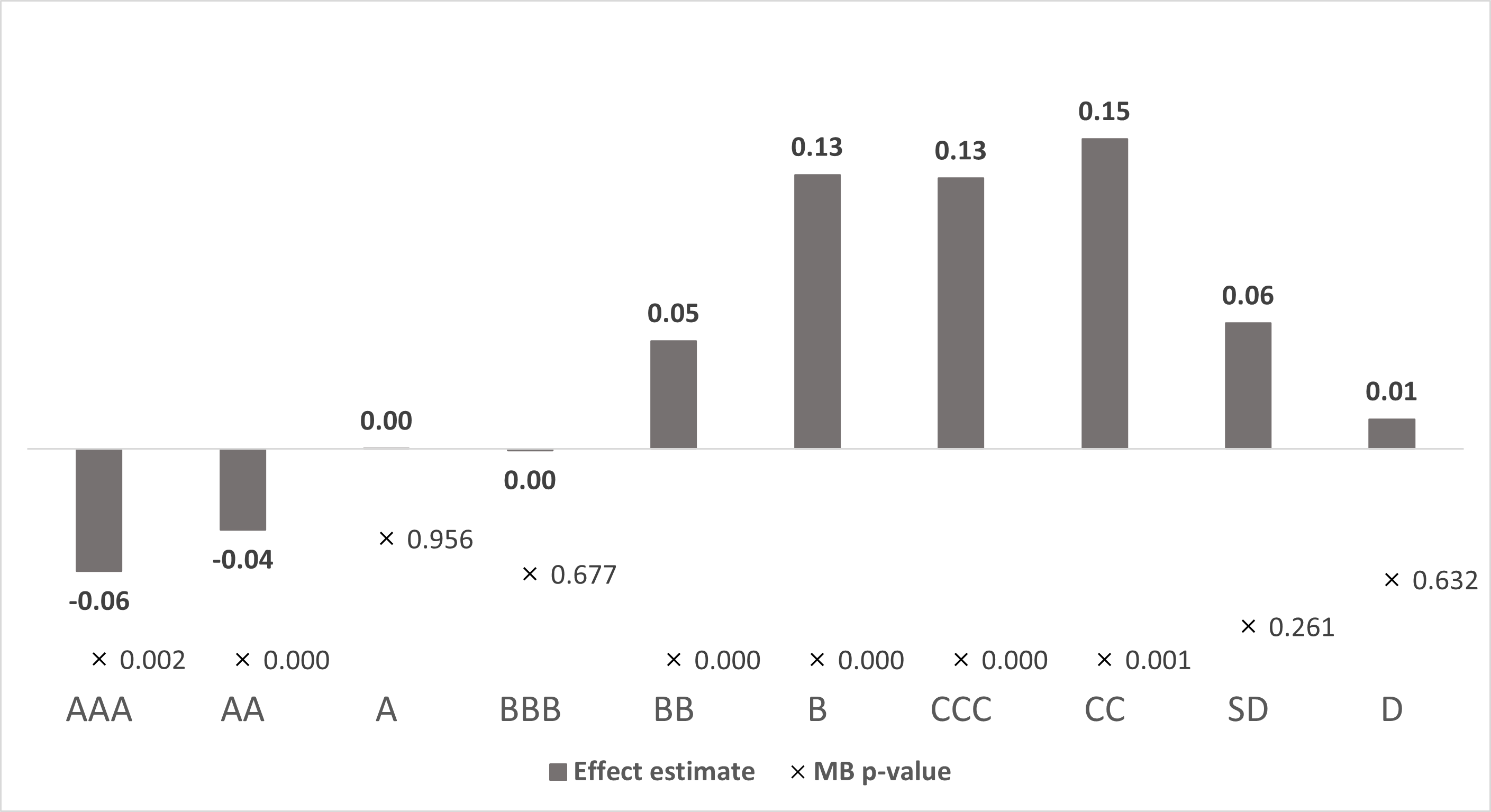}
	\caption{Graphical representation of table  \ref{table_Res_Broad_Rating_Categories} illustrating the heterogeneity of the treatment effect estimates for the ten broad rating categories (gray bars). The numbers have been rounded to two decimal places. For instance, -0.06 for AAA corresponds to -0.0582 in table \ref{table_Res_Broad_Rating_Categories} and indicates that the effect estimate for the broad rating category AAA is a roughly 6pps lower leverage. The values next to the black crosses indicate the respective multiplier bootstrap (MB) p-values (rounded to three decimal places). The position of the black crosses has been selected so as to provide an intuition about the magnitude of the p-values. Note that for ease of reading, we have not added "broad" to the rating category labels.}
	\label{fig:BroadRatingChart}
\end{figure}

Figure \ref{fig:BroadRatingChart} is a graphical representation of the results from table \ref{table_Res_Broad_Rating_Categories}. The shape of the bar chart provides a visual impression about the heterogeneity of the treatment effect and its pronounced pattern following the broad rating categories. From AAA\textsuperscript{broad} to BBB\textsuperscript{broad}, the effect is slightly negative to neutral. From BB\textsuperscript{broad} onward, the effect turns clearly positive (i.e., higher leverage). This is also the dividing line between investment-grade and speculative-grade rating as per the analysis in subsection \ref{EM-Results-InvGSpeG}, which yielded a strong positive effect for speculative-grade rating versus hardly any effect for investment-grade rating. What we interpret as reassuring is the fact that the treatment effect estimates for the individual broad rating categories are very consistent within the two respective ``aggregate categories'' of investment- versus speculative-grade rating. For instance, alternating positive and negative estimates within the speculative categories would appear to be much more counter-intuitive.\\

\begin{figure}[htbp]
	\centering
		\includegraphics[width=1.00\textwidth]{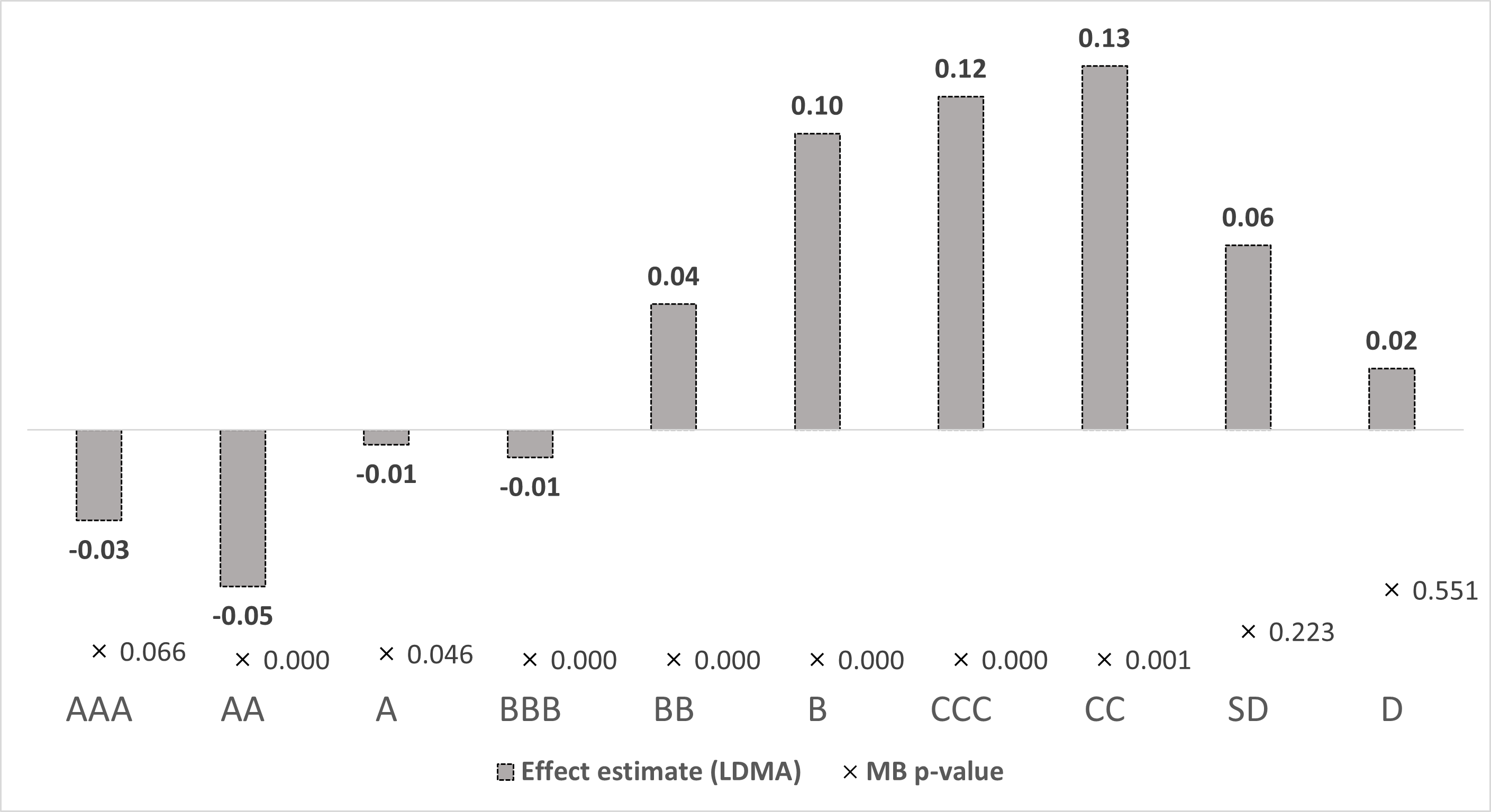}
	\caption{Graphical representation for the effect of the ten broad rating categories on market leverage (LDMA). Similar to the results for book leverage (LDA) in figure \ref{fig:BroadRatingChart}, the chart illustrates the heterogeneity of the treatment effect estimates for the ten broad rating categories (gray bars). Effect estimates have been rounded to two decimal places. The values next to the black crosses indicate the respective multiplier bootstrap (MB) p-values (rounded to three decimal places). The position of the black crosses has been selected so as to provide an intuition about the magnitude of the p-values. Note that for ease of reading, we have not added "broad" to the rating category labels.}
	\label{fig:LDMABroadRatingsChart}
\end{figure}

As a robustness check, we estimate the rating effect by broad category also for the market leverage (LDMA) as defined in equation \ref{eqn:LDMA}. We report here only the graphical representation of the results without commenting them in detail as we consider them very reassuring. Even though AAA\textsuperscript{broad} is slightly lower than AA\textsuperscript{broad}, figure \ref{fig:LDMABroadRatingsChart} for LDMA displays a stark resemblances with the shape in figure \ref{fig:BroadRatingChart}. Notably the overall sequence of effect estimates from negative, roughly neutral to highly positive (and the tapering off at the tail end) resembles the one from our main analysis for LDA.

\subsection{Robustness check: rating effects in a different sample period}
\label{APP-EM-Results-Different-Sample}

As a complementary robustness check for our results from the previous analyses, we consider a second data sample from a different time period. Using double machine learning with the same analytical methodology and data sources as described in subsections \ref{EM-Design}, \ref{EM-Data} and \ref{EM-Results-AnyRating}, we step back in time to the years 2000 to 2004 to arrive at a second data sample of 32'162 company-year observations. With this new data sample, we want to assess our main findings: first, the existence of an effect on leverage from having a rating (versus having no rating); and second, that this rating effect is heterogeneous across rating categories. In particular, we are interested if the second sample confirms the characteristic shape of the rating effect by broad and granular rating category observed in figures \ref{fig:GranularRatingChart} and \ref{fig:BroadRatingChart}. Verifying our findings with this data sample from a different period provides reassurance that the results also hold under potentially different (macro)economic, geopolitical and societal circumstances.\footnote{For instance, the ``dotcom bubble'' burst in 2000 and 2001 saw the terrorist attacks on the World Trade Center; the Euro was introduced in twelve European Union countries in 2002 and 2003 saw the end of Saddam Hussein's rule as Iraqi president; Google's IPO occurred in 2004. More directly relevant to the topic of this paper, \cite{Koller2020} (page 647) observe that the relative increase of companies with speculative grade ratings during 2008 to 2018 was due to newly rated companies entering the debt market, motivated by low interest rates. Thus, the 2000 to 2004 period represents a different environment.} \\

Table \ref{APP-table_Res_General_Rating_11vs5years} compares the results of our main analysis (as per table \ref{table_Res_General_Rating}) in the left column with the results from the second sample period in the right column. The rating effect estimate amounts to 9.6pps, which is 0.8pps higher than the parameter estimate of 8.8pps from the main sample. Compared to the mean leverage of the sample, this corresponds to an impact of 43\% versus 41\% from the main sample. Again, the rating effect is highly significant, both statistically and economically. We interpret this result as adding another piece of evidence confirming the presence of a rating effect.\\

\newpage

\begin{longtable}{lcc}
    \toprule
\textbf{Rating effect} & 2005-2015 & 2000-2004 \\
\textbf{on leverage (LDA)} & n=57'832 & n=32'162 \\
    \midrule
$\theta$ (rating yes/no) & 0.0878 &	0.0962	  \\  
Std. error	&	0.0021 & 0.0029	 \\
t-value & 41.8 & 32.9 \\
p-value	&	0.000 &	0.000 \\
\midrule
\textit{Memo: mean leverage} & \textit{0.212} & \textit{0.224} \\
\textit{Rating effect ($\theta$) vs. mean} & \textit{41\%}	& \textit{43\%} \\
    \bottomrule

\caption{Comparison of results for the estimated causal effect $\theta$ of having or not having a rating on leverage (LDA) for the main data sample from 2005 to 2015 with 57'832 company-year observations compared to a second, different data sample for the years 2000 to 2004 with 32'162 company-year observations. The methodology for the second data sample is the same as for the main one (as described in previous sections), including aggregation of parameter estimates and standard errors over a five-fold split with two repetitions.}
\label{APP-table_Res_General_Rating_11vs5years}
\end{longtable}

\begin{figure}[htbp]
	\centering
		\includegraphics[width=1.00\textwidth]{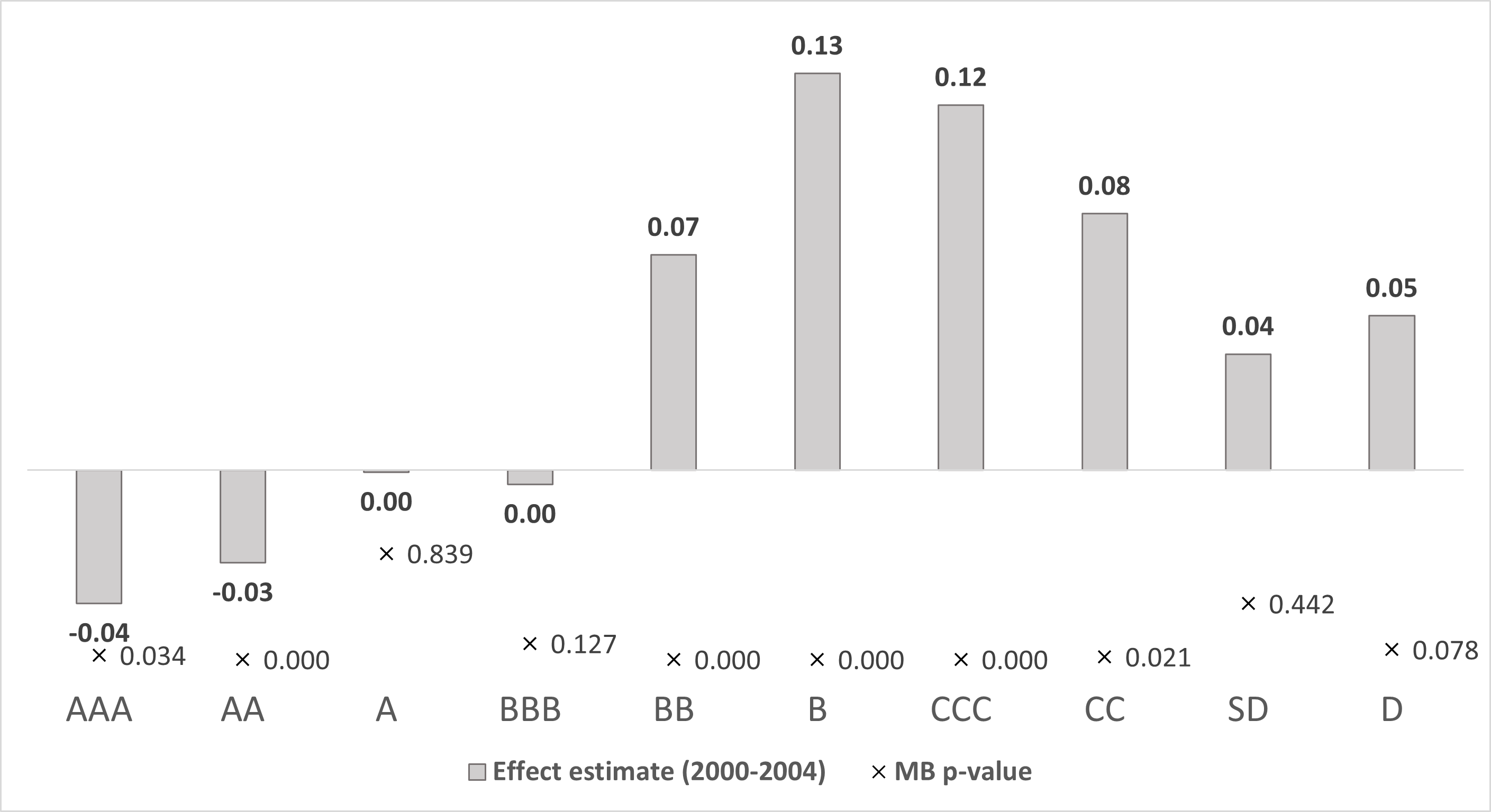}
	\caption{Graphical representation of results for the 2000-2004 five-year period used as robustness check, confirming the heterogeneity of the treatment effect estimates for the ten broad rating categories (light gray bars). The numbers have been rounded to two decimal places. Values below the x-axis indicate negative values (e.g., -0.0007 for A and -0.0046 for B both displayed as 0.00). The values next to the black crosses indicate the respective multiplier bootstrap (MB) p-values (rounded to three decimal places). The position of the black crosses has been selected so as to provide an intuition about the magnitude of the p-values. Note that for ease of reading, we have not added "broad" to the rating category labels.}
	\label{fig:BroadRatingChartFiveYears}
\end{figure}

\begin{figure}[htbp]
	\centering
		\includegraphics[width=1.00\textwidth]{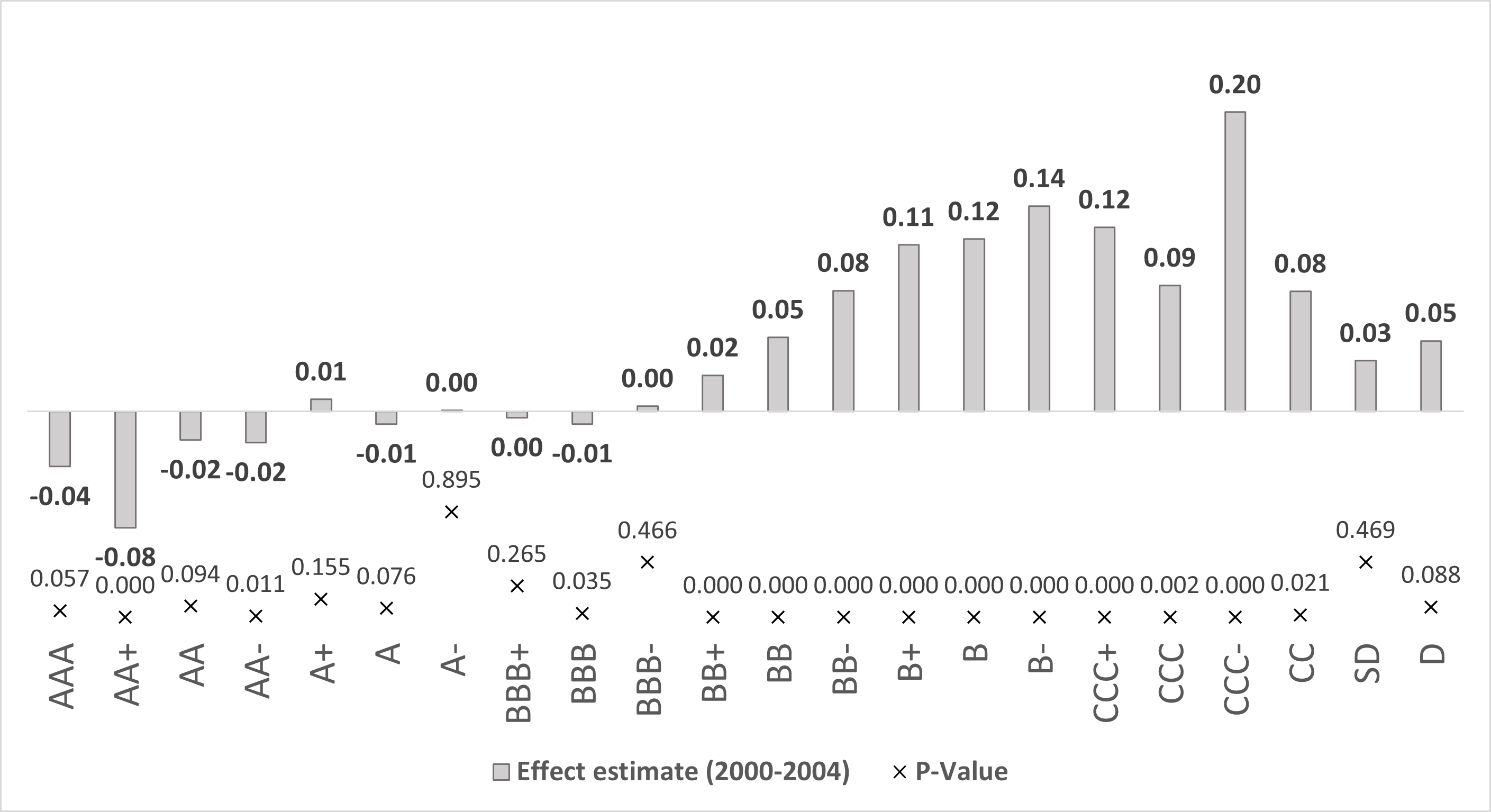}
	\caption{Graphical representation of results for the 2000-2004 five-year period used as robustness check, confirming the heterogeneity of the treatment effect estimates for the 22 granular rating categories (light gray bars). The numbers have been rounded to two decimal places. Values below the x-axis indicate negative values (e.g., -0.0043 displayed as 0.00 for BBB+). The values next to the black crosses indicate the respective multiplier bootstrap (MB) p-values (rounded to three decimal places). The position of the black crosses has been selected so as to provide an intuition about the magnitude of the p-values. Note that for ease of reading, we have not added "straight" to the rating category labels without plus/minus notch qualification.}
	\label{fig:GranularRatingChartFiveYears}
\end{figure}

Figures \ref{fig:BroadRatingChartFiveYears} and \ref{fig:GranularRatingChartFiveYears} are graphical representations of the rating effect estimates from the second data sample for the broad and granular rating categories. The respective multiplier bootstrap p-values are displayed underneath the effect estimates. The shapes in both charts are very similar to the ones resulting from the analysis of the main samples in figures \ref{fig:GranularRatingChart} and \ref{fig:BroadRatingChart}.\\

For the ten broad categories in figure \ref{fig:BroadRatingChartFiveYears}, the effect is slightly negative to neutral from AAA\textsuperscript{broad} to BBB\textsuperscript{broad}. From BB\textsuperscript{broad} onward, the effect turns clearly positive (i.e., higher leverage) and reduces at the tail end in the default categories of SD\textsuperscript{broad} and D\textsuperscript{broad}. As with the main results, the switch from negative/neutral to positive is situated at the dividing line between investment-grade and speculative-grade rating. And again, the individual broad rating treatment effect estimates are very consistent within the two respective ``aggregate categories'' of investment- versus speculative-grade rating.\\

For the 22 granular rating categories in figure \ref{fig:GranularRatingChartFiveYears}, overall results from the second data sample are also very similar to those from the main sample. For the four rating categories without notch qualification (AAA, CC, SD and D), the effect estimates are virtually the same from the granular analysis as compared to the broad analysis. The overall shape of the rating effect curve also resembles the one of the main analysis. Importantly, we again see the gradual increase of the rating effect within the BB category. This confirms our previous observation that the dividing line of the rating impact is not ``sharp'' at the dividing line between investment- (BBB) and speculative-grade (BB) rating.\\

One effect estimate that stands out in figure \ref{fig:GranularRatingChartFiveYears} is the one for the CCC- category. However, similar to what we observed in the main sample, the number of observations in this category is very low (n= 4 in the second sample period versus n=9 in the main sample). A second observation we need to emphasize is the behavior of the estimated rating effects within the A and BBB categories. In the main sample, we found ``concave'' shapes within both categories (please refer to table \ref{table_Res_Granular_Rating_Categories} in the main part of this paper). Specifically, A+ and A- displayed negative effect estimates, while the effect estimate for A\textsuperscript{straight} was positive. The same was true for BBB+ and BBB- relative to BBB\textsuperscript{straight}. In our second sample, the concavity disappears. For A ratings, it actually flips into convexity: A+ and A- display positive effect estimates, while the effect estimate for A\textsuperscript{straight} is negative. And within the category of BBB ratings, BBB+ and BBB\textsuperscript{straight} display a negative impact, while the effect estimate is positive for BBB+. We conclude that the hesitation voiced in the main part of the text to build elaborate theories on such findings is justified. Our ad-hoc interpretation in subsection \ref{EM-Results-GranCat} for the concavity which we found in the main samples could probably serve as example that ``[h]umans are extraordinarily quick to infer that the events they observe are caused by creatures with plans and intentions'' \cite{Blackmore2017} (page 94).\\

\begin{figure}[htbp]
	\centering
		\includegraphics[width=1.00\textwidth]{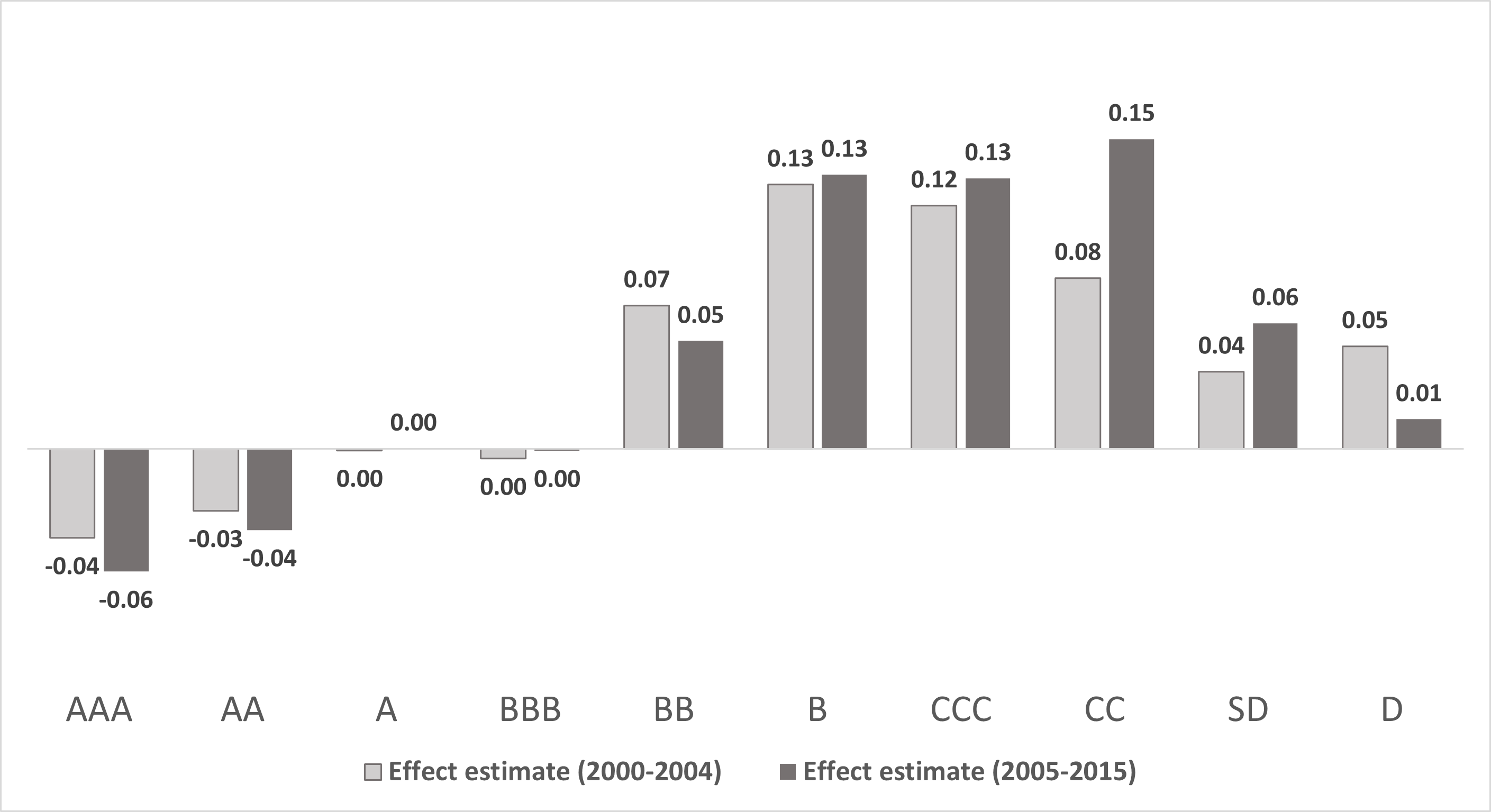}
	\caption{Graphical comparison of the results for the 2005-2015 period from the main analysis (dark gray bars) in this paper with the results from the 2000-2004 period (light gray bars) used as robustness check for the ``effect shape'' of the ten broad rating categories. Effect estimates have been rounded to two decimal places. Values below the x-axis indicate negative values (e.g., -0.0007 for A in the 2000-2004 sample displayed as 0.00). For ease of reading, the chart does not repeat the  respective multiplier bootstrap (MB) p-values (already displayed in previous charts). Also, we have not added "broad" to the rating category labels.}
	\label{fig:CompBroadRatingChartElevenFiveYears}
\end{figure}

\begin{figure}[htbp]
	\centering
		\includegraphics[width=1.00\textwidth]{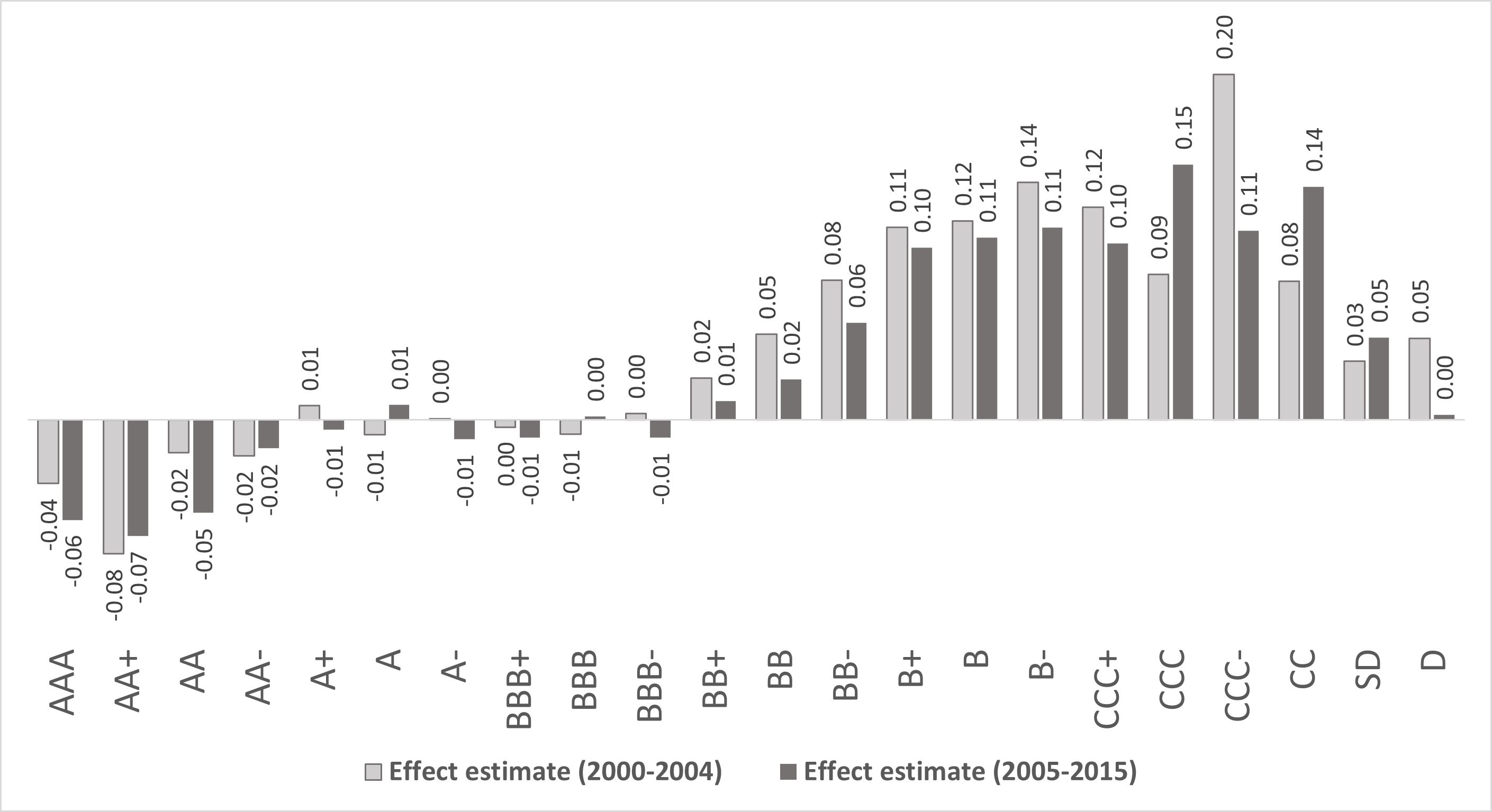}
	\caption{Graphical comparison of the results for the 2005-2015 period from the main analysis (dark gray bars) in this paper with the results from the 2000-2004 period (light grea bars) used as robustness check for the ``effect shape'' of the 22 granular rating categories. Effect estimates have been rounded to two decimal places. Values below the x-axis indicate negative values (e.g., -0.0043 displayed as 0.00 for BBB+ in the 2000-2004 sample). For ease of reading, the chart does not repeat the  respective multiplier bootstrap (MB) p-values (already displayed in previous charts). Also, we have not added "straight" to the rating category labels without plus/minus notch qualification.}
	\label{fig:APP-CompGranularRatingChartElevenFiveYears}
\end{figure}

For ease of comparison, we also plot the effect estimates from the main analysis next to the ones from the second data sample for the broad categories (figure 
\ref{fig:CompBroadRatingChartElevenFiveYears}) and for the granular rating categories (figure  \ref{fig:APP-CompGranularRatingChartElevenFiveYears}). The similarity of the shapes from the main and the second data sample is compelling.\\

\subsection{Robustness check: rating effects when including interest coverage as a covariate}
\label{APP-EM-Results-IntCov}
\label{APP-IntCov}



As described in subsection \ref{EM-Data}, we excluded data items that would allow the random forest as a very flexible learner to back-calculate total debt or equity. However, we still want to verify that the rating effect estimates hold when including selected items that determine credit ratings (or are at least strongly believed to do so). \cite{Koller2020} (pages 645-650) explain that ``credit ratings are primarily related to two financial indicators'' (page 647). One of them is size, which we have already included via items such as the logarithm of sales, the logarithm of assets or the number of employees in our set of covariates. The second is interest coverage, which measures ``a company's ability to comply with its debt service obligations'' (page 648). We therefore include interest coverage ($IntCov$) as defined in \cite{Koller2020} (Exhibit 33.8, left panel, page 649):

\begin{align}
\label{APP-eqn:IntCov}
IntCov_{i,t}=\textit{EBITDA}_{i,t}  / \textit{Interest expenses}_{i,t}
\end{align}

where $EBITDA$ represents earnings before interest, taxes, depreciation and amortization and $interest$ $expenses$ represents the expenses for servicing a company's total financial debt.\footnote{In Compustat, this is the item with code ``xint'' (``Interest and Related Expense - Total'').}\\

We use our empirical sample as described in subsection \ref{EM-Data} and remove company-years with interest expenses of less than USD ten thousand in a given year and arrive at 48'585 company-year observations. We make no change to the double machine learning model as described in subsections \ref{EM-Design} and \ref{EM-Results-AnyRating}. Table \ref{APP-table_Res_General_Rating_IntCov} compares the results for the general rating effect estimate from for our main analysis (as per table \ref{table_Res_General_Rating}) with those from the approach in this subsection which includes interest coverage (``IntCov'') as a feature in the set of covariates. The effect estimate amounts to approximately 7pps including IntCov, or 29\% versus the sample mean leverage of roughly 25\%. This effect estimate is 1.5pps lower than the one from the main analysis, which translates into a 10pps drop in the relative effect magnitude versus the mean leverage (29\% versus 41\% in the main analysis). Still, the rating effect remains clearly present. Key is now to assess the effect heterogeneity and shape of the effect curve in subsequent steps. \\

\newpage

\begin{longtable}{lcc}
    \toprule
\textbf{Rating effect} & Excl. IntCov & Incl. IntCov \\
\textbf{on leverage (LDA)} & n=57'832 & n=48'585 \\
    \midrule
$\theta$ (rating yes/no) & 0.0878 &	0.0731	  \\  
Std. error	&	0.0021 & 0.0021	 \\
t-value & 41.8 & 35.3 \\
p-value	&	0.000 &	0.000 \\
\midrule
\textit{Memo: mean leverage} & \textit{0.212} & \textit{0.249} \\
\textit{Rating effect ($\theta$) vs. mean} & \textit{41\%}	& \textit{29\%} \\
    \bottomrule

\caption{Comparison of results for the estimated causal effect $\theta$ of having a rating (or not) on leverage (LDA) depending on whether interest coverage (``IntCov'') as defined in equation \ref{APP-eqn:IntCov} is excluded or included in the set X of covariates as per equations \ref{eqn:PLR main equation} and \ref{eqn:PLR confounding equation}. The general methodology for ``Incl. IntCov'' is the same as for the main model used throughout this paper (``Excl. IntCov'', as described in previous sections), including aggregation of parameter estimates and standard errors over a five-fold split with two repetitions.}
\label{APP-table_Res_General_Rating_IntCov}
\end{longtable}

\begin{figure}[htbp]
	\centering
		\includegraphics[width=1.00\textwidth]{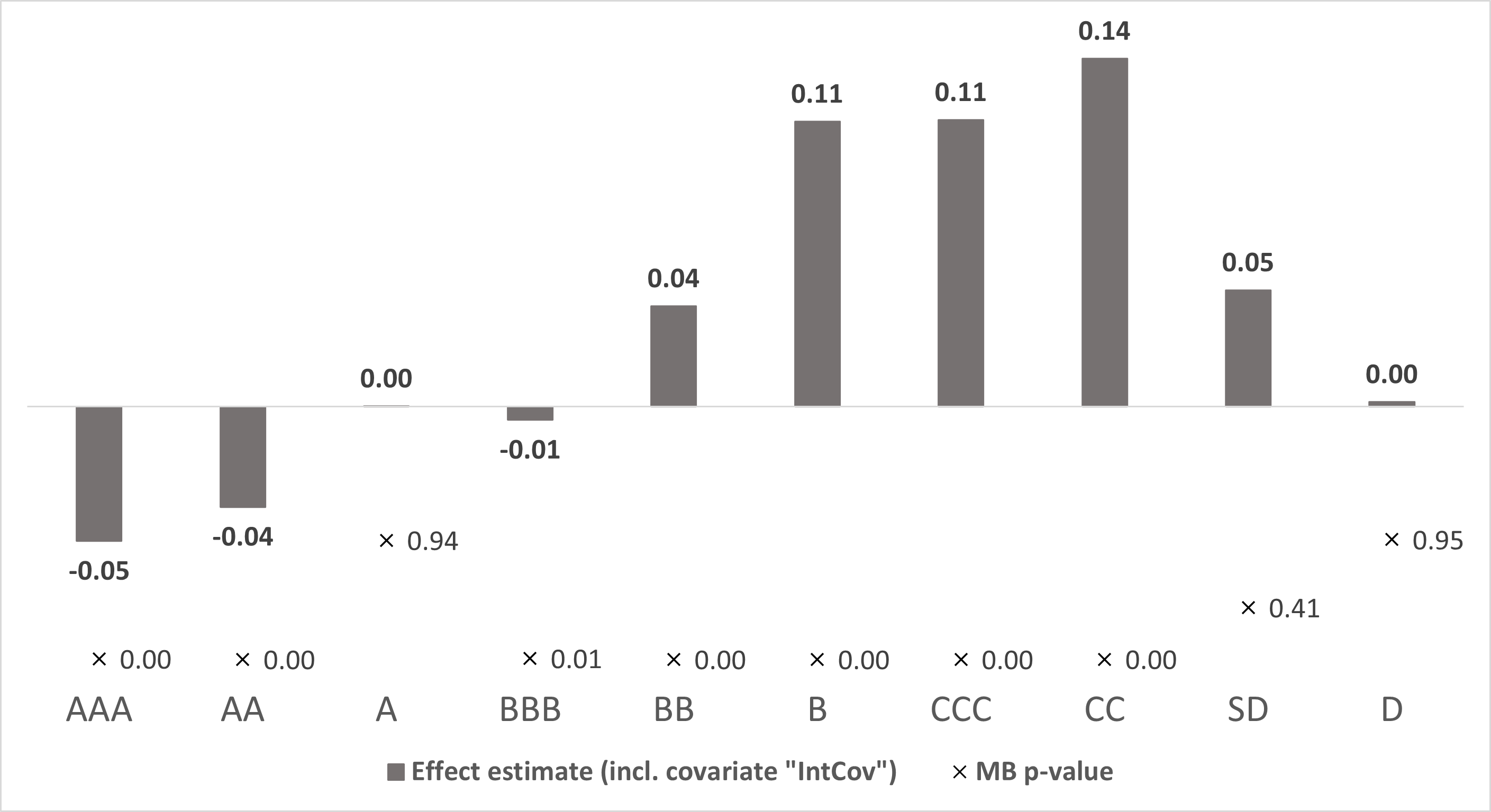}
	\caption{Graphical representation of table  \ref{table_IntCov_Res_Broad_Rating_Categories} illustrating the heterogeneity of the treatment effect estimates for the ten broad rating categories (gray bars). The numbers have been rounded to two decimal places. For instance, -0.05 for AAA corresponds to -0.0532 in table \ref{table_IntCov_Res_Broad_Rating_Categories} and indicates that the effect estimate for the broad rating category AAA is a roughly -5pps lower leverage. The values next to the black crosses indicate the respective multiplier bootstrap (MB) p-values (rounded to three decimal places). The position of the black crosses has been selected so as to provide an intuition about the magnitude of the p-values. Note that for ease of reading, we have not added "broad" to the rating category labels.}
	\label{fig:IntCovBroadRatingChart}
\end{figure}

\begin{figure}[htbp]
	\centering
		\includegraphics[width=1.00\textwidth]{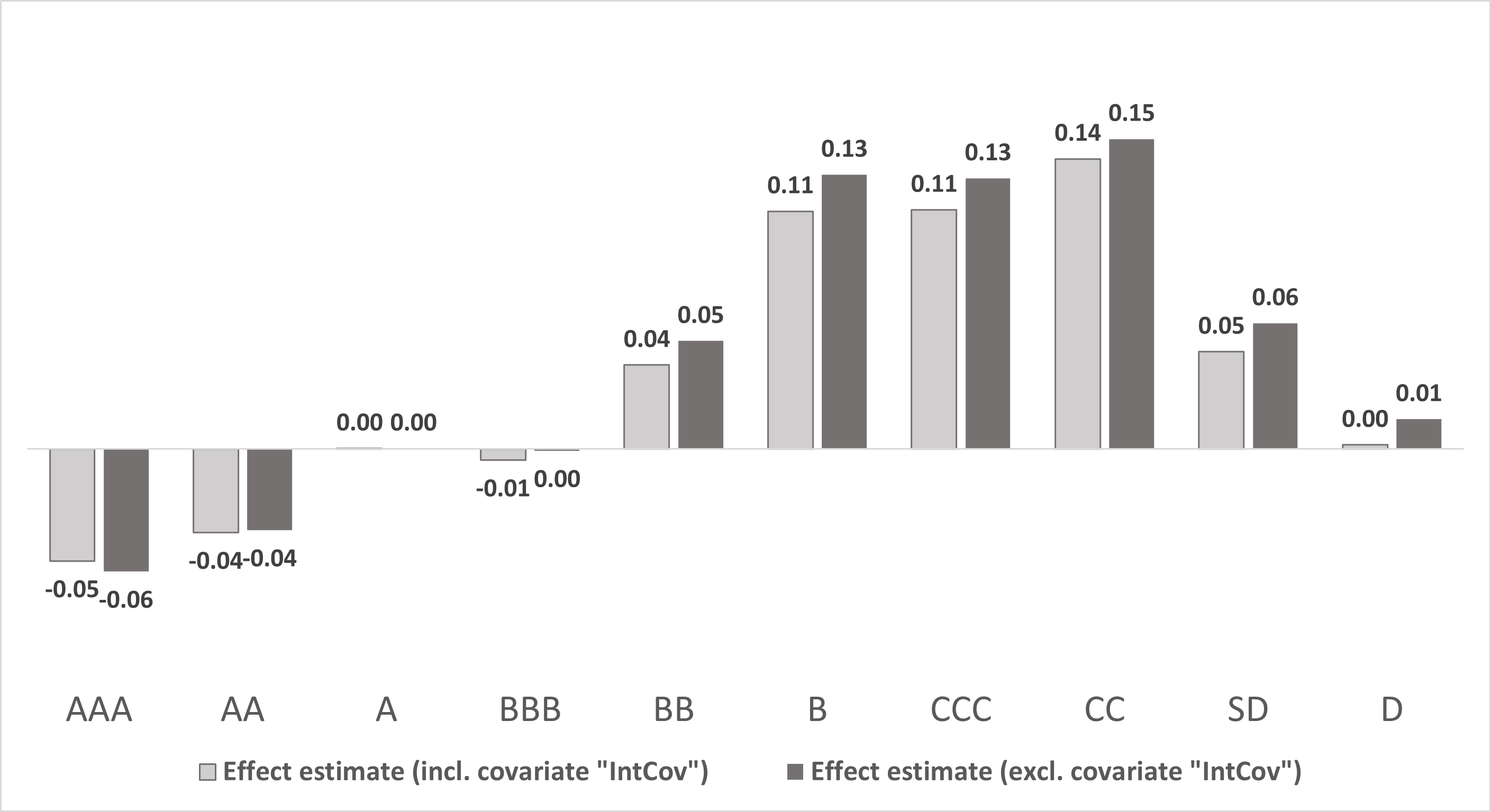}
	\caption{Graphical comparison of the results for the rating effect by broad rating category from the main analysis (excluding IntCov, dark gray bars) in this paper with the results from the analysis including IntCov (light gray bars) used as robustness check for the effect shape. Effect estimates have been rounded to two decimal places. Values below the x-axis indicate negative values (e.g., -0.0009 for BBB excl. covariate ``IntCov'' reported in table \ref{table_Res_Broad_Rating_Categories} is displayed as 0.00 in this figure). For ease of reading, the chart does not repeat the  respective multiplier bootstrap (MB) p-values (already displayed in previous charts). Also, we have not added "broad" to the rating category labels.}
	\label{fig:CompBroadRatingChartIntCov}
\end{figure}

The results for the ten broad rating categories are summarized in table \ref{table_IntCov_Res_Broad_Rating_Categories}. Figure \ref{fig:IntCovBroadRatingChart} provides a graphical representation of their effect heterogeneity and figure \ref{fig:CompBroadRatingChartIntCov} compares the effect shape with the results from the main analysis previously reported in table \ref{table_Res_Broad_Rating_Categories}.\\

\begin{longtable}{lrcccc}
    \toprule
    \multicolumn{6}{l}{\textbf{Broad rating categories with IntCov included as covariate}}  \\
  \midrule
\textbf{Rating effect} & Coef. & MB & MB & RoWo & Bonf \\
\textbf{(on LDA)} & estim. & Std. error & p-val. & p-val. & p-val. \\
    \midrule
\endfirsthead

 \toprule
   \multicolumn{6}{l}{\textbf{Broad rating categories with IntCov included as covariate}}  \\
  \midrule
\textbf{Rating effect} & Coef. & MB & MB & RoWo & Bonf \\
\textbf{(on LDA)} & estim. & Std. error & p-val. & p-val. & p-val. \\
    \midrule
\endhead

\multicolumn{6}{r}{\textit{Continued on next page}}\\ \hline
\endfoot

\hline \hline
\endlastfoot

$\theta^{AAA \ broad}$ 	&	-0.0532	&	0.0187	&	0.004	&	0.024	&	0.044	\\
$\theta^{AA \ broad}$ 	&	-0.0397	&	0.0064	&	0.000	&	0.000	&	0.000	\\
$\theta^{A \ broad}$ 	&	0.0002	&	0.0026	&	0.943	&	0.998	&	1.000	\\
$\theta^{BBB \ broad}$ 	&	-0.0053	&	0.0020	&	0.008	&	0.032	&	0.078	\\
$\theta^{BB \ broad}$ 	&	0.0398	&	0.0024	&	0.000	&	0.000	&	0.000	\\
$\theta^{B \ broad}$ 	&	0.1128	&	0.0031	&	0.000	&	0.000	&	0.000	\\
$\theta^{CCC \ broad}$ 	&	0.1135	&	0.0142	&	0.000	&	0.000	&	0.000	\\
$\theta^{CC \ broad}$ 	&	0.1377	&	0.0429	&	0.001	&	0.007	&	0.013	\\
$\theta^{SD \ broad}$ 	&	0.0461	&	0.0560	&	0.410	&	0.799	&	1.000	\\
$\theta^{D \ broad}$ 	&	0.0019	&	0.0286	&	0.948	&	0.998	&	1.000	\\

\midrule

    \bottomrule

\caption{Results for the estimated causal effect on leverage (LDA) by broad rating category with interest coverage (``IntCov'') included in the set of covariates. Standard errors and corresponding p-values are corrected for simultaneous multiple inference: ``MB'' refers to the multiplier bootstrapping method, ``RoWo'' to the Romano-Wolf procedure and ``Bonf'' to the Bonferroni-correction. Parameter estimates and standard errors are aggregated over a five-fold split with two repetitions. The analytical approach within the double machine learning framework is the same as previously described for the main analyses in this paper.}
\label{table_IntCov_Res_Broad_Rating_Categories}
\end{longtable}

Results for the broad rating categories from the double machine learning specification including interest coverage are very similar to the ones from the main analysis without interest coverage. First, results confirm the heterogeneity of the effects: for the two highest rating categories are again negative, then effect estimates are zero, before increasing to positive for BB ratings and exceeding 10pps for the remainder of ratings excluding SD and D default ratings. Second, the differences between the effect estimates along the rating scale are similar to the one from the main analysis, thus yielding the same overall shape of the effect curve. Figure \ref{fig:CompBroadRatingChartIntCov} illustrates these two points. Third, p-values across the three methodologies employed to account for multiple hypothesis testing are consistent with each other.\\

Figure \ref{fig:CompBroadRatingChartIntCov} also confirms the observation from the result regarding the effect of any rating versus no rating, reported in table \ref{APP-table_Res_General_Rating_IntCov}. There, the absolute effect estimate was roughly 1.5pps lower including interest coverage as an additional covariate than excluding it. The same is true for the individual effect estimates by broad rating category: including interest coverage, most of them are between 1 and 2pps lower than excluding interest coverage, but nevertheless economically relevant and highly significant from a statistical perspective.\\

\begin{figure}[htbp]
	\centering
		\includegraphics[width=1.00\textwidth]{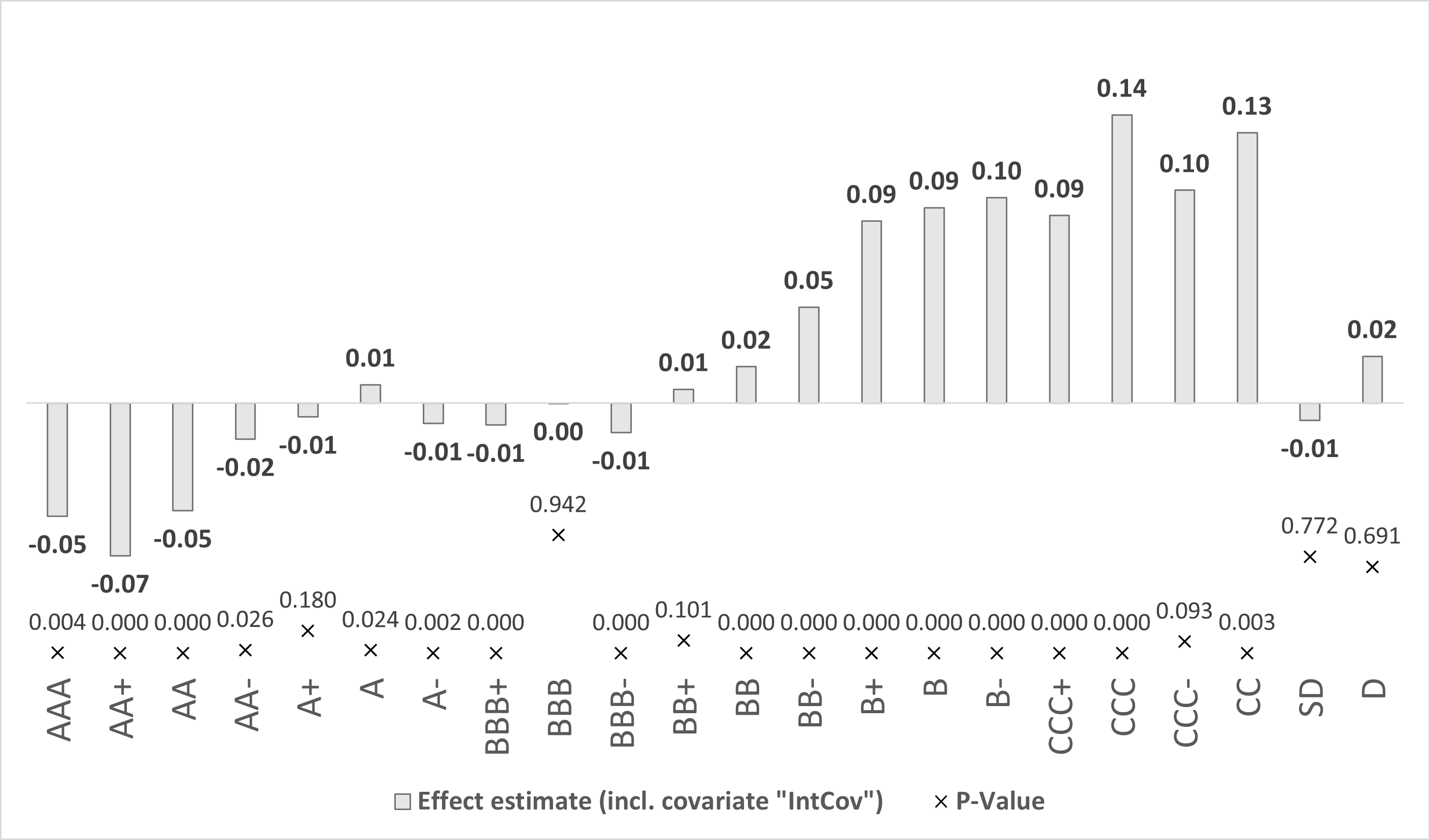}
	\caption{Graphical representation illustrating the heterogeneity of the treatment effect estimates for the 22 granular rating categories (gray bars). The numbers have been rounded to two decimal places. The values next to the black crosses indicate the respective multiplier bootstrap (MB) p-values (rounded to three decimal places). The position of the black crosses has been selected so as to provide an intuition about the magnitude of the p-values. Note that for ease of reading, we have not added "straight" to the rating category labels without ``+/-'' notch qualifications.}
	\label{fig:IntCovGranularRatingChart}
\end{figure}

\begin{figure}[htbp]
	\centering
		\includegraphics[width=1.00\textwidth]{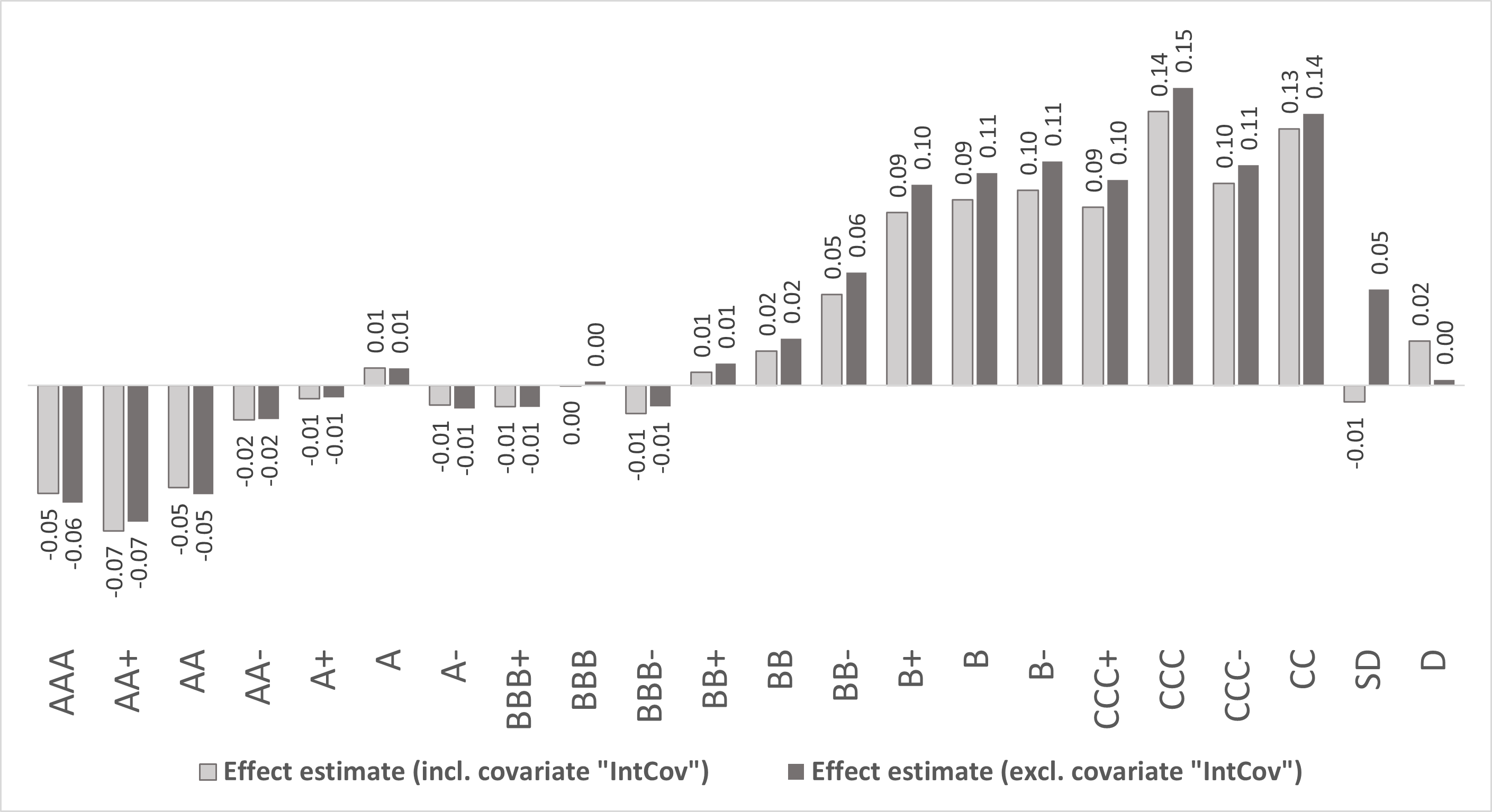}
	\caption{Graphical comparison for the effect estimate of the 22 granular rating categories on leverage (LDA) including ``IntCov'' as covariate feature (light gray bars) versus the results from the main analyses of this paper where ``IntCov'' was not included (dark gray bars). Effect estimates have been rounded to two decimal places. Values below (above) the x-axis indicate a negative (positive) effect (e.g., for BBB incl. IntCov, the effect estimate is -0.0002, while it is +0.0019 for BBB excl. IntCov; both are displayed as 0.00 in the figure). Note that for ease of reading, we have not added "broad" to the rating category labels.}
	\label{fig:APP-IntCovGranularComparison}
\end{figure}

For the 22 granular rating categories, we provide in this subsection with figure \ref{fig:IntCovGranularRatingChart} the effect estimates together with the corresponding multiplier bootstrap p-values and also a graphical comparison of results including versus excluding interest coverage in figure \ref{fig:APP-IntCovGranularComparison}. Also at this granular level, the results including interest coverage are very similar to those from the main analysis without interest coverage, both in terms of magnitude and overall shape. In particular, we also see the gradual rise in effect size over the notch-ratings within the BBB and BB rating classes, which comforts our previous finding that there is no sharp divide in effect between investment-grade and speculative -rade ratings. \\

In conclusion, results from the robustness checks confirm our three main conclusions on the rating effects on leverage: first, ratings affect the leverage ratio. Second, this effect is heterogeneous and depends on the rating category. Third, the transition of the effect size is gradual over the individual, granular categories within BBB and BB and thus does not occur sharp at the switch from investment to speculative grade. \\

\newpage
\footnotesize
\pagebreak
\bibliographystyle{imsart-number}
\bibliography{Literature_HW_P03_v01_221123}

\begin{thebibliography}{120}

\bibitem{Altman1968}
\begin{barticle}[author]
\bauthor{\bsnm{Altman},~\bfnm{Edward~I.}\binits{E.~I.}}
(\byear{1968}).
\btitle{Financial Ratios, Discriminant Analysis and the Prediction of Corporate Bankruptcy}.
\bjournal{The Journal of Finance}
\bvolume{23}
\bpages{589--609}.
\bdoi{10.2307/2978933}
\end{barticle}
\endbibitem

\bibitem{Altman2013}
\begin{bincollection}[author]
\bauthor{\bsnm{Altman},~\bfnm{Edward~I}\binits{E.~I.}}
(\byear{2013}).
\btitle{Predicting financial distress of companies: revisiting the Z-score and ZETA{\textregistered} models}.
In \bbooktitle{Handbook of research methods and applications in empirical finance}
\bpublisher{Edward Elgar Publishing}.
\end{bincollection}
\endbibitem

\bibitem{Amato2004}
\begin{barticle}[author]
\bauthor{\bsnm{Amato},~\bfnm{Jeffery~D.}\binits{J.~D.}} \AND \bauthor{\bsnm{Furfine},~\bfnm{Craig~H.}\binits{C.~H.}}
(\byear{2004}).
\btitle{Are credit ratings procyclical?}
\bjournal{Journal of Banking \& Finance}
\bvolume{28}
\bpages{2641-2677}.
\bnote{Recent Research on Credit Ratings}.
\bdoi{https://doi.org/10.1016/j.jbankfin.2004.06.005}
\end{barticle}
\endbibitem

\bibitem{Amini2021}
\begin{barticle}[author]
\bauthor{\bsnm{Amini},~\bfnm{Shahram}\binits{S.}}, \bauthor{\bsnm{Elmore},~\bfnm{Ryan}\binits{R.}}, \bauthor{\bsnm{{\"O}ztekin},~\bfnm{{\"O}zde}\binits{{\"O}.}} \AND \bauthor{\bsnm{Strauss},~\bfnm{Jack}\binits{J.}}
(\byear{2021}).
\btitle{Can machines learn capital structure dynamics?}
\bjournal{Journal of Corporate Finance}
\bvolume{70}
\bpages{102073}.
\bdoi{https://doi.org/10.1016/j.jcorpfin.2021.102073}
\end{barticle}
\endbibitem

\bibitem{Amrhein2019}
\begin{barticle}[author]
\bauthor{\bsnm{Amrhein},~\bfnm{Valentin}\binits{V.}}, \bauthor{\bsnm{Greenland},~\bfnm{Sander}\binits{S.}} \AND \bauthor{\bsnm{McShane},~\bfnm{Blake}\binits{B.}}
(\byear{2019}).
\btitle{Scientists rise up against statistical significance.}
\bjournal{Nature}
\bvolume{567}
\bpages{305-307}.
\end{barticle}
\endbibitem

\bibitem{Angrist2008}
\begin{bbook}[author]
\bauthor{\bsnm{Angrist},~\bfnm{Joshua~D}\binits{J.~D.}} \AND \bauthor{\bsnm{Pischke},~\bfnm{J{\"o}rn-Steffen}\binits{J.-S.}}
(\byear{2008}).
\btitle{Mostly harmless econometrics}.
\bpublisher{Princeton university press}.
\end{bbook}
\endbibitem

\bibitem{AshbaughSkaife2006}
\begin{barticle}[author]
\bauthor{\bsnm{Ashbaugh-Skaife},~\bfnm{Hollis}\binits{H.}}, \bauthor{\bsnm{Collins},~\bfnm{Daniel~W.}\binits{D.~W.}} \AND \bauthor{\bsnm{LaFond},~\bfnm{Ryan}\binits{R.}}
(\byear{2006}).
\btitle{The effects of corporate governance on firms' credit ratings}.
\bjournal{Journal of Accounting and Economics}
\bvolume{42}
\bpages{203-243}.
\bnote{Conference Issue on Implications of Changing Financial Reporting Standards}.
\bdoi{https://doi.org/10.1016/j.jacceco.2006.02.003}
\end{barticle}
\endbibitem

\bibitem{Athey2018}
\begin{bincollection}[author]
\bauthor{\bsnm{Athey},~\bfnm{Susan}\binits{S.}}
(\byear{2018}).
\btitle{The impact of machine learning on economics}.
In \bbooktitle{The economics of artificial intelligence: An agenda}
\bpages{507--547}.
\bpublisher{University of Chicago Press}.
\end{bincollection}
\endbibitem

\bibitem{Bach2022}
\begin{barticle}[author]
\bauthor{\bsnm{Bach},~\bfnm{Philipp}\binits{P.}}, \bauthor{\bsnm{Chernozhukov},~\bfnm{Victor}\binits{V.}}, \bauthor{\bsnm{Kurz},~\bfnm{Malte~S.}\binits{M.~S.}} \AND \bauthor{\bsnm{Spindler},~\bfnm{Martin}\binits{M.}}
(\byear{2022}).
\btitle{DoubleML -- An Object-Oriented Implementation of Double Machine Learning in R}.
\bdoi{10.48550/ARXIV.2104.03220}
\end{barticle}
\endbibitem

\bibitem{Bach2018}
\begin{bmisc}[author]
\bauthor{\bsnm{Bach},~\bfnm{Philipp}\binits{P.}}, \bauthor{\bsnm{Chernozhukov},~\bfnm{Victor}\binits{V.}} \AND \bauthor{\bsnm{Spindler},~\bfnm{Martin}\binits{M.}}
(\byear{2018}).
\btitle{Valid Simultaneous Inference in High-Dimensional Settings (with the hdm package for R)}.
\bdoi{10.48550/ARXIV.1809.04951}
\end{bmisc}
\endbibitem

\bibitem{BAGHAI2014}
\begin{barticle}[author]
\bauthor{\bsnm{BAGHAI},~\bfnm{RAMIN~P.}\binits{R.~P.}}, \bauthor{\bsnm{SERVAES},~\bfnm{HENRI}\binits{H.}} \AND \bauthor{\bsnm{TAMAYO},~\bfnm{ANE}\binits{A.}}
(\byear{2014}).
\btitle{Have Rating Agencies Become More Conservative? Implications for Capital Structure and Debt Pricing}.
\bjournal{The Journal of Finance}
\bvolume{69}
\bpages{1961-2005}.
\bdoi{https://doi.org/10.1111/jofi.12153}
\end{barticle}
\endbibitem

\bibitem{Baker2002}
\begin{barticle}[author]
\bauthor{\bsnm{Baker},~\bfnm{Malcolm}\binits{M.}} \AND \bauthor{\bsnm{Wurgler},~\bfnm{Jeffrey}\binits{J.}}
(\byear{2002}).
\btitle{Market Timing and Capital Structure}.
\bjournal{The Journal of Finance}
\bvolume{57}
\bpages{1-32}.
\bdoi{https://doi.org/10.1111/1540-6261.00414}
\end{barticle}
\endbibitem

\bibitem{Bancel2004}
\begin{barticle}[author]
\bauthor{\bsnm{Bancel},~\bfnm{Franck}\binits{F.}} \AND \bauthor{\bsnm{Mittoo},~\bfnm{Usha~R.}\binits{U.~R.}}
(\byear{2004}).
\btitle{Cross-Country Determinants of Capital Structure Choice: A Survey of European Firms}.
\bjournal{Financial Management}
\bvolume{33}
\bpages{103--132}.
\end{barticle}
\endbibitem

\bibitem{Barclay2005}
\begin{barticle}[author]
\bauthor{\bsnm{Barclay},~\bfnm{Michael~J.}\binits{M.~J.}} \AND \bauthor{\bsnm{Smith},~\bfnm{Clifford~W.}\binits{C.~W.}}
(\byear{2005}).
\btitle{The Capital Structure Puzzle: The Evidence Revisited}.
\bjournal{Journal of Applied Corporate Finance}
\bvolume{17}
\bpages{8-17}.
\bdoi{https://doi.org/10.1111/j.1745-6622.2005.012\_2.x}
\end{barticle}
\endbibitem

\bibitem{Barclay1999}
\begin{barticle}[author]
\bauthor{\bsnm{Barclay},~\bfnm{Michael~J.}\binits{M.~J.}} \AND \bauthor{\bsnm{Smith~Jr.},~\bfnm{Clifford~W.}\binits{C.~W.}}
(\byear{1999}).
\btitle{THE CAPITAL STRUCTURE PUZZLE: ANOTHER LOOK AT THE EVIDENCE}.
\bjournal{Journal of Applied Corporate Finance}
\bvolume{12}
\bpages{8-20}.
\bdoi{https://doi.org/10.1111/j.1745-6622.1999.tb00655.x}
\end{barticle}
\endbibitem

\bibitem{Becker2011}
\begin{barticle}[author]
\bauthor{\bsnm{Becker},~\bfnm{Bo}\binits{B.}} \AND \bauthor{\bsnm{Milbourn},~\bfnm{Todd}\binits{T.}}
(\byear{2011}).
\btitle{How did increased competition affect credit ratings?}
\bjournal{Journal of Financial Economics}
\bvolume{101}
\bpages{493-514}.
\bdoi{https://doi.org/10.1016/j.jfineco.2011.03.012}
\end{barticle}
\endbibitem

\bibitem{Belloni2013}
\begin{barticle}[author]
\bauthor{\bsnm{Belloni},~\bfnm{Alexandre}\binits{A.}}, \bauthor{\bsnm{Chernozhukov},~\bfnm{Victor}\binits{V.}} \AND \bauthor{\bsnm{Hansen},~\bfnm{Christian}\binits{C.}}
(\byear{2013}).
\btitle{{Inference on Treatment Effects after Selection among High-Dimensional Controls}}.
\bjournal{The Review of Economic Studies}
\bvolume{81}
\bpages{608-650}.
\bdoi{10.1093/restud/rdt044}
\end{barticle}
\endbibitem

\bibitem{Belloni2014}
\begin{barticle}[author]
\bauthor{\bsnm{Belloni},~\bfnm{Alexandre}\binits{A.}}, \bauthor{\bsnm{Chernozhukov},~\bfnm{Victor}\binits{V.}} \AND \bauthor{\bsnm{Hansen},~\bfnm{Christian}\binits{C.}}
(\byear{2014}).
\btitle{High-Dimensional Methods and Inference on Structural and Treatment Effects}.
\bjournal{Journal of Economic Perspectives}
\bvolume{28}
\bpages{29-50}.
\bdoi{10.1257/jep.28.2.29}
\end{barticle}
\endbibitem

\bibitem{Benjamini1995}
\begin{barticle}[author]
\bauthor{\bsnm{Benjamini},~\bfnm{Yoav}\binits{Y.}} \AND \bauthor{\bsnm{Hochberg},~\bfnm{Yosef}\binits{Y.}}
(\byear{1995}).
\btitle{Controlling the False Discovery Rate: A Practical and Powerful Approach to Multiple Testing}.
\bjournal{Journal of the Royal Statistical Society: Series B (Methodological)}
\bvolume{57}
\bpages{289-300}.
\bdoi{https://doi.org/10.1111/j.2517-6161.1995.tb02031.x}
\end{barticle}
\endbibitem

\bibitem{Bera2001}
\begin{barticle}[author]
\bauthor{\bsnm{Bera},~\bfnm{Anil~K.}\binits{A.~K.}} \AND \bauthor{\bsnm{Bilias},~\bfnm{Yannis}\binits{Y.}}
(\byear{2001}).
\btitle{Rao's score, Neyman's C($\alpha$) and Silvey's LM tests: an essay on historical developments and some new results}.
\bjournal{Journal of Statistical Planning and Inference}
\bvolume{97}
\bpages{9-44}.
\bnote{Rao's Score Test}.
\bdoi{https://doi.org/10.1016/S0378-3758(00)00343-8}
\end{barticle}
\endbibitem

\bibitem{Bernstein1993}
\begin{bbook}[author]
\bauthor{\bsnm{Bernstein},~\bfnm{Peter~L}\binits{P.~L.}}
(\byear{1993}).
\btitle{Capital ideas: the improbable origins of modern Wall Street}.
\bpublisher{Simon and Schuster}.
\end{bbook}
\endbibitem

\bibitem{Bertrand2003}
\begin{barticle}[author]
\bauthor{\bsnm{Bertrand},~\bfnm{Marianne}\binits{M.}} \AND \bauthor{\bsnm{Schoar},~\bfnm{Antoinette}\binits{A.}}
(\byear{2003}).
\btitle{{Managing with Style: The Effect of Managers on Firm Policies*}}.
\bjournal{The Quarterly Journal of Economics}
\bvolume{118}
\bpages{1169-1208}.
\bdoi{10.1162/003355303322552775}
\end{barticle}
\endbibitem

\bibitem{Bhojraj2003}
\begin{barticle}[author]
\bauthor{\bsnm{Bhojraj},~\bfnm{Sanjeev}\binits{S.}} \AND \bauthor{\bsnm{Sengupta},~\bfnm{Partha}\binits{P.}}
(\byear{2003}).
\btitle{Effect of Corporate Governance on Bond Ratings and Yields: The Role of Institutional Investors and Outside Directors}.
\bjournal{The Journal of Business}
\bvolume{76}
\bpages{455--475}.
\bdoi{10.1086/344114}
\end{barticle}
\endbibitem

\bibitem{Bishop2006}
\begin{bbook}[author]
\bauthor{\bsnm{Bishop},~\bfnm{Christopher~M}\binits{C.~M.}}
(\byear{2006}).
\btitle{{Pattern recognition and machine learning}}.
\bseries{Information science and statistics}.
\bpublisher{Springer}, \baddress{New York, NY}.
\bnote{Softcover published in 2016}.
\end{bbook}
\endbibitem

\bibitem{Blackmore2017}
\begin{bbook}[author]
\bauthor{\bsnm{Blackmore},~\bfnm{Susan}\binits{S.}}
(\byear{2017}).
\btitle{Consciousness: A very short introduction}.
\bpublisher{Oxford University Press}.
\end{bbook}
\endbibitem

\bibitem{Bowe2014}
\begin{barticle}[author]
\bauthor{\bsnm{Bowe},~\bfnm{Michael}\binits{M.}} \AND \bauthor{\bsnm{Larik},~\bfnm{Waseem}\binits{W.}}
(\byear{2014}).
\btitle{Split Ratings and Differences in Corporate Credit Rating Policy between Moody's and Standard \& Poor's}.
\bjournal{Financial Review}
\bvolume{49}
\bpages{713-734}.
\bdoi{https://doi.org/10.1111/fire.12054}
\end{barticle}
\endbibitem

\bibitem{Brealey2000}
\begin{bbook}[author]
\bauthor{\bsnm{Brealey},~\bfnm{Richard~A}\binits{R.~A.}} \AND \bauthor{\bsnm{Myers},~\bfnm{Stewart~C}\binits{S.~C.}}
(\byear{2000}).
\btitle{Priniciples of Corporate Finance, 6th edition},
\bedition{6th edition} ed.
\bpublisher{McGraw-Hill Higher Education}.
\end{bbook}
\endbibitem

\bibitem{Breiman2001}
\begin{barticle}[author]
\bauthor{\bsnm{Breiman},~\bfnm{Leo}\binits{L.}}
(\byear{2001}).
\btitle{{Statistical Modeling: The Two Cultures (with comments and a rejoinder by the author)}}.
\bjournal{Statistical Science}
\bvolume{16}
\bpages{199 -- 231}.
\bdoi{10.1214/ss/1009213726}
\end{barticle}
\endbibitem

\bibitem{Buehlmann2011}
\begin{bbook}[author]
\bauthor{\bsnm{B{\"u}hlmann},~\bfnm{Peter}\binits{P.}} \AND \bauthor{\bsnm{Van De~Geer},~\bfnm{Sara}\binits{S.}}
(\byear{2011}).
\btitle{Statistics for high-dimensional data: methods, theory and applications}.
\bpublisher{Springer Science \& Business Media}.
\end{bbook}
\endbibitem

\bibitem{Camanho2022}
\begin{barticle}[author]
\bauthor{\bsnm{Camanho},~\bfnm{Nelson}\binits{N.}}, \bauthor{\bsnm{Deb},~\bfnm{Pragyan}\binits{P.}} \AND \bauthor{\bsnm{Liu},~\bfnm{Zijun}\binits{Z.}}
(\byear{2022}).
\btitle{Credit rating and competition}.
\bjournal{International Journal of Finance \& Economics}
\bvolume{27}
\bpages{2873-2897}.
\bdoi{https://doi.org/10.1002/ijfe.2303}
\end{barticle}
\endbibitem

\bibitem{Cantor2004}
\begin{barticle}[author]
\bauthor{\bsnm{Cantor},~\bfnm{Richard}\binits{R.}}
(\byear{2004}).
\btitle{An introduction to recent research on credit ratings}.
\bjournal{Journal of Banking \& Finance}
\bvolume{28}
\bpages{2565-2573}.
\bnote{Recent Research on Credit Ratings}.
\bdoi{https://doi.org/10.1016/j.jbankfin.2004.06.002}
\end{barticle}
\endbibitem

\bibitem{Chaganti1991}
\begin{barticle}[author]
\bauthor{\bsnm{Chaganti},~\bfnm{Rajeswararao}\binits{R.}} \AND \bauthor{\bsnm{Damanpour},~\bfnm{Fariborz}\binits{F.}}
(\byear{1991}).
\btitle{Institutional ownership, capital structure, and firm performance}.
\bjournal{Strategic Management Journal}
\bvolume{12}
\bpages{479-491}.
\bdoi{https://doi.org/10.1002/smj.4250120702}
\end{barticle}
\endbibitem

\bibitem{Chernozhukov2017}
\begin{barticle}[author]
\bauthor{\bsnm{Chernozhukov},~\bfnm{Victor}\binits{V.}}, \bauthor{\bsnm{Chetverikov},~\bfnm{Denis}\binits{D.}}, \bauthor{\bsnm{Demirer},~\bfnm{Mert}\binits{M.}}, \bauthor{\bsnm{Duflo},~\bfnm{Esther}\binits{E.}}, \bauthor{\bsnm{Hansen},~\bfnm{Christian}\binits{C.}} \AND \bauthor{\bsnm{Newey},~\bfnm{Whitney}\binits{W.}}
(\byear{2017}).
\btitle{Double/Debiased/Neyman Machine Learning of Treatment Effects}.
\bjournal{American Economic Review}
\bvolume{107}
\bpages{261-65}.
\bdoi{10.1257/aer.p20171038}
\end{barticle}
\endbibitem

\bibitem{Chernozhukov2018}
\begin{barticle}[author]
\bauthor{\bsnm{Chernozhukov},~\bfnm{Victor}\binits{V.}}, \bauthor{\bsnm{Chetverikov},~\bfnm{Denis}\binits{D.}}, \bauthor{\bsnm{Demirer},~\bfnm{Mert}\binits{M.}}, \bauthor{\bsnm{Duflo},~\bfnm{Esther}\binits{E.}}, \bauthor{\bsnm{Hansen},~\bfnm{Christian}\binits{C.}}, \bauthor{\bsnm{Newey},~\bfnm{Whitney}\binits{W.}} \AND \bauthor{\bsnm{Robins},~\bfnm{James}\binits{J.}}
(\byear{2018}).
\btitle{{Double/debiased machine learning for treatment and structural parameters}}.
\bjournal{The Econometrics Journal}
\bvolume{21}
\bpages{C1-C68}.
\bdoi{10.1111/ectj.12097}
\end{barticle}
\endbibitem

\bibitem{Chernozhukov2013}
\begin{barticle}[author]
\bauthor{\bsnm{Chernozhukov},~\bfnm{Victor}\binits{V.}}, \bauthor{\bsnm{Chetverikov},~\bfnm{Denis}\binits{D.}} \AND \bauthor{\bsnm{Kato},~\bfnm{Kengo}\binits{K.}}
(\byear{2013}).
\btitle{Gaussian approximations and multiplier bootstrap for maxima of sums of high-dimensional random vectors}.
\bjournal{The Annals of Statistics}
\bvolume{41}
\bpages{2786--2819}.
\bdoi{10.1214/13-AOS1161}
\end{barticle}
\endbibitem

\bibitem{Chernozhukov2014}
\begin{barticle}[author]
\bauthor{\bsnm{Chernozhukov},~\bfnm{Victor}\binits{V.}}, \bauthor{\bsnm{Chetverikov},~\bfnm{Denis}\binits{D.}} \AND \bauthor{\bsnm{Kato},~\bfnm{Kengo}\binits{K.}}
(\byear{2014}).
\btitle{{Gaussian approximation of suprema of empirical processes}}.
\bjournal{The Annals of Statistics}
\bvolume{42}
\bpages{1564 -- 1597}.
\bdoi{10.1214/14-AOS1230}
\end{barticle}
\endbibitem

\bibitem{Chernozhukov2017a}
\begin{barticle}[author]
\bauthor{\bsnm{CHERNOZHUKOV},~\bfnm{VICTOR}\binits{V.}} \AND \bauthor{\bsnm{FERN\'ANDEZ-VAL},~\bfnm{IV\'AN}\binits{I.}}
(\byear{2017}).
\btitle{14.382 L1. LEAST SQUARES, ADAPTIVE PARTIALLING-OUT, SIMULTANEOUS INFERENCE}.
\bjournal{Massachusetts Instituteof Technology: MIT OpenCourseWare, https://ocw.mit.edu. License: Creative Commons BY-NC-SA}.
\end{barticle}
\endbibitem

\bibitem{Chernozhukov2022}
\begin{bmisc}[author]
\bauthor{\bsnm{Chernozhukov},~\bfnm{Victor}\binits{V.}}, \bauthor{\bsnm{Hansen},~\bfnm{Christian}\binits{C.}}, \bauthor{\bsnm{Spindler},~\bfnm{Martin}\binits{M.}} \AND \bauthor{\bsnm{Syrgkanis},~\bfnm{Vsilis}\binits{V.}}
(\byear{2022}).
\btitle{Applied Causal Inference Powered by ML and AI}.
\bhowpublished{Forthcoming (draft dated November 06, 2021)}.
\end{bmisc}
\endbibitem

\bibitem{DeHaan2011}
\begin{bincollection}[author]
\bauthor{\bsnm{De~Haan},~\bfnm{Jakob}\binits{J.}} \AND \bauthor{\bsnm{Amtenbrink},~\bfnm{Fabian}\binits{F.}}
(\byear{2011}).
\btitle{Credit rating agencies}.
In \bbooktitle{Handbook of Central Banking, Financial Regulation and Supervision}
\bpublisher{Edward Elgar Publishing}.
\end{bincollection}
\endbibitem

\bibitem{DeAngelo1980}
\begin{barticle}[author]
\bauthor{\bsnm{DeAngelo},~\bfnm{Harry}\binits{H.}} \AND \bauthor{\bsnm{Masulis},~\bfnm{Ronald~W.}\binits{R.~W.}}
(\byear{1980}).
\btitle{Optimal capital structure under corporate and personal taxation}.
\bjournal{Journal of Financial Economics}
\bvolume{8}
\bpages{3-29}.
\bdoi{https://doi.org/10.1016/0304-405X(80)90019-7}
\end{barticle}
\endbibitem

\bibitem{EBA2015}
\begin{bmisc}[author]
\bauthor{\bsnm{{European Banking Authority (EBA)}}}
(\byear{2015}).
\btitle{Amended Mapping of S\&P Global Ratings' credit assessments under the Standardised Approach}.
\bhowpublished{https://www.eba.europa.eu/sites/default/files/documents/10180/2733281/d878f14e-ef82-47be-be63-ab10fc353af2/(Mapping\%20Report\%20-\%20S\%20and\%20P).pdf (accessed 22 February 2023)}.
\end{bmisc}
\endbibitem

\bibitem{Fama1997}
\begin{barticle}[author]
\bauthor{\bsnm{Fama},~\bfnm{Eugene~F.}\binits{E.~F.}} \AND \bauthor{\bsnm{French},~\bfnm{Kenneth~R.}\binits{K.~R.}}
(\byear{1997}).
\btitle{Industry costs of equity}.
\bjournal{Journal of Financial Economics}
\bvolume{43}
\bpages{153-193}.
\bdoi{https://doi.org/10.1016/S0304-405X(96)00896-3}
\end{barticle}
\endbibitem

\bibitem{Faulkender2005}
\begin{barticle}[author]
\bauthor{\bsnm{Faulkender},~\bfnm{Michael}\binits{M.}} \AND \bauthor{\bsnm{Petersen},~\bfnm{Mitchell~A.}\binits{M.~A.}}
(\byear{2005}).
\btitle{{Does the Source of Capital Affect Capital Structure?}}
\bjournal{The Review of Financial Studies}
\bvolume{19}
\bpages{45-79}.
\bdoi{10.1093/rfs/hhj003}
\end{barticle}
\endbibitem

\bibitem{Fedorov2013}
\begin{bbook}[author]
\bauthor{\bsnm{Fedorov},~\bfnm{Valerii~V}\binits{V.~V.}} \AND \bauthor{\bsnm{Leonov},~\bfnm{Sergei~L}\binits{S.~L.}}
(\byear{2013}).
\btitle{Optimal design for nonlinear response models}.
\bpublisher{CRC Press}.
\end{bbook}
\endbibitem

\bibitem{Feng2008}
\begin{barticle}[author]
\bauthor{\bsnm{Feng},~\bfnm{D.}\binits{D.}}, \bauthor{\bsnm{Gourieroux},~\bfnm{C.}\binits{C.}} \AND \bauthor{\bsnm{Jasiak},~\bfnm{J.}\binits{J.}}
(\byear{2008}).
\btitle{The ordered qualitative model for credit rating transitions}.
\bjournal{Journal of Empirical Finance}
\bvolume{15}
\bpages{111-130}.
\bdoi{https://doi.org/10.1016/j.jempfin.2006.12.003}
\end{barticle}
\endbibitem

\bibitem{Fischer1989}
\begin{barticle}[author]
\bauthor{\bsnm{Fischer},~\bfnm{Edwin~O.}\binits{E.~O.}}, \bauthor{\bsnm{Heinkel},~\bfnm{Robert}\binits{R.}} \AND \bauthor{\bsnm{Zechner},~\bfnm{Josef}\binits{J.}}
(\byear{1989}).
\btitle{Dynamic Capital Structure Choice: Theory and Tests}.
\bjournal{The Journal of Finance}
\bvolume{44}
\bpages{19-40}.
\bdoi{https://doi.org/10.1111/j.1540-6261.1989.tb02402.x}
\end{barticle}
\endbibitem

\bibitem{Fonseca2020}
\begin{barticle}[author]
\bauthor{\bsnm{Fonseca},~\bfnm{Peter Vaz~da}\binits{P.~V.~d.}}, \bauthor{\bsnm{Savelli},~\bfnm{Andrea~Decourt}\binits{A.~D.}} \AND \bauthor{\bsnm{Juca},~\bfnm{Michele~Nascimento}\binits{M.~N.}}
(\byear{2020}).
\btitle{A Systematic Review of the Influence of Taxation on Corporate Capital Structure}.
\bjournal{International Journal of Economics \& Business Administration (IJEBA)}
\bvolume{8}
\bpages{155--178}.
\end{barticle}
\endbibitem

\bibitem{Frank2009}
\begin{barticle}[author]
\bauthor{\bsnm{Frank},~\bfnm{Murray~Z.}\binits{M.~Z.}} \AND \bauthor{\bsnm{Goyal},~\bfnm{Vidhan~K.}\binits{V.~K.}}
(\byear{2009}).
\btitle{Capital Structure Decisions: Which Factors Are Reliably Important?}
\bjournal{Financial Management}
\bvolume{38}
\bpages{1-37}.
\bdoi{https://doi.org/10.1111/j.1755-053X.2009.01026.x}
\end{barticle}
\endbibitem

\bibitem{Garavaglia1991}
\begin{binproceedings}[author]
\bauthor{\bsnm{Garavaglia},~\bfnm{S.}\binits{S.}}
(\byear{1991}).
\btitle{An application of a counter-propagation neural network: simulating the Standard and Poor's Corporate Bond Rating system}.
In \bbooktitle{Proceedings First International Conference on Artificial Intelligence Applications on Wall Street}
\bpages{278,279,280,281,282,283,284,285,286,287}.
\bpublisher{IEEE Computer Society}, \baddress{Los Alamitos, CA, USA}.
\bdoi{10.1109/AIAWS.1991.236588}
\end{binproceedings}
\endbibitem

\bibitem{Golbayani2020}
\begin{bmisc}[author]
\bauthor{\bsnm{Golbayani},~\bfnm{Parisa}\binits{P.}}, \bauthor{\bsnm{Wang},~\bfnm{Dan}\binits{D.}} \AND \bauthor{\bsnm{Florescu},~\bfnm{Ionut}\binits{I.}}
(\byear{2020}).
\btitle{Application of Deep Neural Networks to assess corporate Credit Rating}.
\bdoi{10.48550/ARXIV.2003.02334}
\end{bmisc}
\endbibitem

\bibitem{Graham2001}
\begin{barticle}[author]
\bauthor{\bsnm{Graham},~\bfnm{John~R}\binits{J.~R.}} \AND \bauthor{\bsnm{Harvey},~\bfnm{Campbell~R}\binits{C.~R.}}
(\byear{2001}).
\btitle{The theory and practice of corporate finance: evidence from the field}.
\bjournal{Journal of Financial Economics}
\bvolume{60}
\bpages{187-243}.
\bnote{Complementary Research Methodologies: The InterPlay of Theoretical, Empirical and Field-Based Research in Finance}.
\bdoi{https://doi.org/10.1016/S0304-405X(01)00044-7}
\end{barticle}
\endbibitem

\bibitem{Graham2011}
\begin{barticle}[author]
\bauthor{\bsnm{Graham},~\bfnm{John~R.}\binits{J.~R.}} \AND \bauthor{\bsnm{Leary},~\bfnm{Mark~T.}\binits{M.~T.}}
(\byear{2011}).
\btitle{A Review of Empirical Capital Structure Research and Directions for the Future}.
\bjournal{Annual Review of Financial Economics}
\bvolume{3}
\bpages{309-345}.
\bdoi{10.1146/annurev-financial-102710-144821}
\end{barticle}
\endbibitem

\bibitem{Grunert2012}
\begin{barticle}[author]
\bauthor{\bsnm{Grunert},~\bfnm{Jens}\binits{J.}} \AND \bauthor{\bsnm{Norden},~\bfnm{Lars}\binits{L.}}
(\byear{2012}).
\btitle{Bargaining power and information in SME lending}.
\bjournal{Small Business Economics}
\bvolume{39}
\bpages{401--417}.
\bdoi{10.1007/s11187-010-9311-6}
\end{barticle}
\endbibitem

\bibitem{Halevy2009}
\begin{barticle}[author]
\bauthor{\bsnm{Halevy},~\bfnm{Alon}\binits{A.}}, \bauthor{\bsnm{Norvig},~\bfnm{Peter}\binits{P.}} \AND \bauthor{\bsnm{Pereira},~\bfnm{Fernando}\binits{F.}}
(\byear{2009}).
\btitle{The Unreasonable Effectiveness of Data}.
\bjournal{IEEE Intelligent Systems}
\bvolume{24}
\bpages{8-12}.
\bdoi{10.1109/MIS.2009.36}
\end{barticle}
\endbibitem

\bibitem{Haerdle2000}
\begin{bbook}[author]
\bauthor{\bsnm{H\"ardle},~\bfnm{Wolfgang~Karl}\binits{W.~K.}}, \bauthor{\bsnm{Liang},~\bfnm{Hua}\binits{H.}} \AND \bauthor{\bsnm{Gao},~\bfnm{Jiti}\binits{J.}}
(\byear{2000}).
\btitle{Partially Linear Models}.
\bdoi{10.1007/978-3-642-57700-0}
\end{bbook}
\endbibitem

\bibitem{Hastie2009}
\begin{bbook}[author]
\bauthor{\bsnm{Hastie},~\bfnm{Trevor}\binits{T.}}, \bauthor{\bsnm{Tibshirani},~\bfnm{Robert}\binits{R.}}, \bauthor{\bsnm{Friedman},~\bfnm{Jerome~H}\binits{J.~H.}} \AND \bauthor{\bsnm{Friedman},~\bfnm{Jerome~H}\binits{J.~H.}}
(\byear{2009}).
\btitle{The elements of statistical learning: data mining, inference, and prediction}
\bvolume{2}.
\bpublisher{Springer}.
\end{bbook}
\endbibitem

\bibitem{Hawawini2002}
\begin{barticle}[author]
\bauthor{\bsnm{Hawawini},~\bfnm{Gabriel}\binits{G.}} \AND \bauthor{\bsnm{Viallet},~\bfnm{Claude}\binits{C.}}
(\byear{2002}).
\btitle{Finance for Executives Managing for Value Creation, 2nd Edition}.
\bjournal{South-Western Cengage Learning, OH, USA}.
\end{barticle}
\endbibitem

\bibitem{Hernan2020}
\begin{bbook}[author]
\bauthor{\bsnm{Hern\'{a}n},~\bfnm{Miguel~A.}\binits{M.~A.}} \AND \bauthor{\bsnm{Robins},~\bfnm{James~M.}\binits{J.~M.}}
(\byear{2020}).
\btitle{Causal Inference: What If.}
\bpublisher{Boca Raton: Chapman \& Hall/CRC.}
\end{bbook}
\endbibitem

\bibitem{Holland1986}
\begin{barticle}[author]
\bauthor{\bsnm{Holland},~\bfnm{Paul~W.}\binits{P.~W.}}
(\byear{1986}).
\btitle{Statistics and Causal Inference}.
\bjournal{Journal of the American Statistical Association}
\bvolume{81}
\bpages{945-960}.
\bdoi{10.1080/01621459.1986.10478354}
\end{barticle}
\endbibitem

\bibitem{Holm1979}
\begin{barticle}[author]
\bauthor{\bsnm{Holm},~\bfnm{Sture}\binits{S.}}
(\byear{1979}).
\btitle{A Simple Sequentially Rejective Multiple Test Procedure}.
\bjournal{Scandinavian Journal of Statistics}
\bvolume{6}
\bpages{65--70}.
\end{barticle}
\endbibitem

\bibitem{Hu2019}
\begin{barticle}[author]
\bauthor{\bsnm{Hu},~\bfnm{Xiaolu}\binits{X.}}, \bauthor{\bsnm{Huang},~\bfnm{Haozhi}\binits{H.}}, \bauthor{\bsnm{Pan},~\bfnm{Zheyao}\binits{Z.}} \AND \bauthor{\bsnm{Shi},~\bfnm{Jing}\binits{J.}}
(\byear{2019}).
\btitle{Information asymmetry and credit rating: A quasi-natural experiment from China}.
\bjournal{Journal of Banking \& Finance}
\bvolume{106}
\bpages{132-152}.
\bdoi{https://doi.org/10.1016/j.jbankfin.2019.06.003}
\end{barticle}
\endbibitem

\bibitem{Huang2004}
\begin{barticle}[author]
\bauthor{\bsnm{Huang},~\bfnm{Zan}\binits{Z.}}, \bauthor{\bsnm{Chen},~\bfnm{Hsinchun}\binits{H.}}, \bauthor{\bsnm{Hsu},~\bfnm{Chia-Jung}\binits{C.-J.}}, \bauthor{\bsnm{Chen},~\bfnm{Wun-Hwa}\binits{W.-H.}} \AND \bauthor{\bsnm{Wu},~\bfnm{Soushan}\binits{S.}}
(\byear{2004}).
\btitle{Credit rating analysis with support vector machines and neural networks: a market comparative study}.
\bjournal{Decision Support Systems}
\bvolume{37}
\bpages{543-558}.
\bnote{Data mining for financial decision making}.
\bdoi{https://doi.org/10.1016/S0167-9236(03)00086-1}
\end{barticle}
\endbibitem

\bibitem{Imbens2015}
\begin{bbook}[author]
\bauthor{\bsnm{Imbens},~\bfnm{Guido~W}\binits{G.~W.}} \AND \bauthor{\bsnm{Rubin},~\bfnm{Donald~B}\binits{D.~B.}}
(\byear{2015}).
\btitle{Causal inference in statistics, social, and biomedical sciences}.
\bpublisher{Cambridge University Press}.
\end{bbook}
\endbibitem

\bibitem{Ioannidis2005}
\begin{barticle}[author]
\bauthor{\bsnm{Ioannidis},~\bfnm{John P.~A.}\binits{J.~P.~A.}}
(\byear{2005}).
\btitle{Why Most Published Research Findings Are False}.
\bjournal{PLOS Medicine}
\bvolume{2}
\bpages{null}.
\bdoi{10.1371/journal.pmed.0020124}
\end{barticle}
\endbibitem

\bibitem{James2013}
\begin{bbook}[author]
\bauthor{\bsnm{James},~\bfnm{Gareth}\binits{G.}}, \bauthor{\bsnm{Witten},~\bfnm{Daniela}\binits{D.}}, \bauthor{\bsnm{Hastie},~\bfnm{Trevor}\binits{T.}} \AND \bauthor{\bsnm{Tibshirani},~\bfnm{Robert}\binits{R.}}
(\byear{2013}).
\btitle{An introduction to statistical learning}
\bvolume{112}.
\bpublisher{Springer}.
\end{bbook}
\endbibitem

\bibitem{Jensen1986}
\begin{barticle}[author]
\bauthor{\bsnm{Jensen},~\bfnm{Michael~C.}\binits{M.~C.}}
(\byear{1986}).
\btitle{Agency Costs of Free Cash Flow, Corporate Finance, and Takeovers}.
\bjournal{The American Economic Review}
\bvolume{76}
\bpages{323--329}.
\end{barticle}
\endbibitem

\bibitem{Jensen1976}
\begin{barticle}[author]
\bauthor{\bsnm{Jensen},~\bfnm{Michael~C}\binits{M.~C.}} \AND \bauthor{\bsnm{Meckling},~\bfnm{William~H}\binits{W.~H.}}
(\byear{1976}).
\btitle{Theory of the firm: Managerial behavior, agency costs and ownership structure}.
\bjournal{Journal of financial economics}
\bvolume{3}
\bpages{305--360}.
\end{barticle}
\endbibitem

\bibitem{Karlsen2021}
\begin{bmastersthesis}[author]
\bauthor{\bsnm{Karlsen},~\bfnm{Kristian}\binits{K.}} \AND \bauthor{\bsnm{Mathisen},~\bfnm{Nils}\binits{N.}}
(\byear{2021}).
\btitle{Capital Structure and Machine Learning Techniques in Scandinavia}
\btype{Master's thesis},
\bpublisher{NTNU}.
\end{bmastersthesis}
\endbibitem

\bibitem{Kemper2013}
\begin{barticle}[author]
\bauthor{\bsnm{Kemper},~\bfnm{Kristopher~J.}\binits{K.~J.}} \AND \bauthor{\bsnm{Rao},~\bfnm{Ramesh~P.}\binits{R.~P.}}
(\byear{2013}).
\btitle{Do Credit Ratings Really Affect Capital Structure?}
\bjournal{Financial Review}
\bvolume{48}
\bpages{573-595}.
\bdoi{https://doi.org/10.1111/fire.12016}
\end{barticle}
\endbibitem

\bibitem{Kim2020}
\begin{barticle}[author]
\bauthor{\bsnm{Kim},~\bfnm{Hyeongjun}\binits{H.}}, \bauthor{\bsnm{Cho},~\bfnm{Hoon}\binits{H.}} \AND \bauthor{\bsnm{Ryu},~\bfnm{Doojin}\binits{D.}}
(\byear{2020}).
\btitle{Corporate Default Predictions Using Machine Learning: Literature Review}.
\bjournal{Sustainability}
\bvolume{12}.
\bdoi{10.3390/su12166325}
\end{barticle}
\endbibitem

\bibitem{Kisgen2006}
\begin{barticle}[author]
\bauthor{\bsnm{Kisgen},~\bfnm{Darren~J.}\binits{D.~J.}}
(\byear{2006}).
\btitle{Credit Ratings and Capital Structure}.
\bjournal{The Journal of Finance}
\bvolume{61}
\bpages{1035-1072}.
\bdoi{https://doi.org/10.1111/j.1540-6261.2006.00866.x}
\end{barticle}
\endbibitem

\bibitem{Kisgen2009}
\begin{barticle}[author]
\bauthor{\bsnm{Kisgen},~\bfnm{Darren~J.}\binits{D.~J.}}
(\byear{2009}).
\btitle{Do Firms Target Credit Ratings or Leverage Levels?}
\bjournal{Journal of Financial and Quantitative Analysis}
\bvolume{44}
\bpages{1323-1344}.
\bdoi{10.1017/S002210900999041X}
\end{barticle}
\endbibitem

\bibitem{Kisgen2019}
\begin{barticle}[author]
\bauthor{\bsnm{Kisgen},~\bfnm{Darren~J.}\binits{D.~J.}}
(\byear{2019}).
\btitle{The impact of credit ratings on corporate behavior: Evidence from Moody's adjustments}.
\bjournal{Journal of Corporate Finance}
\bvolume{58}
\bpages{567-582}.
\bdoi{https://doi.org/10.1016/j.jcorpfin.2019.07.002}
\end{barticle}
\endbibitem

\bibitem{Koller2020}
\begin{bbook}[author]
\bauthor{\bsnm{Koller},~\bfnm{T.}\binits{T.}}, \bauthor{\bsnm{Goedhart},~\bfnm{M.}\binits{M.}} \AND \bauthor{\bsnm{Wessels},~\bfnm{D.}\binits{D.}}
(\byear{2020}).
\btitle{Valuation: Measuring and Managing the Value of Companies (7th edition)}.
\bseries{Wiley Finance}.
\bpublisher{Wiley}.
\end{bbook}
\endbibitem

\bibitem{Kumar2017}
\begin{barticle}[author]
\bauthor{\bsnm{Kumar},~\bfnm{Satish}\binits{S.}}, \bauthor{\bsnm{Colombage},~\bfnm{Sisira}\binits{S.}} \AND \bauthor{\bsnm{Rao},~\bfnm{Purnima}\binits{P.}}
(\byear{2017}).
\btitle{Research on capital structure determinants: a review and future directions}.
\bjournal{International Journal of Managerial Finance}
\bvolume{13}
\bpages{106--132}.
\bdoi{10.1108/IJMF-09-2014-0135}
\end{barticle}
\endbibitem

\bibitem{Kwon1997}
\begin{barticle}[author]
\bauthor{\bsnm{Kwon},~\bfnm{Y.}\binits{Y.}}, \bauthor{\bsnm{Han},~\bfnm{I.}\binits{I.}} \AND \bauthor{\bsnm{Lee},~\bfnm{K.}\binits{K.}}
(\byear{1997}).
\btitle{Ordinal Pairwise Partitioning (OPP) Approach to Neural Networks Training in Bond rating}.
\bjournal{Intelligent Systems in Accounting, Finance and Management}
\bvolume{6}
\bpages{23--40}.
\end{barticle}
\endbibitem

\bibitem{Lang1989}
\begin{barticle}[author]
\bauthor{\bsnm{Lang},~\bfnm{Larry H.~P.}\binits{L.~H.~P.}}, \bauthor{\bsnm{Stulz},~\bfnm{RenéM.}\binits{R.}} \AND \bauthor{\bsnm{Walkling},~\bfnm{Ralph~A.}\binits{R.~A.}}
(\byear{1989}).
\btitle{Managerial performance, Tobin's Q, and the gains from successful tender offers}.
\bjournal{Journal of Financial Economics}
\bvolume{24}
\bpages{137-154}.
\bdoi{https://doi.org/10.1016/0304-405X(89)90075-5}
\end{barticle}
\endbibitem

\bibitem{LIVINGSTON2010}
\begin{barticle}[author]
\bauthor{\bsnm{LIVINGSTON},~\bfnm{MILES}\binits{M.}}, \bauthor{\bsnm{WEI},~\bfnm{JIE~(DIANA)}\binits{J.~D.}} \AND \bauthor{\bsnm{ZHOU},~\bfnm{LEI}\binits{L.}}
(\byear{2010}).
\btitle{Moody's and S\&P Ratings: Are They Equivalent? Conservative Ratings and Split Rated Bond Yields}.
\bjournal{Journal of Money, Credit and Banking}
\bvolume{42}
\bpages{1267-1293}.
\bdoi{https://doi.org/10.1111/j.1538-4616.2010.00341.x}
\end{barticle}
\endbibitem

\bibitem{SPFSL2022}
\begin{bmisc}[author]
\bauthor{\bsnm{{Standard \& Poor's Financial Services LLC}}}
(\byear{2022}).
\btitle{Guide to Credit Rating Essentials. What are credit ratings and how do they work?}
\bhowpublished{https://www.spglobal.com/ratings/\_division-assets/pdfs/guide\_to\_credit\_rating\_essentials\_digital.pdf (accessed 02 December 2022)}.
\end{bmisc}
\endbibitem

\bibitem{SPFSL2022a}
\begin{bmisc}[author]
\bauthor{\bsnm{{Standard \& Poor's Financial Services LLC}}}
(\byear{2022}).
\btitle{How We Rate Nonfinancial Corporate Entities}.
\bhowpublished{https://www.spglobal.com/ratings/\_division-assets/pdfs/041019\_howweratenonfinancialcorporateentities.pdf (accessed 03 December 2022)}.
\end{bmisc}
\endbibitem

\bibitem{Lovell2008}
\begin{barticle}[author]
\bauthor{\bsnm{Lovell},~\bfnm{Michael~C.}\binits{M.~C.}}
(\byear{2008}).
\btitle{A Simple Proof of the FWL Theorem}.
\bjournal{The Journal of Economic Education}
\bvolume{39}
\bpages{88-91}.
\bdoi{10.3200/JECE.39.1.88-91}
\end{barticle}
\endbibitem

\bibitem{Mackey2018}
\begin{binproceedings}[author]
\bauthor{\bsnm{Mackey},~\bfnm{Lester}\binits{L.}}, \bauthor{\bsnm{Syrgkanis},~\bfnm{Vasilis}\binits{V.}} \AND \bauthor{\bsnm{Zadik},~\bfnm{Ilias}\binits{I.}}
(\byear{2018}).
\btitle{Orthogonal Machine Learning: Power and Limitations}.
In \bbooktitle{Proceedings of the 35th International Conference on Machine Learning}
(\beditor{\bfnm{Jennifer}\binits{J.}~\bsnm{Dy}} \AND \beditor{\bfnm{Andreas}\binits{A.}~\bsnm{Krause}}, eds.).
\bseries{Proceedings of Machine Learning Research}
\bvolume{80}
\bpages{3375--3383}.
\bpublisher{PMLR}.
\end{binproceedings}
\endbibitem

\bibitem{Marsland2015}
\begin{bbook}[author]
\bauthor{\bsnm{Marsland},~\bfnm{Stephen}\binits{S.}}
(\byear{2015}).
\btitle{Machine Learning: An Algorithmic Perspective},
\bedition{Second edition} ed.
\bpublisher{Chapman and Hall/CRC}.
\end{bbook}
\endbibitem

\bibitem{Matthies2013}
\begin{btechreport}[author]
\bauthor{\bsnm{Matthies},~\bfnm{Alexander~B.}\binits{A.~B.}}
(\byear{2013}).
\btitle{Empirical research on corporate credit-ratings: A literature review}
\btype{SFB 649 Discussion Paper} No. \bnumber{2013-003},
\baddress{Berlin}.
\end{btechreport}
\endbibitem

\bibitem{McShane2019}
\begin{barticle}[author]
\bauthor{\bsnm{McShane},~\bfnm{Blakeley~B.}\binits{B.~B.}}, \bauthor{\bsnm{Gal},~\bfnm{David}\binits{D.}}, \bauthor{\bsnm{Gelman},~\bfnm{Andrew}\binits{A.}}, \bauthor{\bsnm{Robert},~\bfnm{Christian}\binits{C.}} \AND \bauthor{\bsnm{Tackett},~\bfnm{Jennifer~L.}\binits{J.~L.}}
(\byear{2019}).
\btitle{Abandon Statistical Significance}.
\bjournal{The American Statistician}
\bvolume{73}
\bpages{235-245}.
\bdoi{10.1080/00031305.2018.1527253}
\end{barticle}
\endbibitem

\bibitem{Miller1977}
\begin{barticle}[author]
\bauthor{\bsnm{Miller},~\bfnm{Merton~H.}\binits{M.~H.}}
(\byear{1977}).
\btitle{Debt and Taxes}.
\bjournal{The Journal of Finance}
\bvolume{32}
\bpages{261--275}.
\bdoi{10.2307/2326758}
\end{barticle}
\endbibitem

\bibitem{Mintzberg2009}
\begin{barticle}[author]
\bauthor{\bsnm{Mintzberg},~\bfnm{Henry}\binits{H.}}, \bauthor{\bsnm{Lampel},~\bfnm{Joseph}\binits{J.}} \AND \bauthor{\bsnm{Ahlstrand},~\bfnm{Bruce}\binits{B.}}
(\byear{2009}).
\btitle{Strategy Safari: A Guided Tour Through the Jungles of Strategic Management}.
\end{barticle}
\endbibitem

\bibitem{MIS2022}
\begin{bmisc}[author]
\bauthor{\bsnm{{Moody's Investor Service (MIS)}}}
(\byear{2022}).
\btitle{Rating Symbols and Definitions}.
\bhowpublished{https://www.moodys.com/researchdocumentcontentpage.aspx?docid=PBC\_79004 (accessed 04 December 2022)}.
\end{bmisc}
\endbibitem

\bibitem{Modigliani1958}
\begin{barticle}[author]
\bauthor{\bsnm{Modigliani},~\bfnm{Franco}\binits{F.}} \AND \bauthor{\bsnm{Miller},~\bfnm{Merton~H.}\binits{M.~H.}}
(\byear{1958}).
\btitle{The Cost of Capital, Corporation Finance and the Theory of Investment}.
\bjournal{The American Economic Review}
\bvolume{48}
\bpages{261--297}.
\end{barticle}
\endbibitem

\bibitem{Morgan2015}
\begin{bbook}[author]
\bauthor{\bsnm{Morgan},~\bfnm{Stephen~L}\binits{S.~L.}} \AND \bauthor{\bsnm{Winship},~\bfnm{Christopher}\binits{C.}}
(\byear{2015}).
\btitle{Counterfactuals and causal inference}.
\bpublisher{Cambridge University Press}.
\end{bbook}
\endbibitem

\bibitem{Murphy2022}
\begin{bbook}[author]
\bauthor{\bsnm{Murphy},~\bfnm{K.~P.}\binits{K.~P.}}
(\byear{2022}).
\btitle{Probabilistic Machine Learning: An Introduction}.
\bseries{Adaptive Computation and Machine Learning series}.
\bpublisher{MIT Press}.
\end{bbook}
\endbibitem

\bibitem{Myers1984a}
\begin{barticle}[author]
\bauthor{\bsnm{Myers},~\bfnm{Stewart~C}\binits{S.~C.}}
(\byear{1984}).
\btitle{The Capital Structure Puzzle}.
\bjournal{THE JOURNAL OF FINANCE}
\bvolume{39}.
\end{barticle}
\endbibitem

\bibitem{Neal2020}
\begin{barticle}[author]
\bauthor{\bsnm{Neal},~\bfnm{Brady}\binits{B.}}
(\byear{2020}).
\btitle{Introduction to causal inference from a machine learning perspective}.
\bjournal{Course Lecture Notes (draft dated Dec 17, 2020)}.
\end{barticle}
\endbibitem

\bibitem{Neyman1979}
\begin{barticle}[author]
\bauthor{\bsnm{Neyman},~\bfnm{Jerzy}\binits{J.}}
(\byear{1979}).
\btitle{C($\alpha$) Tests and Their Use}.
\bjournal{Sankhy\=a: The Indian Journal of Statistics, Series A (1961-2002)}
\bvolume{41}
\bpages{1--21}.
\end{barticle}
\endbibitem

\bibitem{Novartis2023}
\begin{bmisc}[author]
\bauthor{\bsnm{Novartis}}
(\byear{2023}).
\btitle{New Novartis: Pure-Play Innovative Medicines Company (J.P. Morgan Healthcare Conference January 9, 2023)}.
\bhowpublished{$https://www.novartis.com/sites/novartis_com/files/2023-new-novartis-pure-play-innovative-medicines-company.pdf$ (accessed January 13, 2023)}.
\end{bmisc}
\endbibitem

\bibitem{Partnoy2006}
\begin{barticle}[author]
\bauthor{\bsnm{Partnoy},~\bfnm{Frank}\binits{F.}}
(\byear{2006}).
\btitle{How and why credit rating agencies are not like other gatekeepers}.
\end{barticle}
\endbibitem

\bibitem{Pearl2009}
\begin{bbook}[author]
\bauthor{\bsnm{Pearl},~\bfnm{Judea}\binits{J.}}
(\byear{2009}).
\btitle{Causality}.
\bpublisher{Cambridge university press}.
\end{bbook}
\endbibitem

\bibitem{Pearl2019}
\begin{barticle}[author]
\bauthor{\bsnm{Pearl},~\bfnm{Judea}\binits{J.}}
(\byear{2019}).
\btitle{The Seven Tools of Causal Inference, with Reflections on Machine Learning}.
\bjournal{Commun. ACM}
\bvolume{62}
\bpages{54-60}.
\bdoi{10.1145/3241036}
\end{barticle}
\endbibitem

\bibitem{Peters2017}
\begin{bbook}[author]
\bauthor{\bsnm{Peters},~\bfnm{Jonas}\binits{J.}}, \bauthor{\bsnm{Janzing},~\bfnm{Dominik}\binits{D.}} \AND \bauthor{\bsnm{Sch{\"o}lkopf},~\bfnm{Bernhard}\binits{B.}}
(\byear{2017}).
\btitle{Elements of causal inference: foundations and learning algorithms}.
\bpublisher{The MIT Press}.
\end{bbook}
\endbibitem

\bibitem{FitchRatings2022}
\begin{bmisc}[author]
\bauthor{\bsnm{{Fitch Ratings}}}
(\byear{2022}).
\btitle{Rating Definitions}.
\bhowpublished{https://www.fitchratings.com/research/ structured-finance/rating-definitions-21-03-2022 (accessed 04 December 2022)}.
\end{bmisc}
\endbibitem

\bibitem{FitchRatings2022a}
\begin{bmisc}[author]
\bauthor{\bsnm{{Fitch Ratings}}}
(\byear{2022}).
\btitle{The Rating Process - How Fitch Assigns Credit Ratings}.
\bhowpublished{https://www.fitchratings.com/research/corporate-finance/the-rating-process-how-fitch-assigns-credit-ratings-24-02-2022 (accessed 04 December 2022)}.
\end{bmisc}
\endbibitem

\bibitem{Robinson1988}
\begin{barticle}[author]
\bauthor{\bsnm{Robinson},~\bfnm{P.~M.}\binits{P.~M.}}
(\byear{1988}).
\btitle{Root-N-Consistent Semiparametric Regression}.
\bjournal{Econometrica}
\bvolume{56}
\bpages{931--954}.
\bdoi{10.2307/1912705}
\end{barticle}
\endbibitem

\bibitem{Romano2005}
\begin{barticle}[author]
\bauthor{\bsnm{Romano},~\bfnm{Joseph~P}\binits{J.~P.}} \AND \bauthor{\bsnm{Wolf},~\bfnm{Michael}\binits{M.}}
(\byear{2005}).
\btitle{Exact and Approximate Stepdown Methods for Multiple Hypothesis Testing}.
\bjournal{Journal of the American Statistical Association}
\bvolume{100}
\bpages{94-108}.
\bdoi{10.1198/016214504000000539}
\end{barticle}
\endbibitem

\bibitem{Romano2005a}
\begin{barticle}[author]
\bauthor{\bsnm{Romano},~\bfnm{Joseph~P.}\binits{J.~P.}} \AND \bauthor{\bsnm{Wolf},~\bfnm{Michael}\binits{M.}}
(\byear{2005}).
\btitle{Stepwise Multiple Testing as Formalized Data Snooping}.
\bjournal{Econometrica}
\bvolume{73}
\bpages{1237-1282}.
\bdoi{https://doi.org/10.1111/j.1468-0262.2005.00615.x}
\end{barticle}
\endbibitem

\bibitem{Rosenbaum2020}
\begin{bbook}[author]
\bauthor{\bsnm{Rosenbaum},~\bfnm{P.~R.}\binits{P.~R.}}
(\byear{2020}).
\btitle{Design of Observational Studies}.
\bseries{Springer Series in Statistics}.
\bpublisher{Springer International Publishing}.
\end{bbook}
\endbibitem

\bibitem{Ross1977}
\begin{barticle}[author]
\bauthor{\bsnm{Ross},~\bfnm{Stephen~A.}\binits{S.~A.}}
(\byear{1977}).
\btitle{The Determination of Financial Structure: The Incentive-Signalling Approach}.
\bjournal{The Bell Journal of Economics}
\bvolume{8}
\bpages{23--40}.
\bdoi{10.2307/3003485}
\end{barticle}
\endbibitem

\bibitem{Ross2002}
\begin{bbook}[author]
\bauthor{\bsnm{Ross},~\bfnm{Stephen~A}\binits{S.~A.}}, \bauthor{\bsnm{Westerfield},~\bfnm{Randolph}\binits{R.}} \AND \bauthor{\bsnm{Jaffe},~\bfnm{Jeffrey~F}\binits{J.~F.}}
(\byear{2002}).
\btitle{Corporate finance, 6th edition},
\bedition{6th edition} ed.
\bpublisher{Irwin/McGraw-Hill}.
\end{bbook}
\endbibitem

\bibitem{Rubin1980}
\begin{barticle}[author]
\bauthor{\bsnm{Rubin},~\bfnm{Donald~B}\binits{D.~B.}}
(\byear{1980}).
\btitle{Randomization analysis of experimental data: The Fisher randomization test comment}.
\bjournal{Journal of the American statistical association}
\bvolume{75}
\bpages{591--593}.
\end{barticle}
\endbibitem

\bibitem{Rubin2005}
\begin{barticle}[author]
\bauthor{\bsnm{Rubin},~\bfnm{Donald~B}\binits{D.~B.}}
(\byear{2005}).
\btitle{Causal Inference Using Potential Outcomes}.
\bjournal{Journal of the American Statistical Association}
\bvolume{100}
\bpages{322-331}.
\bdoi{10.1198/016214504000001880}
\end{barticle}
\endbibitem

\bibitem{Swoboda1994}
\begin{bbook}[author]
\bauthor{\bsnm{Swoboda},~\bfnm{Peter}\binits{P.}}
(\byear{1994}).
\btitle{Betriebliche Finanzierung},
\bedition{3rd edition} ed.
\bpublisher{Physica-Verlag}.
\end{bbook}
\endbibitem

\bibitem{Taddy2022}
\begin{bbook}[author]
\bauthor{\bsnm{Taddy},~\bfnm{M.}\binits{M.}}
(\byear{2022}).
\btitle{ISE Modern Business Analytics}.
\bpublisher{McGraw-Hill Education}.
\end{bbook}
\endbibitem

\bibitem{VanderWeele2018}
\begin{barticle}[author]
\bauthor{\bsnm{VanderWeele},~\bfnm{Tyler~J}\binits{T.~J.}} \AND \bauthor{\bsnm{Mathur},~\bfnm{Maya~B}\binits{M.~B.}}
(\byear{2018}).
\btitle{{SOME DESIRABLE PROPERTIES OF THE BONFERRONI CORRECTION: IS THE BONFERRONI CORRECTION REALLY SO BAD?}}
\bjournal{American Journal of Epidemiology}
\bvolume{188}
\bpages{617-618}.
\bdoi{10.1093/aje/kwy250}
\end{barticle}
\endbibitem

\bibitem{Hayek1975}
\begin{barticle}[author]
\bauthor{\bparticle{von} \bsnm{Hayek},~\bfnm{Friedrich~August}\binits{F.~A.}}
(\byear{1975}).
\btitle{The Pretence of Knowledge}.
\bjournal{The Swedish Journal of Economics}
\bvolume{77}
\bpages{433--442}.
\end{barticle}
\endbibitem

\bibitem{Wallis2019}
\begin{bincollection}[author]
\bauthor{\bsnm{Wallis},~\bfnm{Mark}\binits{M.}}, \bauthor{\bsnm{Kumar},~\bfnm{Kuldeep}\binits{K.}} \AND \bauthor{\bsnm{Gepp},~\bfnm{Adrian}\binits{A.}}
(\byear{2019}).
\btitle{Credit rating forecasting using machine learning techniques}.
In \bbooktitle{Managerial Perspectives on Intelligent Big Data Analytics}
\bpages{180--198}.
\bpublisher{IGI Global}.
\end{bincollection}
\endbibitem

\bibitem{Wasserbacher2022}
\begin{barticle}[author]
\bauthor{\bsnm{Wasserbacher},~\bfnm{Helmut}\binits{H.}} \AND \bauthor{\bsnm{Spindler},~\bfnm{Martin}\binits{M.}}
(\byear{2022}).
\btitle{Machine learning for financial forecasting, planning and analysis: recent developments and pitfalls}.
\bjournal{Digital Finance}
\bvolume{4}
\bpages{63--88}.
\bdoi{10.1007/s42521-021-00046-2}
\end{barticle}
\endbibitem

\bibitem{Welch2004}
\begin{barticle}[author]
\bauthor{\bsnm{Welch},~\bfnm{Ivo}\binits{I.}}
(\byear{2004}).
\btitle{Capital Structure and Stock Returns}.
\bjournal{Journal of Political Economy}
\bvolume{112}
\bpages{106-132}.
\bdoi{10.1086/379933}
\end{barticle}
\endbibitem

\bibitem{White2013}
\begin{barticle}[author]
\bauthor{\bsnm{White},~\bfnm{Lawrence~J.}\binits{L.~J.}}
(\byear{2013}).
\btitle{Credit Rating Agencies: An Overview}.
\bjournal{Annual Review of Financial Economics}
\bvolume{5}
\bpages{93-122}.
\bdoi{10.1146/annurev-financial-110112-120942}
\end{barticle}
\endbibitem

\bibitem{Wooldridge2010}
\begin{bbook}[author]
\bauthor{\bsnm{Wooldridge},~\bfnm{J.~M.}\binits{J.~M.}}
(\byear{2010}).
\btitle{Econometric Analysis of Cross Section and Panel Data, second edition}.
\bseries{The MIT Press}.
\bpublisher{MIT Press}.
\end{bbook}
\endbibitem

\bibitem{Wright2017}
\begin{barticle}[author]
\bauthor{\bsnm{Wright},~\bfnm{Marvin~N.}\binits{M.~N.}} \AND \bauthor{\bsnm{Ziegler},~\bfnm{Andreas}\binits{A.}}
(\byear{2017}).
\btitle{ranger: A Fast Implementation of Random Forests for High Dimensional Data in C++ and R}.
\bjournal{Journal of Statistical Software}
\bvolume{77}.
\bdoi{10.18637/jss.v077.i01}
\end{barticle}
\endbibitem

\bibitem{Zaid2020}
\begin{barticle}[author]
\bauthor{\bsnm{Zaid},~\bfnm{Mohammad A.~A.}\binits{M.~A.~A.}}, \bauthor{\bsnm{Wang},~\bfnm{Man}\binits{M.}}, \bauthor{\bsnm{T.~F.~Abuhijleh},~\bfnm{Sara}\binits{S.}}, \bauthor{\bsnm{Issa},~\bfnm{Ayman}\binits{A.}}, \bauthor{\bsnm{W.~A.~Saleh},~\bfnm{Mohammed}\binits{M.}} \AND \bauthor{\bsnm{Ali},~\bfnm{Farman}\binits{F.}}
(\byear{2020}).
\btitle{Corporate governance practices and capital structure decisions: the moderating effect of gender diversity}.
\bjournal{Corporate Governance: The International Journal of Business in Society}
\bvolume{20}
\bpages{939--964}.
\bdoi{10.1108/CG-11-2019-0343}
\end{barticle}
\endbibitem

\end{thebibliography}

\end{document}